\documentclass[chap]{thesis}
\usepackage[hang]{caption}      % to indent subsequent lines of captions
\usepackage{cite,citesort}
\usepackage{graphicx,amsfonts,amsmath}

\def \BEA {\begin{eqnarray}}
\def \EEA {\end{eqnarray}}
\def \BC {\begin{cases}}
\def \EC {\end{cases}}
\def \ve {\varepsilon}
\def \d {\delta}
\def \l {\lambda}
\def \wt {\tilde{\omega}}
\def \Tt {\tilde{T}}
\def \tt {\tilde{t}}
\def \w {\omega}

\def \a {\alpha}
\def \b {\beta}
\def \g {\gamma}
\def \at {\tilde{a}}
\def \af {\hat{a}}
\def \la {\langle}
\def \ra {\rangle}
\def \D {\Delta}

\def \i {\imath}
\def \eqN {\overset{N}{=}}
\def \eq1 {\overset{1}{=}}

\def \p {\partial}

\def \eb {\bar{e}}

\def \T {\theta}
\def \tinf {\theta_{\infty}}
\def \y {\check{y}}
\def \f {\hat{f}}

\begin{document}

%%%%%%%%%%%%%%%%%%%%%%%%%%%%%%%%%%%%%%%%%%%%%%%%%%%%%%%%%%%%%%%%%%%
%                                                                 %
%                            TITLE PAGE                           %
%                            PhD Thesis                           %
%                                                                 %
%%%%%%%%%%%%%%%%%%%%%%%%%%%%%%%%%%%%%%%%%%%%%%%%%%%%%%%%%%%%%%%%%%%

% Supply information for use on title page:
%
\thesistitle{\bf Characterization of thermalized Fermi-Pasta-Ulam chains}
\author{Boris Gershgorin}
\degree{Doctor of Philosophy}
\department{Mathematics}
\signaturelines{5}     %max number of signature lines is 7
\thadviser{Yuri V. Lvov}

%\cothadviser{David Cai} % If you have 2 thesis advisers
\memberone{David Cai}
\membertwo{Gregor Kovacic}
\memberthree{Peter Kramer}
\memberfour{Victor Roytburd}                   
\submitdate{June 2007\\(For Graduation August 2007)}
\copyrightyear{2007}   % if omitted, current year is used.

% Print titlepage and other prefatory material:
%
\titlepage
\abstitlepage          % required by Office of Graduate School (1 copy)
\copyrightpage         % optional
\tableofcontents
\listoffigures         % required if there are figures
   % titlepage material for PhD thesis
%%%%%%%%%%%%%%%%%%%%%%%%%%%%%%%%%%%%%%%%%%%%%%%%%%%%%%%%%%%%%%%%%%%
%                                                                 %
%                         ACKNOWLEDGEMENT                         %
%                                                                 %
%%%%%%%%%%%%%%%%%%%%%%%%%%%%%%%%%%%%%%%%%%%%%%%%%%%%%%%%%%%%%%%%%%%

\specialhead{ACKNOWLEDGMENT}

I would like to express my deep gratitude to my advisor Professor Yuri Lvov.
Without his enthusiasm, inspiration, wide knowledge, and logical thinking this thesis would not have been possible.
I also would like to thank Dr. Lvov for his friendly help and support during my five years in this PhD program.

I wish to thank Professor David Cai from Courant Institute whose motivation, thoughtful comments and deep knowledge of mathematics inspired me.

I express my gratitude to Professor Bernard Fleishman for encouraging me to apply to RPI and for his caring support during the PhD program.

I am grateful to all the Professors from the Department of Mathematical Sciences at RPI who taught me a great deal about applied mathematics.
In particular, thank you Dr. Kovacic, Dr. Roytburd, Dr. Kramer, Dr. Schwendeman and Dr. Isaacson.

Professor Sergey Nazarenko of the University of Warwick gave me the opportunity to learn more about the physical side of applied mathematics.

I appreciate the kind invitation and warm hospitality of Professor Paul Milewski from the University of Wisconsin, Professor Lenya Ryzhik from the University of Chicago,
and to Professor Vadim Zharnitsky from the University of Illinois, Urbana-Champaign during my visits to their respective Universities.

Dr. Naoto Yakoyama offered useful remarks and friendly support.

I am also very grateful to Dawnmarie Robens, Graduate Student Coordinator, and to Michele Kronau, Assistant to the Chair, for their help and for making the
environment in our department so warm and friendly.

I am thankful to my relatives who were very helpful: Myra and Alvin White, Miriam Schaffer, Naomi and James Collins, Ron Schaefer, Harvey and Lynn Kalish.

Most of all I am grateful to my parents and my brother for their endless love, support, and patience.
Their love and care were a source of energy for me during all these years.

%%%%%%%%%%%%%%%%%%%%%%%%%%%%%%%%%%%%%%%%%%%%%%%%%%%%%%%%%%%%%%%%%%%
%                                                                 %
%                            ABSTRACT                             %
%                                                                 %
%%%%%%%%%%%%%%%%%%%%%%%%%%%%%%%%%%%%%%%%%%%%%%%%%%%%%%%%%%%%%%%%%%%

\specialhead{ABSTRACT}

The Fermi-Pasta-Ulam (FPU) chains of particles in \textit{thermal equilibrium} are studied from both wave-interaction and particle-interaction points of view.
It is shown that, even in a strongly nonlinear regime, the chain in thermal equilibrium can be effectively described by a system of weakly
interacting \textit{renormalized} nonlinear waves.
These waves possess (i) the Rayleigh-Jeans distribution and (ii) zero correlations between waves, just as noninteracting free waves would.
This renormalization is achieved through a set of canonical transformations.
The renormalized linear dispersion of these renormalized waves is obtained and shown to be in excellent agreement with numerical experiments.
Moreover, a dynamical interpretation of the renormalization of the dispersion relation is provided via a self-consistency, mean-field argument.
It turns out that this renormalization arises mainly from the trivial resonant wave interactions, i.e., interactions with no momentum exchange.
Furthermore, using a multiple time-scale, statistical averaging method, we show that the interactions of near-resonant waves give rise to the broadening of the
resonance peaks in the frequency spectrum of renormalized modes.
The theoretical prediction for the resonance width for the thermalized $\beta$-FPU chain is found to be
in very good agreement with its numerically measured value.
Moreover, we show that the dynamical scenario for thermalized $\beta$-FPU chains is spatially highly localized discrete breathers riding chaotically on spatially
extended, renormalized waves.
We present numerical evidence of existence of discrete breathers in thermal equilibrium. 

\section{Thesis outline}
The thesis is organized as follows.
In Chapter~\ref{sect_fpu_nonlinear}, we provide a historic reference to nonlinear science and the FPU problem in particular.
In Chapter~\ref{sect_mechanics}, we present a short overview of the theory of Hamiltonian mechanics.
The methods of Hamiltonian mechanics will lead us to the dynamical description of the FPU chains.
In Chapter~\ref{sect_stat_phys}, a brief introduction to the equilibrium statistical mechanics is given.
There, we define the notion of microcanonical and canonical ensembles.
In Chapter~\ref{sect_WT}, we present a formalism used in wave turbulence that we will later apply to the FPU system in order to give it a wave description.
In Chapter~\ref{sect_symplectic}, we describe symplectic integrators, which are numerical algorithms that conserve the symplectic structure of Hamiltonian systems.
In Chapter~\ref{sect_chaos}, we give an introduction to chaos and discuss a few examples that demonstrate the chaotic behavior of nonlinear systems.
In Chapter~\ref{sect_renormalization}, we rewrite the $\beta$-FPU chain as an interacting four-wave Hamiltonian system.
We demonstrate how to describe a strongly nonlinear system as a system of waves that resemble free waves in terms of
the power spectrum and vanishing correlations between waves.
We show how to construct the corresponding renormalized variables with the renormalized linear dispersion.
In Chapter~\ref{sect_numerical}, we study the dynamics of the chain numerically and find excellent agreement between the renormalized dispersion,
obtained analytically (in Chapter~\ref{sect_renormalization}) and numerically.
In Chapter~\ref{sect_dispersion}, we describe the resonance manifold analytically and illustrate its controlling role in long-time averaged dynamics
using numerical simulation.
In Chapter~\ref{sect_self_consistency}, we derive an approximation for the renormalization factor for the linear dispersion using a self-consistency condition.
In Chapter~\ref{sect_width}, we study the broadening effect of frequency peaks  and predict analytically the form of the spatiotemporal spectrum for
the $\beta$-FPU chain.
We also provide the comparison of our prediction with the numerical experiments.
In Chapter~\ref{sect_breathers}, we discuss the energy localization in the $\beta$-FPU chain in the form of discrete breathers.
It was previously know that discrete breathers arise in the transient to thermal equilibrium under certain initial conditions.
Here, we numerically demonstrate that discrete breathers also persist in thermal equilibrium.
We present the conclusions in Chapter~\ref{sect_conclusions}.

\chapter{Fermi-Pasta-Ulam problem as a part of nonlinear science}
\label{sect_fpu_nonlinear}
In this chapter, we provide a general motivation for studying nonlinear science.
In particular, we discuss the Fermi-Pasta-Ulam problem, its history, formulation, and various implication in physics and mathematics.
\section{Nonlinear macroscopic systems}
Study of nonlinear systems is a rich and fast growing direction in science~\cite{Campbell_NS}.
Perhaps the key to the endless rich discoveries in nonlinear science arises from the fact that there are no
general methods and approaches that can be used to universally characterize \textit{any}
nonlinear system.
Indeed, it is hard to imagine that a general algorithm can be developed to solve typical nonlinear equations.
Since most of the phenomena that one finds in nature are nonlinear, one needs to look for special analytical methods of resolving particular nonlinear problems.
For example, the simplest model of a physical system that everyone studies in school, the mathematical pendulum, is a nonlinear system.
As we all know, in this case the \textit{linearization} provides a very good approximation of the pendulum motion in the small-amplitude regime.
Thus, one of the methods of treating a nonlinear system is to reduce it in some sense to a ``close'' linear system.
However, the main drawback of this simplification is in losing all the rich nonlinear phenomena after linearizing.
In addition, linearization does not always provide a physically meaningful approximation.
Consider a water flow in the pipe, when the velocity of the water is high and the flow is characterized as turbulent.
Then we can hardly approximate the flow by a linear laminar water motion.
Similarly, the formation of eddies and other spatially localized structures in the ocean can not be explained in the framework of spatially extended
linear dispersive waves.

Furthermore, many physical systems are not only nonlinear, they also have a large number of degrees of freedom.
For example, the number of the air molecules in a room is of the order of Avagadro number, which is approximately equal to $N_A\approx 6.02\times 10^{23}$.
Of the same order is the number of atoms in a crystal of a semiconductor.
In biology, the DNA consists of millions of base pairs of molecules.
In general, the study of many-body systems, or the so-called macroscopic systems, is a subject of statistical mechanics, which will be briefly introduced in the
following chapter.
All these and many other examples suggest that the study of many-body nonlinear systems can have important ramification in mathematics, physics, engineering,
and biology.
In order to characterize a behavior of a complicated physical system, one can first study its simplified model, which carries major features of its ``parental''
system while it is simple enough to be treated analytically or/and numerically.
The celebrated Fermi-Pasta-Ulam (FPU) chain is one such system.
%One of such systems is probably the most studied problem in nonlinear science --- .
\section{History}
\label{sect_fpu_history}
\begin{figure}
\includegraphics[scale=0.7]{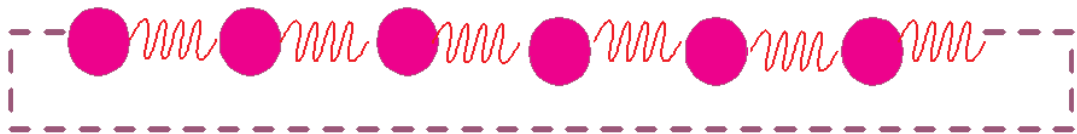}
\caption{FPU chain.}
\label{Fig_model}
\end{figure}

The study of discrete one-dimensional chains of particles with the nearest-neighbor interactions provides insight to the dynamics of various physical and biological
systems, such as crystals, wave systems, and biopolymers~\cite{fpu50,Peyrard,Toda}.
Let us first introduce the FPU model, which is going to be the main object of study in this thesis.
Consider a one-dimensional chain of identical particles coupled with identical \textit{nonlinear} springs as shown in Fig.~\ref{Fig_model}.
This chain of oscillators was first introduced in the numerical experiment designed by physicist Enrico Fermi,
computer scientist John Pasta, and mathematician Stan Ulam.
Their last names gave the acronym FPU.
The significance of the discovery made by FPU is manifested in a large number of theories and different branches of nonlinear science that
appeared in the last fifty years.
Here, we only present a concise description of the history of the problem, various attempts to resolve it, and some of its applications.
More detailed information can be found in~\cite{Ford}.

In the early 1950s, the computer MANIAC I (Mathematical Analyzer Numerical Integrator And Calculator) was built and awaiting a significant question to resolve.
FPU proposed to use MANIAC I to integrate a one dimensional many-particle system, i.e., the FPU chain described above.
The motivation for this experiment was to verify the fundamental beliefs of the statistical mechanics such as equipartition of energy among the degrees of freedom
and ergodicity.
Equipartition of energy means that all the degrees of freedom of the system contain the same amount of energy in average.
And the ergodic hypothesis states that the time average of an observable is equivalent to the phase space average, i.e., average over phase space
of the same nonlinear system taken at one time.
The FPU chain with the third order nonlinearity in the potential of the springs was simulated by a programmer named Mary Tsingou.
The results were extremely surprising.
Despite the expected equipartition of energy, a very regular almost periodic behavior was observed.
The system was initiated with only the first fundamental mode excited and after some time ($\sim$197 longest linear periods) the energy of the first fundamental
mode recovered to within $3\%$ of its initial value.
This phenomenon was called ``FPU recurrence''.
In 1955, the report of this experiment was distributed among the limited number of researches~\cite{FPU}.
However, before this preprint was published, Fermi died.
This sudden death prevented the preprint from being  published --- Pasta and Ulam could not publish the paper with Fermi's name on it since he had neither read or
approved the manuscript.
On the other hand, they could not publish it without his name since he was one of the creators of the whole idea.
The report was published with the collected papers of Fermi only a decade later.

The paradoxical behavior of the FPU chain initiated various attempts to resolve the problem.
Since the system exhibited a near-integrable behavior, the attempts to find a ``solution'' were made via approximating the system by completely integrable systems.
(Note that the system is called completely integrable if it can be described by a Hamiltonian that is a function of the momenta only~\cite{Ott}.)
Martin Kruskal and Norman Zabusky noticed that the continuum approximation of the discrete FPU problem is a famous Korteweg-deVries (KdV) equation
\BEA
u_t+uu_x+u_{xxx}=0.
\EEA
See Appendix~\ref{app_kdv} for the details on how to obtain the KdV equation using the small amplitude, long wavelength approximation of the FPU chain.
The KdV equation is completely integrable and, moreover, possesses special solutions called \textit{solitons}~\cite{KruskalZabusky}.

Solitons were first discovered in water canal by Scottish engineer John Scott Russell in 1834.
He was watching horses pull a barge along the Union Canal in Edinburgh when the rope to the barge broke.
The barge suddenly dipped into the water, which created a stable wave that set off up the canal with very little change in shape.
Russell rode his horse along the canal for several kilometers watching the wave's progress.
However, the significance of solitons in physics was only understood after the solitons were found to be the solutions of
PDE's that describe physical phenomena, such as the KdV equation.

Now we return to the connection of the KdV equation and the FPU chain.
The numerical integration of the KdV equation with periodic boundary condition and the initial condition in the form of the one cycle of the cosine
(the first Fourier mode as in FPU experiment) revealed an interesting dynamical behavior.
The cosine first transformed into a number of spatially localized pulses, solitons, which after some time superposed back to the initial cosine.
Although, this approach is only an approximation and does not provide a rigorous treatment of the FPU phenomenon, it gives an intuitive explanation of the
observed phenomenon.
However, the discovery of soliton that is a pure example of a coherent structure was significant by itself.
There are numerous examples of coherent structures in nature~\cite{Campbell_NS}: from the giant ($\sim10^8$ meters) Red Spot in the atmosphere of Jupiter to the
microstructures ($\sim 10^{-9}$ meters) in crystals.
After the solitons were found in nature, the whole class of completely integrable nonlinear partial differential equations became a
rapidly growing research subject in mathematical physics.

Further studies of the FPU model revealed that higher strength of nonlinearity induces irregular dynamics as opposed to the ordered, almost integrable
behavior observed in the initial experiment.
Thus, the study of another characteristic of many nonlinear systems, \textit{chaos}, was also influenced significantly as a result of the FPU discovery.
The existence of a certain threshold of the total system energy, called \textit{stochasticity threshold}, which roughly separates the regimes with near-integrable
and chaotic dynamics, was studied both numerically and analytically~\cite{Chirikov1,Chirikov2}.
If the total energy of the system is below the stochasticity threshold, then the recurrent behavior is observed as in the FPU experiment.
However, when the total system energy increases and takes values above the stochasticity threshold, the dynamics of the chain becomes chaotic, and eventually
the total energy becomes equally distributed among all the degrees of freedom as was expected in the FPU experiment.
A number of questions still remain open. The transition from the near-integrable to chaotic regime is not well understood.
In particular, the behavior of the stochasticity threshold is not known in the thermodynamic limit (i.e., when the number of degrees of freedom goes to infinity).
Moreover, the route to thermalization (and hence energy equipartition) in the chaotic regime is not fully characterized.

The near-integrable behavior of the FPU system with weak nonlinearity is closely intertwined with the celebrated Kolmogorov-Arnold-Moser theorem~\cite{Ott}.
The theorem essentially says that for a small perturbation of the non-degenerate integrable system most of the invariant tori survive.
That is to say, a small perturbation of the non-degenerate integrable system exhibits a near-integrable behavior.
If, however, the strength of the perturbation is increased, the invariant tori do break and a transition to chaos is observed.
The theorem was first formulated by A. Kolmogorov in 1954 (without a proof) and then proved independently by V. Arnold in 1963
(for analytic Hamiltonian systems) and by J. Moser in 1962 (for twist maps --- the area preserving maps with a phase twist that is radius-dependent).

\chapter{Hamiltonian Mechanics of discrete systems}
\label{sect_mechanics}
In this chapter, we discuss the classical mechanical approach of studying the behavior of a system of particles using the dynamical properties of the system.
A full and complete study of the classical mechanics can be found in~\cite{LL1,Licht}.
Suppose the system consists of $N$ particles, which are described by $N$ coordinates $q_j$ and $N$ momenta $p_j$, where
each component is a $d$-dimensional vector and $j$ is an integer in the range from $1$ to $N$. In this thesis, we will focus on a one dimensional system.
If the system does not interact with anything --- an isolated system as we will discuss in Chapter~\ref{sect_stat_phys} --- then it can be fully characterized by a
Hamiltonian $H(p,q)$, which is the total energy of the system.
The dynamics of such a system is governed by the following canonical equations of motion
\BEA
\BC
\dot q_j=\displaystyle{\frac{\partial H}{\partial p_j}},\\
\dot p_j=-\displaystyle{\frac{\partial H}{\partial q_j}}.
\EC\label{pqdot}
\EEA
It is often desirable to study the system in some new variables $(P,Q)$ different from the initial ones, $(p,q)$.
The transformation $(p,q)\rightarrow(P,Q)$ is called \textit{canonical} if it preserves the form of the canonical equations of motion,
i.e., Eq.~(\ref{pqdot}).

For the weakly nonlinear systems, it is often convenient to make a transformation from the physical space to the Fourier space since in the Fourier space the
system can be viewed as weakly interacting waves.
The discrete Fourier transformation is defined via
\BEA
\BC
Q_k=\displaystyle{\frac{1}{\sqrt{N}}}\sum_{j=1}^Nq_je^{\frac{2\pi \i kj}{N}},\\
P_k=\displaystyle{\frac{1}{\sqrt{N}}}\sum_{j=1}^Np_je^{\frac{2\pi \i kj}{N}}.
\EC\label{PQ}
\EEA
The Fourier transform of the real data, e.g., real $q_j$, has the following symmetry property
\BEA
Q_{N-k}=Q_k^*.\label{PQ_sym}
\EEA
Using Eq.~(\ref{PQ_sym}), we show that transformation~(\ref{PQ}) is canonical.
\BEA
\dot{Q}_k&=&\displaystyle{\frac{1}{\sqrt{N}}}\sum_{j=1}^N\dot{q}_je^{\frac{2\pi \i kj}{N}}=
\displaystyle{\frac{1}{\sqrt{N}}}\sum_{j=1}^N \frac{\partial H}{\partial p_j} e^{\frac{2\pi \i kj}{N}}=
\displaystyle{\frac{1}{\sqrt{N}}}\sum_{j,l=1}^N \frac{\partial H}{\partial P_l} \frac{\partial P_l}{\partial p_j} e^{\frac{2\pi \i kj}{N}}\nonumber\\
&=&\displaystyle{\frac{1}{N}}\sum_{j,l=1}^N \frac{\partial H}{\partial P_l} e^{\frac{2\pi \i (k+l)j}{N}}=\frac{\partial H}{\partial P_{N-k}}=
\frac{\partial H}{\partial P_k^*}
\EEA
Similarly, we can show that
\BEA
\dot{P}_k=-\frac{\partial H}{\partial Q_k^*}
\EEA
Yet another convenient transformation, which is often made, is a transformation to the so called \textit{normal modes}.
This is a linear transformation given by the following formula
\BEA
a_k=\frac{P_k-\i\w_kQ_k}{\sqrt{2\w_k}},\label{a}
\EEA
where $\w_k$ is an arbitrary positive function.
Next, we show that transformation~(\ref{a}) is canonical if and only if
\BEA
\w_k=\w_{N-k}.\label{w_canonical}
\EEA
We have
\BEA
\frac{\p a_k}{\p t}&=&\frac{1}{\sqrt{2\w_k}}\left(\frac{\p P_k}{\p t}-\i\w_k\frac{\p Q_k}{\p t} \right)=
\frac{1}{\sqrt{2\w_k}}\left(-\frac{\p H}{\p Q_k^*}-\i\w_k\frac{\p H}{\p P_k^*} \right)\nonumber\\
&=&\frac{1}{\sqrt{2\w_k}}\left(-\Big(\frac{\p H}{\p a_k^*}\frac{\p a_k^*}{\p Q_k^*}+\frac{\p H}{\p a_{N-k}}\frac{\p a_{N-k}}{\p Q_k^*}\Big)-
\i\w_k\Big(\frac{\p H}{\p a_k^*}\frac{\p a_k^*}{\p P_k^*}+\frac{\p H}{\p a_{N-k}}\frac{\p a_{N-k}}{\p P_k^*}\Big) \right)\nonumber\\
&=&\frac{1}{\sqrt{2\wt_k}}\left(\frac{\p H}{\p a_k^*}\Big(-\i\sqrt\frac{\w_k}{2}-\i\sqrt\frac{\w_k}{2}\Big)+
\frac{\p H}{\p a_{N-k}}\Big(\i\sqrt\frac{\w_{N-k}}{2}-\i\frac{\w_k}{\sqrt{2\w_{N-k}}}\Big)\nonumber
\right)
\EEA
Then the equation of motion takes the canonical form
\BEA
\i\dot{a}_k=\frac{\p H}{\p a_k^*},\label{adot}
\EEA
if and only if property~(\ref{w_canonical}) is satisfied.
Here we have presented a few formal facts from Hamiltonian mechanics.
We will use these transformations for studying the dynamical properties of the FPU chains.

\chapter{Equilibrium statistical physics}
\label{sect_stat_phys}
Statistical physics is a branch of physics that studies systems with a large number of degrees of freedom (macroscopic systems).
In principle, Newton's laws of motion, provide a way of describing the dynamics of all the particles in a given system.
However, this is practically impossible for most physical macroscopic systems (for example, gas in a room) due to an enormous number of equations that have
to be solved.
Even if it can be solved, the macroscopic behavior of these trajectories requires a different conceptual framework.
%Therefore, different methods should be used to characterize macroscopic systems.
Statistical physics provides such tools --- it uses the probabilistic approach, which is applicable in the case of a large number of particles in a system.
\section{Isolated systems and subsystems}
If a given system does not interact with any other systems, then it is referred to as an \textit{isolated} system.
From our everyday experience we know that an isolated system reaches \textit{equilibrium} state, i.e., the state when there are no temporal changes in any
characteristics of the system.
For example, if we pour some boiling water in the thermos then add a few ice cubes there and close the thermos, after some time the isolated system water+ice
will equilibrate:
the ice will melt and all the water in the thermos will have a uniform temperature, i.e., \textit{thermal equilibrium} will be reached.
And if instead of the boiling water it was hot tea, the tea substance will also be uniformly distributed if we wait long enough.
This shows that not only thermal but also chemical equilibrium is achieved.
As another example of equilibration, we consider air in a room.
Suppose, some perfume is sprayed in one corner of the room and then the room is left closed for some time.
Eventually, the air will mix and the perfume will be uniformly distributed over the whole room.
In these examples the use of the notion of an isolated system is consistent with the following facts.
In the first case, the thermos does not allow any thermal or chemical exchanges between the system water+ice with the outer world.
In the second case, walls, doors, windows, ceiling, and floor isolate the air in the room from the air outside the room so that the essential part of the perfume
stays inside the room for a long time.

Another type of system that is studied in statistical mechanics, is a system that is considered a part of a much large system.
In this case this smaller system is referred to as a \textit{subsystem} and the larger system is referred to as a \textit{thermal bath}.
It is important to point out that a subsystem should have a large number of degrees of freedom by itself.
This makes the resolution of the dynamical equations of motion for each individual particle of the subsystem practically impossible.
Therefore, a statistical approach should be applied to describe a subsystem.
As an example, we can again consider the above described system water+ice in a thermos.
However, now suppose that the thermos is not ideal and allows slow heat exchange between its content and the outer world.
Suppose that this thermos with water+ice is left in a large room with a freezing temperature.
Here, water+ice is a subsystem and the cold air in the large room is a thermal bath and the thermos plays a role of the interface between the subsystem and its
thermal bath.
First, the system water+ice will thermalize but then in a much longer time (which depends on the goodness of the thermos) the water inside the thermos will freeze.
Eventually, the content of the thermos will reach the temperature of the air in the cold room, i.e., of the thermal bath.
This example demonstrates that depending on the physical situation and the observation time scales,
the system water+ice can be either regarded as an isolated system or as a subsystem of a much larger system.
If the thermos is good at isolating its content from the outer world, then the system water+ice can be treated as isolated.
However, if the thermos is not perfect and the observation time is much longer than the time necessary to exchange heat between the content of the thermos and
the cold air in the room, then the system water+ice is treated as a subsystem, which interacts with the thermal bath.
\section{Liouville's theorem}
In order to provide a statistical description of the system in thermal equilibrium we introduce a notion of a \textit{phase space}.
Suppose a system has $N$ degrees of freedom, then its state at any time is described by $N$ coordinates and $N$ momenta, $q_j$ and
$p_j$, respectively, with an integer $j=\{1,...,N\}$, and the phase space is $2N$-dimensional space of points $(q,p)$.
Time evolution of the system produces a trajectory $(q(t),p(t))$ in the phase space.
Since the motion of a large system is very complex due to various interactions among the particles, the trajectory of the system after a long enough time
should ``cover'' all the possible states of the system in the phase space.
Mathematically it can be formalized by introducing the probability density function (pdf) of all the possible states in the phase space.
Consider a small volume element in the phase that contains all the points $(q,p)$ such that $q_j\in(q_j^0,q_j^0+dq_j)$ and $p_j\in(p_j^0,p_j^0+dp_j)$ for some
point $(q^0,p^0)$.
Denote $dw$ as a probability for the system to be in this small volume in the phase space.
The probability $dw$ can be expressed in terms of its pdf $\rho$ via
\BEA
dw=\rho(q^0,p^0)dqdp.\nonumber
\EEA
Here $\rho$ is normalized, so that
\BEA
\int\rho~dqdp=1.\label{norma1}
\EEA
Let us derive one of the fundamental properties of the pdf of an isolated system, Liouville's theorem, which states that the the pdf $\rho$ is constant along
the phase trajectories.
Formally, this statement comes from the conservation of the ``number'' of points in the phase space
\BEA
\frac{\partial \rho}{\partial t}+\mbox{div}(\rho v)=0,\label{continuous}
\EEA
where $v$ is the ``velocity''.
In our case, the velocity is given by $v=(\dot{q},\dot{p})^T$.
In the stationary state, we have
\BEA
\frac{\partial \rho}{\partial t}=0.\nonumber
\EEA
Then Eq.~(\ref{continuous}) becomes
\BEA
\sum_{j=1}^N\left[\frac{\partial}{\partial q_j}(\rho\dot{q}_j)+\frac{\partial}{\partial p_j}(\rho\dot{p}_j)\right]=0.\nonumber
\EEA
And finally, taking into account canonical equations of motion, Eq.~(\ref{pqdot}),
we obtain that the probability density function stays constant along the trajectories
\BEA
\frac{d\rho}{dt}=0,\label{Liouville}
\EEA
where the so called Lagrangian derivative is defined via
\BEA
\frac{d}{dt}=\frac{\p}{\p t}+\sum_{j=1}^N\left[\dot{q}_j\frac{\partial}{\partial q_j}+\dot{p}_j\frac{\partial}{\partial p_j}\right]
\EEA
A direct consequence of Liouville's theorem is that $\rho(q,p)$ is an integral of motion, and its logarithm is an \textit{additive} integral of motion.
If the system consists of two subsystems with pdf's $\rho_1$ and $\rho_2$ then the pdf of the whole system $\rho_{12}$ is a product of individual pdf's of
the subsystems, if the subsystems can be regarded as statistically independent.
Therefore, we have
\BEA
\ln\rho_{12}=\ln\rho_1+\ln\rho_2.\label{independent}
\EEA
There are in general seven independent additive integrals of motion as we know from mechanics.
They are energy, three components of momentum and three components of angular momentum.
Suppose we observe our system from a coordinate axis that is attached to the center of mass of the system, i.e., in this coordinate the total momentum vanishes.
By a similar trick, we can eliminate the total angular momentum.
Then, the only integral of motion of an isolated system would be its energy $E(q,p)$.
In the phase space, the motion of the isolated system becomes restricted to the surface that is given by the equation
\BEA
E(q,p)=E_0,\nonumber
\EEA
where $E_0$ is the value of the total system energy.
In the thermal equilibrium, we assume that all the points of this energy surface are equally probable.
Then the pdf takes form
\BEA
\rho=\mbox{const}\cdot\delta(E(q,p)-E_0).\label{ro_micro}
\EEA
The \textit{ensemble} of identical systems that are described by the probability measure~(\ref{ro_micro}) is called the \textit{microcanonical ensemble}.
\section{Entropy and the second law of thermodynamics}
Consider a small subsystem of a large isolated system in equilibrium.
It turns out that the pdf of the subsystem in thermal equilibrium can be obtained analytically.
In order to do this we have to introduce entropy and temperature of the system.
These quantities are related to each other and we start with entropy.
As was noted above, the only additive integral of motion of an isolated system is its total energy.
Therefore, for the subsystems of this isolated system we have
\BEA
\ln\rho_s=a_s+bE_s(q,p),\label{linear}
\EEA
from where we conclude that $\rho$ is a function of energy only.
Then the total energy of the subsystem will be concentrated around a constant value $\bar{E}$ and the fluctuations around $\bar{E}$ of the total energy of the
subsystem
will be very small.
Denote $\Delta q\Delta p$ to be the volume of the phase space that consists of the phase points with energies close to $\bar{E}$
\BEA
\rho(\bar{E})\Delta q\Delta p=1.\label{phase_volume}
\EEA
This volume $\Delta q\Delta p$ corresponds to the ``number'' of points of the phase space that have total energy close to $\bar{E}$ and
the trajectory of the system should most of the time be inside the volume $\Delta q\Delta p$.
Entropy is defined via
\BEA
S=\ln\frac{\Delta q\Delta p}{(2\pi\hbar)^N},\label{entropy_def}
\EEA
where $\hbar=6.58211915(56)\times10^{-16}~\mbox{eV}\cdot \mbox{s}$ is Planck's constant and the constant $(2\pi\hbar)^N$ ensures that entropy is a dimensionless
quantity. Actually this factor has a deeper physical meaning --- it is the smallest phase volume of a pair of degrees of freedom.
Intuitively, entropy is a measure of disorder in the system --- the higher values it takes, the more phase points have the energies close to $\bar{E}$.
In other words, with the same total energy, the state of the system with higher entropy is more disordered, i.e., has larger phase space volume to be in, than the
state with the lower entropy.
Let us obtain the expression for the entropy via the pdf $\rho$.
Substituting $\Delta q\Delta p$ from Eq.~(\ref{phase_volume}) into the definition of entropy~(\ref{entropy_def}), we have
\BEA
S=-\ln[(2\pi\hbar)^N\rho(\bar{E})].\label{entropy2}
\EEA
Next, using Eq.~(\ref{linear}) we obtain
\BEA
S=-\ln(2\pi\hbar)^N-(a+b\bar{E})=-\ln(2\pi\hbar)^N-\la\ln\rho(E)\ra=-\la\ln[(2\pi\hbar)^N\rho(E)]\ra,\nonumber
\EEA
where $\la\dots\ra$ denotes ensemble averaging over the probability measure given by $\rho$
Thus we have derived an equivalent definition of entropy
\BEA
S=-\la\ln[(2\pi\hbar)^N\rho]\ra=-\int\rho\ln [(2\pi\hbar)^N\rho]dqdp\label{entropy_ave},
\EEA

From all observations it is known that if an isolated system is not in equilibrium then it will approach equilibrium as time progresses.
This general law of nature is formalized by the \textit{entropy law} or \textit{the second law of thermodynamics}:\\
\textbf{If an isolated system is not in equilibrium at some moment of time then at subsequent moments of time its entropy will most probably monotonically increase.}
We have to point out that the statement of the entropy law as it is given here is not in contradiction with the time reversibility of the dynamical equations
of motion, since here we only talk about the most probable state of the system.
However, the issue with time reversibility of the dynamical equations and time irreversibility of the experimentally observed property of entropy, i.e.,
that entropy \textit{never} decreases (if we disregard micro-fluctuations) is a deeper open issue and we will not pursue this question here.
For our purpose, we will only need the fact that in thermal equilibrium entropy has its maximum.
\section{Temperature and Gibbs measure}
\label{sect_temperature}
Let us introduce one of the most important thermodynamic quantities, i.e., \textit{temperature}.
Suppose an isolated system consists of two subsystems, which are in equilibrium with each other.
Denote $E_1$ and $E_2$ to be total energies of each subsystem and $S_1$ and $S_2$ to be entropies of each subsystem.
For the total energy and entropy we have
\BEA
E&=&E_1+E_2,\nonumber\\
S&=&S_1(E_1)+S_2(E_2).\nonumber
\EEA
Since $E$ is a constant and $E_2=E-E_1$, total entropy $S$ is a function of $E_1$ only.
The necessary condition for $S$ to have maximum in thermal equilibrium is
\BEA
\frac{dS}{dE_1}=\frac{dS_1}{dE_1}+\frac{dS_2}{dE_2}\frac{dE_2}{dE_1}=\frac{dS_1}{dE_1}-\frac{dS_2}{dE_2}=0
\EEA
This property of the entropy of the subsystems is easily generalizable to any number of subsystems.
Therefore, we can define the temperature $T$ as
\BEA
\frac{dS}{dE}=\frac{1}{T}.\label{temperature}
\EEA
The temperatures of the two systems in thermal equilibrium are equal
\BEA
T_1=T_2.\nonumber
\EEA

Now we obtain the pdf of the subsystem, which is equilibrium with its thermal bath.
Suppose that the subsystem together with the thermal bath is an isolated system.
Then the microcanonical distribution for this combined system is described by
\BEA
dw=\mbox{const}\cdot\delta(E+E'-E^{(0)})~dqdpdq'dp',\label{micro_all}
\EEA
where $E$, $E'$, and $E^{(0)}$ are the energies of the subsystem, of the thermal bath, and the combined system, respectively.
Then the probability density function for the subsystem becomes
\BEA
\rho=\mbox{const}\int\delta(E+E'-E^{(0)})~dq'dp'.\label{rho_pq}
\EEA
The integrand in Eq.~(\ref{rho_pq}) depends only on $E'$, therefore, we can change the integration variables from $q',p'$ to $E'$ via
\BEA
dq'dp'\rightarrow \frac{\Delta q'\Delta p'}{\Delta E'}dE'.\label{change}
\EEA
Note that $\Delta E'$ is the length of the energy interval that corresponds to the phase space volume given by $\Delta q'\Delta p'$.
Using the entropy definition given in Eq.~(\ref{entropy_def}) we obtain
\BEA
\rho=\mbox{const}\int\frac{e^{S'}}{\Delta E'}\delta(E+E'-E^{(0)})dE'.\label{rho_E1}
\EEA
In equilibrium, the fluctuations of $E'$ are very small and the pdf of the energy of the thermal bath as a function of $E'$ has a sharp peak around its average value.
And the width of this peak $\Delta E'$ is practically independent of the energy of the subsystem $E$.
Therefore, we can use the value $E'=E^{(0)}$ in $\Delta E'$ and after integration Eq.~(\ref{rho_E1}) becomes
\BEA
\rho=\mbox{const}\cdot e^{S'}\vline_{E'=E^{(0)}-E}.\label{rho_E2}
\EEA
Since $E$ is small compared to $E^{(0)}$, we use the Taylor expansion with small parameter $E$ in $S'(E^{(0)}-E)$
\BEA
S'(E^{(0)}-E)=S'(E^{(0)})-E\frac{dS'(E^{(0)})}{dE^{(0)}}.\label{Taylor_S}
\EEA
After combining Eqs.~(\ref{rho_E2}) and~(\ref{Taylor_S}) and using the definition of temperature~(\ref{temperature})
we obtain the following form for the pdf of the subsystem
\BEA
\rho(q,p)=Z\exp\left(-\frac{E(q,p)}{T}\right),\label{Gibbs_def}
\EEA
and the so called \textit{partition function} $Z$ is defined from the normalization condition~(\ref{norma1})
\BEA
Z=\int \exp\left(-\frac{E(q,p)}{T}\right)~dqdp.\label{partitionZ}
\EEA
Distribution given by Eq.~(\ref{Gibbs_def}) was found by Gibbs in 1901 and is referred to as the \textit{Gibbs distribution} or \textit{canonical distribution}
or more commonly Boltzman distribution.

The explicit form of the pdf given by Eq.~(\ref{Gibbs_def}) is very convenient when the average characteristics of the various dynamical quantities have to be
computed.
As a simple example, we consider an atom with mass $m$ and Hamiltonian
\BEA
H=\frac{1}{2m}p^2.\label{1atomH}
\EEA
According to Eq.~(\ref{Gibbs_def}), the probability measure has the form
\BEA
dw=\frac{1}{Z}\exp\left(-\frac{1}{2mT}p^2\right)dp.
\EEA
The average value of the kinetic energy of the atom can be easily computed,
\BEA
\left\langle\frac{p^2}{2m}\right\rangle=\frac{\int p^2\exp\left(-\frac{1}{2mT}p^2\right)dp}{2m\int \exp\left(-\frac{1}{2mT}p^2\right)dp}=\frac{T}{2}.\label{p2_equip}
\EEA
This results is generalized in the \textit{equipartition theorem}, which states that for the system in thermal equilibrium described by a quadratic Hamiltonian
the following relationship holds
\BEA
\left\la x_m\frac{\p H}{\p x_n}\right\ra=\delta^m_nT,\label{equipartition}
\EEA
where $x_j$ is a degree of freedom, i.e., one of $q_j$ or $p_j$.

However it is necessary to point out that practically both microcanonical and canonical distributions become identical when the number of degrees of freedom goes to
infinity.
The only difference between these distributions arises when one computes the fluctuations of the total energy around its average value.
In the canonical distribution, these fluctuations are non-zero.
On the contrary, in the microcanonical distribution, they are zero by definition.
And as the number of degrees of freedom grows, the fluctuations of the total energy decrease as $\sim1/\sqrt{N}$~\cite{Reichl}.
From the practical point of view, the calculations using the canonical distribution are much easier mathematically.
We will use both notions of the microcanonical and canonical distribution when we study the statistical behavior of the FPU system from the
wave point of view and we will see the equivalence of both approaches.

\chapter{Wave Turbulence}
\label{sect_WT}
Wave turbulence theory (WT) studies a statistical state of a system of nonlinear dispersive waves that weakly interact with each other,
and their dynamics is described statistically.
WT has been used for almost eighty years to provide a statistical description of various physical systems.
Peierls initiated the methods of WT in~\cite{Peierls}, in which the kinetic equation for phonons in solids was obtained.
Among other examples of WT are ocean, atmosphere, plasmas and Bose-Einstein condensates~\cite{ZLF,Hasselmann,Pushkarev,Benney}.
Perhaps the key discoveries in WT are made in application to oceanography~\cite{ZLF,Zakharov1,Zakharov2} by V.E. Zakharov et al.
There, it was argued that systems of dispersive waves develop a Kolmogorov type of turbulence in non-equilibrium state as opposed to the thermalized state
as it was studied before.
\textit{Kolmogorov-Zakharov non-equilibrium spectra} that predict cascades of various excitations, play a central role in the modern development of WT.
These spectra arise in the systems that are driven away from equilibrium by forcing and damping.
The non-equilibrium situations are the main focus of WT.
However, in this thesis we only discuss the thermal equilibrium state of the FPU system.
Nevertheless, we will use the ideas and methods that are commonly used in WT.
Therefore, we provide a brief description of WT here.

In the general setting of WT, dispersive waves are governed by a Hamiltonian, such as
\BEA
H=\int\w_k|a_k|^2~dk+\frac{1}{2}\int T^{kl}_{ms}a_k^*a_l^*a_ma_s\delta^{kl}_{ms}~dkdldmds,\label{WTH}
\EEA
where $a_k(t)$ describes the evolution of the $k$th wave mode in time, $\w_k$ is a linear dispersion,
and $T^{kl}_{ms}$ is an interaction tensor coefficient, which is considered to be small in the case of the weak coupling.
The formal procedure for obtaining the Hamiltonian of type~(\ref{WTH}) is given in Chapter~\ref{sect_mechanics} and a more comprehensive discussion is
provided in~\cite{ZLF}.
In Chapter~\ref{sect_dispersion}, we will derive a discrete form of Hamiltonian~(\ref{WTH}) for the $\beta$-FPU chain [Eq.~(\ref{H_a})].
In Eq.~(\ref{WTH}), we consider only the fourth order interactions among the waves.
The $\beta$-FPU system that we will study is of the same type with quartic potential interactions.
WT aims to derive the kinetic equation for the power spectrum defined by
\BEA
n_k=\langle|a_k|^2\rangle,\label{WTn}
\EEA
where $\la\dots\ra$ stands for averaging over an ensemble of initial data.
This is usually achieved by combining the statistical description of the wave field $a_k$ with its dynamic description using the Hamiltonian~(\ref{WTH}).

It is assumed that the wave field $a_k$ is \textit{near-Gaussian} if the nonlinear interactions are weak.
The near-Gaussian assumption leads to the following approximations of the fourth and sixth order correlators
\BEA
\la a_k^*a_l^*a_\a a_\b\ra&=&n_kn_l(\d^k_\a\d^l_\b+\d^k_\b\d^l_\a),\label{WT4corr}\\
\la a_k^*a_l^*a_m^*a_\a a_\b a_\g \ra&=&n_kn_ln_m(\d^k_\a(\d^l_\b\d^m_\g+\d^l_\g\d^m_\b)+\d^k_\b(\d^l_\a\d^m_\g+\d^l_\g\d^m_\a)\label{WT6corr}\\
& &~~~~~~~~~~~+\d^k_\g(\d^l_\a\d^m_\b+\d^l_\b\d^m_\a)).\nonumber
\EEA
These approximations are crucial in making the closure in the hierarchy of equations for each order of small parameter.

Now we outline the main steps in deriving the four-wave kinetic equation.
The dynamical evolution of $a_k(t)$ is described by the following equation
\BEA
\i \dot{a}_k=\frac{\delta H}{\delta a_k^*}=\w_ka_k+\int T^{kl}_{ms}a_l^*a_ma_s\delta^{kl}_{ms}~dldmds.\label{WTeq}
\EEA
We compute the time evolution of the wave action
\BEA
\dot{n}_k=2\Im\int T^{kl}_{ms}J^{kl}_{ms}\d^{kl}_{ms}~dldmds,\label{WTJ}
\EEA
where
\BEA
J^{kl}_{ms}=\la a_k^*a_l^*a_ma_s\ra.\label{WT_J}
\EEA
Now, we use the approximation of the fourth order correlator and in the zeroth order we obtain
\BEA
\dot{n}_k=2n_k\Im\int T^{kl}_{kl}n_l~dl.\label{WTn0}
\EEA
In Eq.~(\ref{WTn0}), the integrand is real and therefore the RHS is equal to zero.
Therefore, in order to find the higher order contribution, we should not split the fourth order correlator and should consider its time evolution as a whole.
We obtain the following dynamic equation
\BEA
\left(\i\frac{\p}{\p t}+(\w_k+\w_l-\w_m-\w_s)\right)J^{kl}_{ms}=2\bar{T}^{kl}_{ms}\big(n_mn_s(n_k+n_l)-n_kn_l(n_m+n_s)\big).
\EEA
Now we use the notion of the separation of linear and nonlinear time scales.
We suppose that the integral contribution of the fast oscillations vanishes as time increases, which yields
\BEA
J^{kl}_{ms}=\frac{2\bar{T}^{kl}_{ms}\big(n_mn_s(n_k+n_l)-n_kn_l(n_m+n_s)\big)}{\w_k+\w_l-\w_m-\w_s+\i\d},
\EEA
where the small term $\i\d$ was added.
Physically, the term $\i\d$ represents friction, which is always present in any realistic system.
However, this theoretic treatment has a profound effect on wave statistical dynamics and it is consistent with the fact that the resonances dominate the long time
dynamics.
Next, we use the following equality
\BEA
\Im\left(\frac{1}{\w+\i\d}\right)=-\pi\d(w).
\EEA
Finally, we obtain the four-wave kinetic equation
\BEA
\dot{n}_k=2\pi\int|T^{kl}_{ms}|^2\big(n_mn_s(n_k+n_l)-n_kn_l(n_m+n_s)\big)\d^{kl}_{ms}\d(\w^{kl}_{ms})~dldmds,
\EEA
where $\w^{kl}_{ms}=\w_k+\w_l-\w_m-\w_s$.
We note that the four-wave resonance conditions arise here: only those quartets of waves that are on or very close to the resonance manifold provide
effectively mix of energy through the four wave interactions.
And the resonance manifold is described by the set of wave numbers that satisfy
\BEA
\BC
k+l=m+s,\\
\w_k+\w_l=\w_m+\w_s.
\EC
\label{WT4w}
\EEA

We will study the interaction of waves in the thermalized state of the FPU system also using near-Gaussian assumption and notion of separation of linear
and nonlinear time-scales.
We will also study the resonances manifold of the FPU chain, which is described by the discrete version of the Eq.~(\ref{WT4w})

\chapter{Symplectic integrators for Hamiltonian systems}
\label{sect_symplectic}
With the growing power of computers, it becomes more feasible to numerically simulate
many-body systems with a large number of degrees of freedom.
However, usual numerical algorithms such as Euler scheme or Runge-Kutta methods do not conserve the total energy of a Hamiltonian system for long time simulations.
Hamiltonian systems possess symplectic structures, which we define below.
Therefore, efficient and precise algorithms that capture this crucial characteristic of Hamiltonian system, are needed.
Here, we describe a class of such algorithms called \textit{symplectic integrators} and provide all the necessary formulas and parameters for one
particular method, i.e., the sixth order Yoshida method~\cite{Yoshida}.

Suppose, we study a system of $N$ particles.
At any given time the position of the system is described by the set of $d\cdot N$ coordinates and $d\cdot N$ momenta, which we
will denote as a pair $(q,p)$ assuming that both components are $d$-dimensional vectors.
In Hamiltonian mechanics, the evolution of the system is given by the Hamiltonian function $H(p,q)$ via
the canonical equations of motion~(\ref{pqdot}).
These equations define a time flow of the phase space --- to find the position $(q_\tau,p_\tau)$ of the system at time $\tau$
one can integrate Eq.~(\ref{pqdot}) up to $t=\tau$ with the initial conditions $(q_0,p_0)$ given at $t=0$.
By definition, the time flow of the phase space is symplectic if it preserves the differential form
\BEA
\w=dp\wedge dq.\label{sympl_wedge}
\EEA
In order to construct the symplectic integrator, we provide a formal description of the Hamiltonian flow given by Eqs.~(\ref{pqdot}).
Define $z=(p,q)$ and the Poisson bracket $\{\cdot,\cdot\}$ as
\BEA
\{f,g\}=\frac{\partial f}{\partial q}\frac{\partial g}{\partial p}-\frac{\partial f}{\partial p}\frac{\partial g}{\partial q}.\label{sympl_poisson}
\EEA
Then Eqs.~(\ref{pqdot}) can be written in the form
\BEA
\dot z=\{z,H(z)\}.\label{sympl_zdot1}
\EEA
By introducing the notation $D_H=\{\cdot,H(\cdot)\}$ for the differential operator, Eq.~(\ref{sympl_zdot1}) can be rewritten as
\BEA
\dot z=D_Hz,\label{sympl_zdot2}
\EEA
and its formal solution $t=\tau$ is given by
\BEA
z(\tau)=e^{\tau D_H}z(0).\label{sympl_zt}
\EEA
Since the total energy is conserved, we can write
\BEA
H(z(\tau))=H(z(0)).\label{HtH0}
\EEA
Suppose, the total energy of the system is a sum of the kinetic energy that is a function of $p$ only and a potential energy that is a function of $q$ only
\BEA
H(q,p)=T(p)+V(q).\label{sympl_HTV}
\EEA
Then the corresponding differential operators are denoted as
\BEA
\BC
D_T=\frac{\partial T}{\partial p}\frac{\partial}{\partial q},\\
D_V=-\frac{\partial V}{\partial q}\frac{\partial}{\partial p},
\EC\label{D_TV}
\EEA
and thus the formal solution~(\ref{sympl_zt}) becomes
\BEA
z(\tau)=e^{\tau(D_T+D_V)}z(0).\label{sympl_ztTV}
\EEA
Since the differential operators $D_T$ and $D_V$ are non-commutative, exponential of the sum of two operators in Eq.~(\ref{sympl_ztTV}) is \textit{not}
equal to the product of exponentials of the individual components.
Instead, for any non-commutative differential operators $A$ and $B$, the approximate relationship holds
\BEA
e^{\tau(A+B)}=e^{\tau A}e^{\tau B}+o(\tau^2).\label{sympl_exp1}
\EEA
Our goal is to build the $n$-th order symplectic scheme, and in order to achieve this we generalize Eq.~(\ref{sympl_exp1}) and construct the expansion of the form
\BEA
e^{\tau(A+B)}=e^{c_1\tau A}e^{d_1\tau B}e^{c_2\tau A}e^{d_2\tau B}\times\dots\times e^{c_k\tau A}e^{d_k\tau B}+o(\tau^{n+1}),\label{sympl_expn}
\EEA
where $c_1\dots c_k$ and $d_1\dots d_k$ are real numbers.
Now, the solution for $z(\tau)$ is approximated by
\BEA
z'(\tau)=\left(\prod_{j=1}^k e^{c_j\tau A}e^{d_j\tau B} \right)z(0),\label{sympl_cut}
\EEA
Note that Eq.~(\ref{sympl_cut}) provides a \textit{symplectic mapping} in the phase space, i.e., because it consists of a series of
elementary symplectic mappings $e^{c_j\tau A}$ and $e^{d_j\tau B}$.
Moreover, using Taylor expansions of the operators~(\ref{D_TV}) up to the first order in $\tau$, the corresponding  elementary mappings are explicitly computable
\BEA
q^{(j)}&=&q^{(j-1)}+\tau c_j\frac{\partial T}{\partial p}(p^{(j-1)}),\label{sympl_dq}\\
p^{(j)}&=&p^{(j-1)}-\tau d_j\frac{\partial V}{\partial q}(q^{(j)}),\label{sympl_dp}
\EEA
for $j=1$ to $j=k$.
Naturally, the question arises whether the expansion of the form~(\ref{sympl_expn}) exists for any order $n$.
It turns out that for every \textit{even} order $n$ there exists at least one set of exact coefficients $c_1\dots c_k$ and $d_1\dots d_k$ so that Eq.~(\ref{sympl_cut})
provides an order $n$ approximate solution for the dynamical equation~(\ref{sympl_zdot1}).
The general approach of finding the coefficients $c_j$ and $d_j$ is as follows.
We expand the LHS of Eq.~(\ref{sympl_expn}) in powers of $\tau$ up to the order $n$.
Then we equate the coefficients of the corresponding powers on both sides of Eq.~(\ref{sympl_expn}) and obtain a system of equations for $c_j$ and $d_j$.
The resulting coefficients for $n=6$ are given by~\cite{Yoshida}
\BEA
 c_1&=&0.392256805238780~~~~~d_1= 0.784513610477560,\nonumber\\
 c_2&=&0.510043411918458~~~~~d_2= 0.235573213359357,\nonumber\\
 c_3&=&-0.471053385409757~~~d_3=-1.177679984178870,\nonumber\\
 c_4&=&0.068753168252518~~~~~d_4= 1.315186320683906,\nonumber\\
 c_5&=&0.068753168252518~~~~~d_5=-1.177679984178870,\nonumber\\
 c_6&=&-0.471053385409757~~~d_6= 0.235573213359357,\nonumber\\
 c_7&=&0.510043411918458~~~~~d_7= 0.784513610477560,\nonumber\\
 c_8&=&0.392256805238780~~~~~d_8= 0.0,\nonumber
\EEA
To summarize, here we presented the numerical algorithm for solving the Hamiltonian equations of motion~(\ref{pqdot}) using the symplectic
 algorithm given by Eqs.~(\ref{sympl_dq}) and~(\ref{sympl_dp}).
Note that strictly speaking a symplectic algorithm may not preserve energy as well as an explicit Runge-Kutta method for a fixed time step $\tau$.
However, it is superior to an explicit Runge-Kutta method since the system energy is bounded when computed by a symplectic algorithm in contrast to the
unbounded energy when computed by an explicit Runge-Kutta method.
This property of a symplectic algorithm becomes important when a long time (statistical) behavior of a Hamiltonian system is studied.

\chapter{Chaos in dynamical systems}
\label{sect_chaos}
Here we present a few general facts from the theory of \textit{chaos}, which provides a language to describe random, turbulent, irregular behavior
of the dynamical systems.
A systematic and comprehensive discussion of chaos can be found, e.g., in~\cite{Ott}.
Chaos arises in many dynamical systems.
It was first noticed by H. Poincar\'{e} when he studied the motion of the system of three celestial bodies
and realized that very complex trajectories could arise.
Thereafter, such scientists as G. Birkoff, M. Cartwright and J. Littlewood, S. Smale and A. Kolmogorov and many others investigated various aspects of chaotic dynamics.
However, the importance of chaos was really appreciated for understanding real complex physical phenomena only when the computational power became widely available.
\section{Logistic map}
We start with an example of discrete maps in which both regular and chaotic behaviors can be observed.
The \textit{logistic map} is given by the following nonlinear transformation
\BEA
x_{n+1}=f(x_n),
\EEA
where
\BEA
f(x)=4\lambda x(1-x).\label{chaos_logistic}
\EEA
and $\lambda$ is a controlling (bifurcation) parameter.
The map has two fixed points $x_1=0$ and $x_2=1-1/4\lambda$.
For the parameter range $\lambda>1/4$, the fixed point $x_1$ is unstable.
For $\l\in(1/4,3/4)$, the fixed point $x_2$ is stable and for all other $\l$ it is unstable.
Here, the fact that the fixed point is stable means that a trajectory that is initiated at any point except for the set of point of the Lebesgue measure zero
will eventually converge to the fixed point. Then this fixed point is the \textit{attractor} of the map.
However, the attractor of the map can be more complicated than just a point.
The attractor, i.e., the set of points to which the set of all possible initial points converges (or becomes attracted), can be a very complicated object such
as a fractal, i.e., whose Hausdorf dimension~\cite{Froyland} is not an integer (then it is called a \textit{strange attractor}).
We will observe the route of the logistic map to chaos with the change of the parameter $\l$.

\begin{figure}
\includegraphics[scale=0.8]{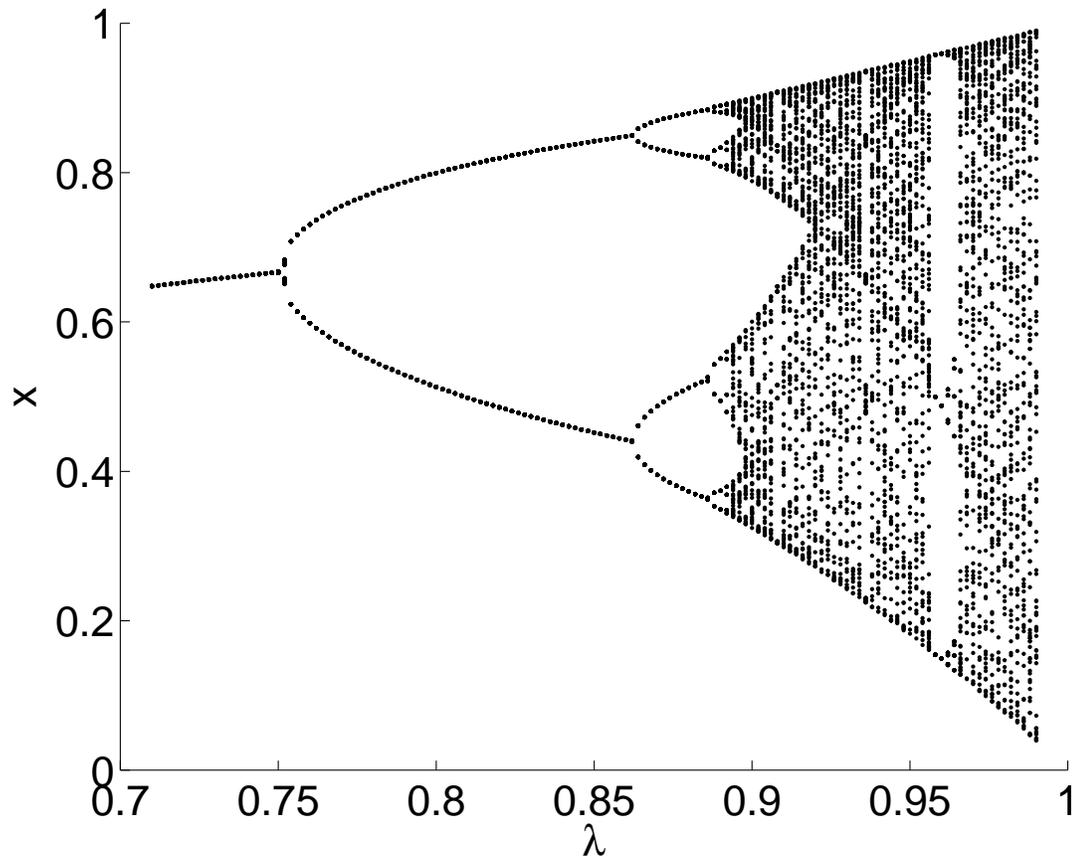}
\caption{Attractor of the logistic map as a function of the parameter $\l$.}
\label{Fig_bifur}
\end{figure}

In Fig.~\ref{Fig_bifur}, we show the attractor of the logistic map~(\ref{chaos_logistic}) as a function of $\l$.
We observe that when $\l$ crosses the value $\l=3/4$, the system encounters a \textit{pitch-fork} or \textit{period doubling bifurcation}~\cite{Froyland} ---
the single line that represents the fixed point solution splits into two lines that represent a period two orbit.
Then the attractor becomes two points instead of one in the case of a fixed point.

After $\l$ is increased further, the bifurcation happens again but now to each of the two branches.
After the $n$th bifurcation, the length of the period is $2^n$.
As $n$ goes to infinity, the period becomes infinitely large and the system becomes chaotic.
Here, this happens at $\l_{\infty}\approx 0.89286...$
For most $\l>\l_{\infty}$ we have a case of a strange attractor.
However, periodic orbits still exist even for $\l>\l_{\infty}$.
But most of the orbits are \textit{chaotic} for $\l>\l_{\infty}$.

In general, a chaotic orbit is characterized by the following properties
\begin{itemize}
    \item it is not periodic,
    \item it is not attracting to a periodic limiting set,
    \item it has sensitive dependence on the initial conditions.
\end{itemize}
A convenient diagnostic of the sensitive dependence on the initial conditions (or small orbit perturbations) is the Lyapunov exponent.
Lyapunov exponent gives an average exponential rate of divergence of the two trajectories with close initial conditions.
If we denote $dx_n=x_{n+1}-x_n$ then $h$ is a Lyapunov exponent if
\BEA
dx_n\sim\exp(nh)dx_0
\EEA
Thus we come to the strict definition of a Lyapunov exponent
\BEA
h=\lim_{n\rightarrow\infty}\frac{1}{N}\ln|\Lambda_N|,\label{lyapunov_def}
\EEA
where
\BEA
\Lambda_n=f'(x_N)f'(x_{N-1})\dots f'(x_1)
\EEA
We note that if $h>0$, then two initially close trajectories will diverge with the rate $h$, i.e., the case $h>0$ indicates chaos.
\begin{figure}
\includegraphics[scale=0.8]{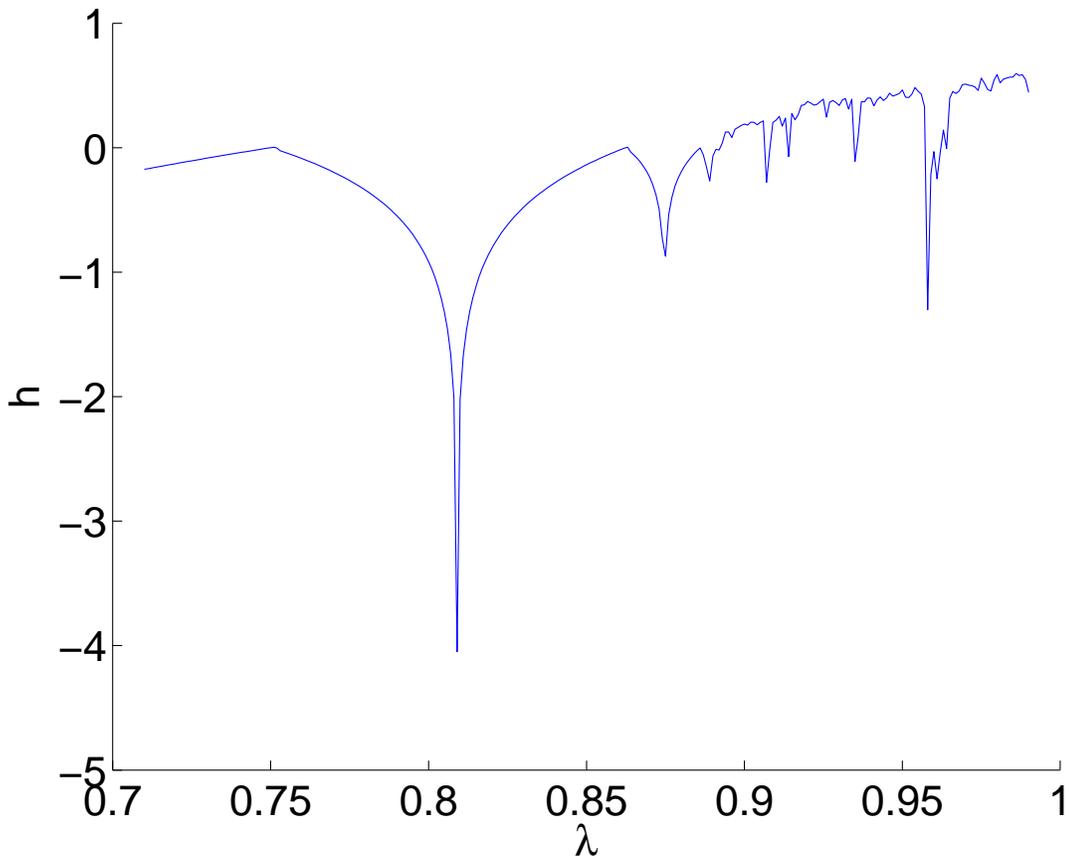}
\caption{Lyapunov exponent of the logistic map as a function of the parameter $\l$.}
\label{Fig_lyap}
\end{figure}
In Fig.~\ref{Fig_lyap}, we demonstrate the dependency of the Lyapunov exponent of the logistic map on the controlling parameter $\l$.
Note that the periodic orbits correspond to the dips ($h<0$) in the plot and the bifurcation points correspond to the zero values of the Lyapunov exponents.
For the continuous trajectories we have the following generalization of the Lyapunov exponent
\BEA
h=\int\ln|M'(x)|d\mu(x),
\EEA
where $|M'(x)|$ is a jacobian of the transformation $x_{n+1}=M(x_n)$ in the continuous limit $dx_n\rightarrow0$ and
$\mu(x)$ is an \textit{invariant measure}~\cite{Ott}.

To summarize, we have considered an example of a discrete map.
We have observed that by changing the controlling parameter $\l$, the orbits of the logistic map exhibit a change of behavior from regular with
a periodic limiting cycle to chaotic with a strange attractor.
Using this example, the important notions of the theory of chaos such as attractor and Lyapunov exponent are introduced.
\section{Lorenz model}
We continue our discussion of chaotic systems with a continuous autonomous dynamical system --- the Lorenz model.
The model was originally obtained by truncating the Fourier expansion of the Navier-Stokes equations.
This model also exhibits a change of behavior from regular to chaotic as the controlling parameter changes.
The model is given by the following equations
\BEA
\dot{x}&=&-\sigma(x-y),\\
\dot{y}&=&(r-z)x-y,\\
\dot{z}&=&xy-bz,
\EEA
where $x$, $y$, and $z$ are real functions of time and $\sigma$, $r$, and $b$ are real positive parameters.
We follow~\cite{Froyland} and choose $\sigma=10$ and $b=8/3$ for our numerical experiments.
This leaves $r$ to be the only parameter that is varied.

\begin{figure}
\includegraphics[scale=0.8]{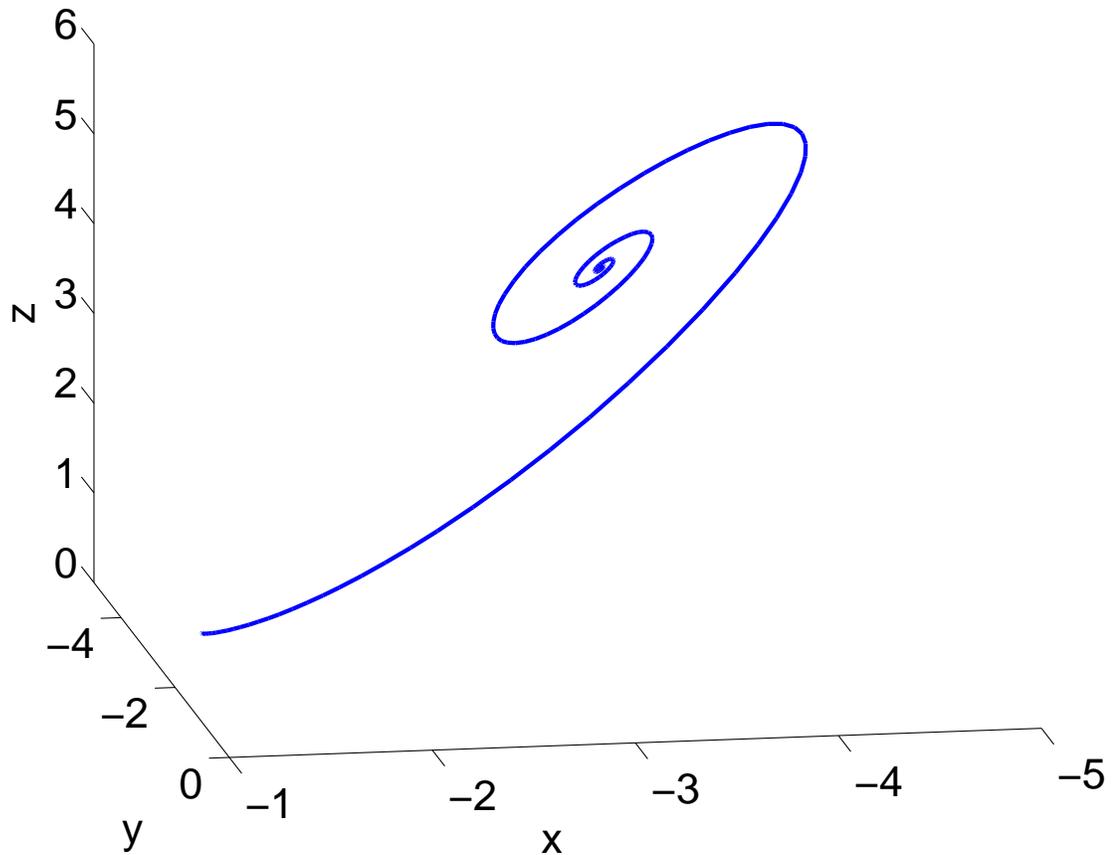}
\caption{Orbit converges to a stable point $C_2$ for $r=5$.}
\label{Fig_stable}
\end{figure}

\begin{figure}
\includegraphics[scale=0.8]{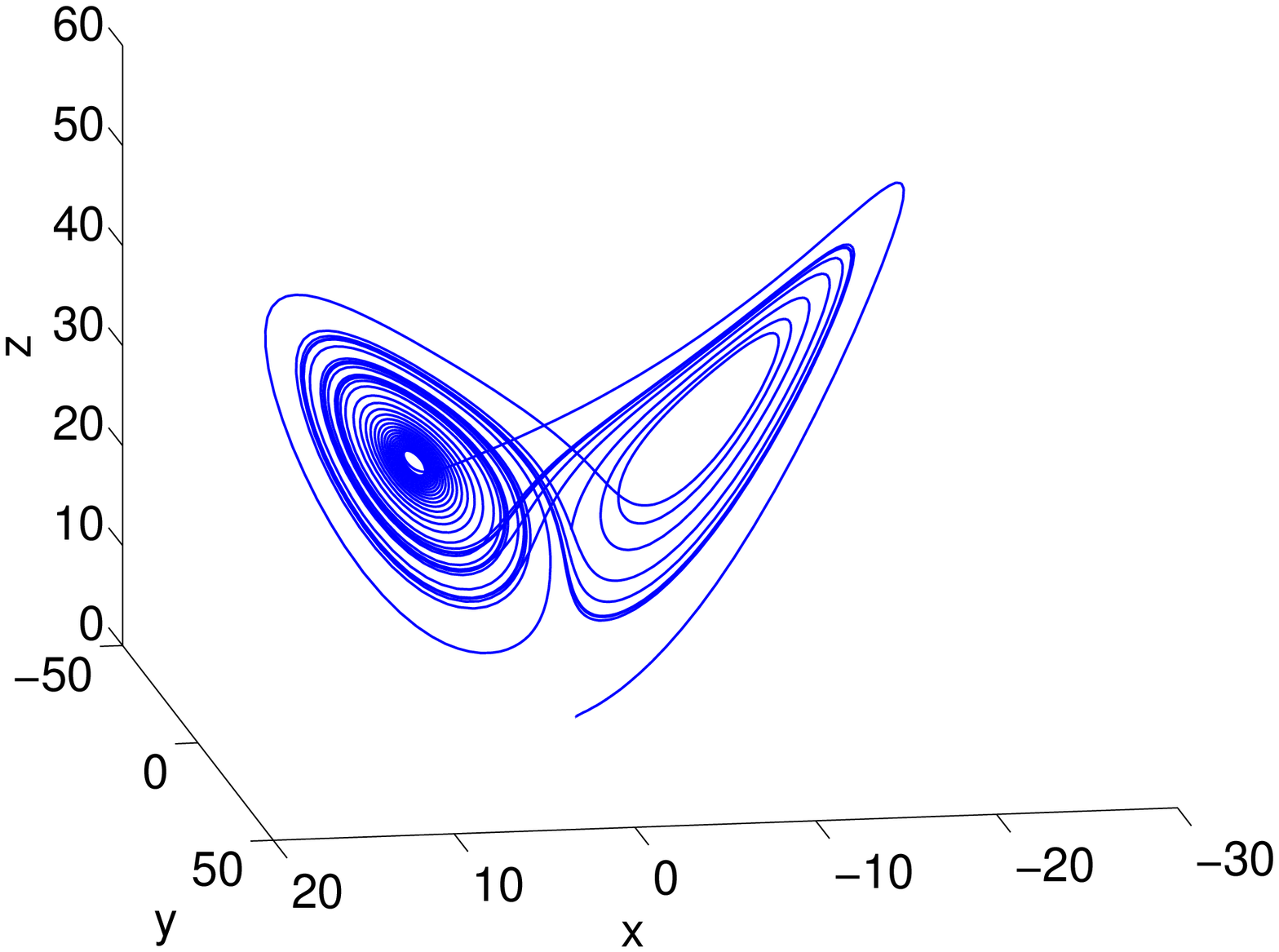}
\caption{Chaotic behavior of the an orbit for $r=30$.}
\label{Fig_unstable}
\end{figure}

\begin{figure}
\includegraphics[scale=0.8]{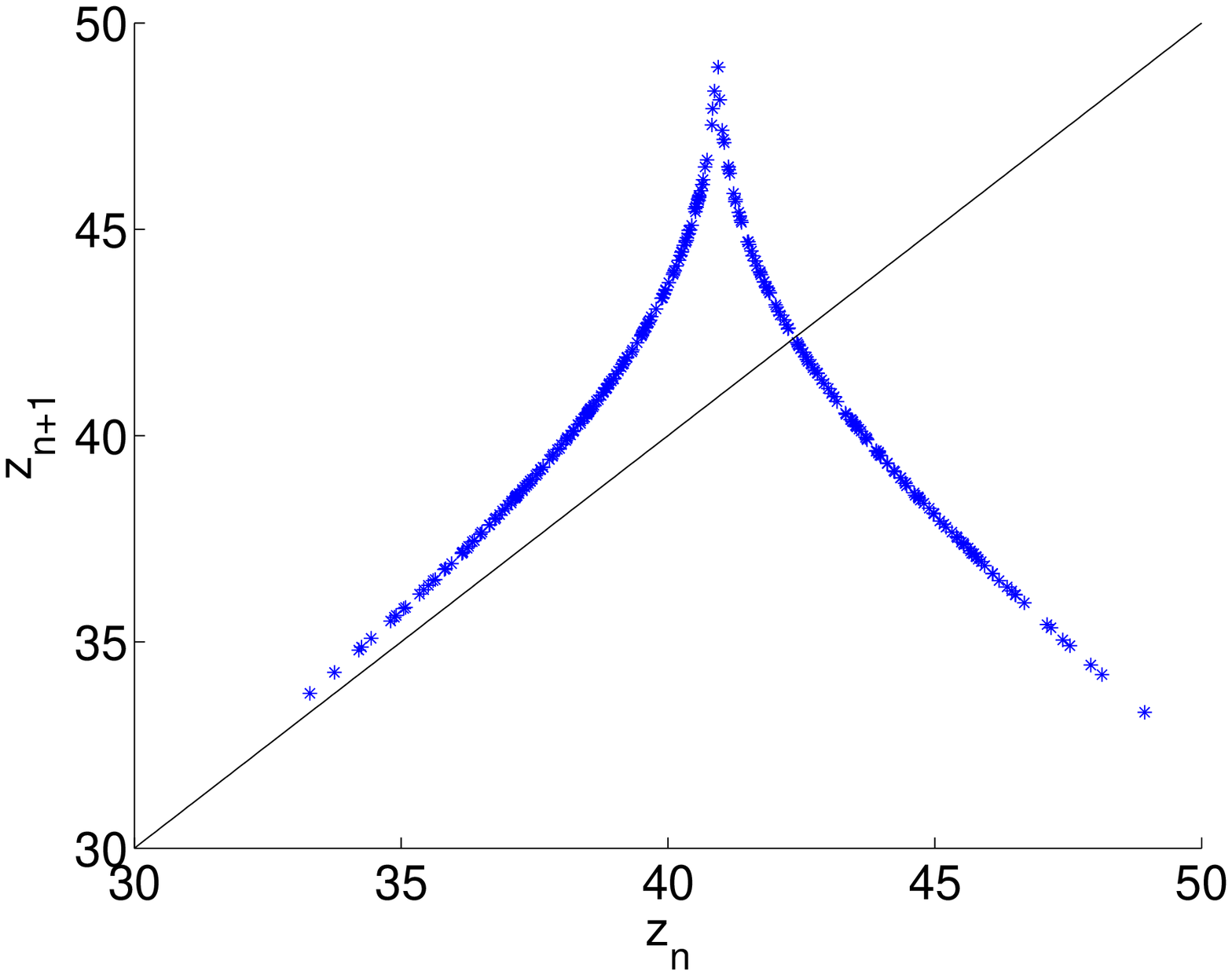}
\caption{$z_{n+1}$ as a function of $z_n$ for the Lorenz model. The black line with slope 1 is shown for comparison.}
\label{Fig_tent}
\end{figure}

For $0<r<1$, the origin is the only attracting fixed point with all the orbits approaching the origin.
For $r>1$, two stable fixed points arise
\BEA
C_{1,2}=\big(\pm\sqrt{b(r-1)},\pm\sqrt{b(r-1)},r-1\big).
\EEA
On the other hand, the origin loses its stability at $r=1$.
In Fig.~\ref{Fig_stable}, we observe an orbit that converges to the fixed point $C_2$ for $r=5$.
If we increase $r$ further then we find that for $r>27.74...$, the fixed points $C_{1,2}$ become unstable and the behavior of most orbits become chaotic.
The form of the nonlinearity prevents the orbits to diverge to infinity.
In Fig.~\ref{Fig_unstable}, we observe a typical behavior of an orbit for $r=30$.
Since the fixed points $C_{1,2}$ are now unstable, the orbit spirals outward around one of the fixed points.
Then it reaches the vicinity of the other fixed point and spirals out again and so on.
The behavior of most of the orbits in this parameter range is chaotic.
In order to see that we can consider some discrete subset of the continuous trajectory and show that this discrete subset actually produces a chaotic mapping.
In particular, we consider a trajectory $z(t)$ and construct a sequence $z_n$ of the local maxima of $z(t)$.
Then this sequence $z_{n}\rightarrow z_{n+1}$ is approximately a one-dimensional map function, which is shown in Fig.~\ref{Fig_tent}.
Note that the slope of this resulting one-dimensional map is greater than 1 (compare with a black line in Fig.~\ref{Fig_tent}).
Therefore, $|dz_{n+1}/dz_n|>1$ and the Lyapunov exponent $h$ is positive which indicates chaos.
To complete the discussion we just add that the windows of periodic motion appear if the parameter $r$ is increased.

This example demonstrates that a system can exhibit both regular and chaotic behaviors and by changing a control parameter one can achieve both.
\section{Sensitive dependence on initial conditions}
\label{sect_sensitive}
One of the characteristic properties of chaotic dynamical systems is the sensitive dependence on the initial conditions.
Suppose we have a dynamical system
\BEA
\frac{dx}{dt}=F[x(t)],\label{DS}
\EEA
where $x(t)$ is a vector solution.
Then if the system is chaotic, any two trajectories that are initiated at two very close points will exponentially diverge from each other.
This can be formulated more rigorously in the following way.
Consider two points $x_1(0)$ and $x_2(0)$ such that $|x_1(0)-x_2(0)|$ is small.
Then if the system~(\ref{DS}) is chaotic, then the difference $|x_1(t)-x_2(t)|$ will grow with time exponentially, i.e., positive Lyapunov exponent,
where $x_j(t)$ is the solution of Eq.~(\ref{DS}) with the initial value $x_j(0)$.

Therefore, even a small perturbation in the initial condition yields a completely different solution of a dynamical system in a chaotic regime.
This property has a direct impact on the numerical computations.
We always introduce some roundoff and truncation errors in both defining the initial conditions and then in computing the trajectory itself.
Therefore, for chaotic systems, we can never obtain a numerical solution, which is close to the exact solution to the equation with a given initial data.
If we use slightly perturbed initial data or different numerical integration schemes in the computations, for any reasonable longtime trajectory, i.e., $T=O(h^{-1})$,
we will obtain a completely different numerical solution.

Let us demonstrate the sensitive dependence on the numerical noise using the $\beta$-FPU system.
First, we show that the $\beta$-FPU system is indeed chaotic, i.e., the Lyapunov exponent is always positive.
\begin{figure}
\includegraphics[scale=0.8]{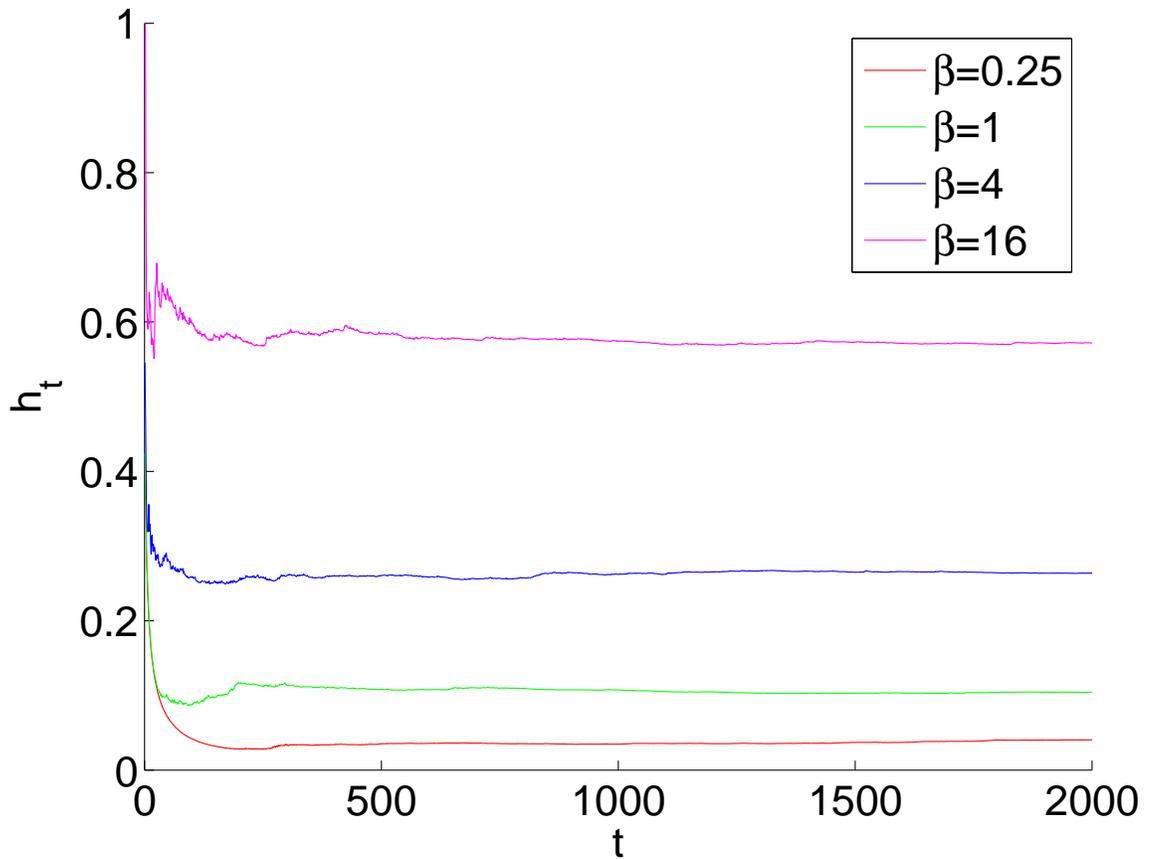}
\caption{Positive Lyapunov exponents.}
\label{Fig_fpu_lyap}
\end{figure}
In Fig.~\ref{Fig_fpu_lyap}, we present the comparison of the successive approximations to the  Lyapunov exponents of the $\beta$-FPU system for
four different values of the nonlinearity strength $\beta=\{0.25,1,4,16\}$ and same system energy $H=100$.
The algorithm for computing the Lyapunov exponent of the $\b$-FPU system is given in Appendix~\ref{app_lyap}.
Note that although the Lyapunov exponents are positive for any strength of nonlinearity, larger values of the Lyapunov exponent corresponds to larger values of
nonlinearity.
Therefore, we expect more chaotic behavior for strongly nonlinear systems.
In Fig.~\ref{Fig_qt}, we compare the trajectories of the $\beta$-FPU system with $N=128$ particles, simulated with the same parameters as in Fig.~\ref{Fig_fpu_lyap},
 but with different time steps.
The red curves correspond to $dt=0.01$ and the blue curves correspond to $dt=0.02$.
Different time steps imply that different numerical noises were introduced during computation, which should result in the divergence of these two trajectories due to
the chaotic nature of the $\beta$-FPU.
We make at least two observations after examining Fig.~\ref{Fig_qt}.
(i) The $\beta$-FPU system is indeed chaotic since the two trajectories diverge resulting in the positive Lyapunov exponent, $h>0$.
(ii) For higher values of nonlinearity strength $\beta$ the system is more chaotic, meaning that this divergence occurs faster since the Lyapunov exponent
is larger for stronger nonlinearity.
\begin{figure}
\includegraphics[height=8in, width=6in]{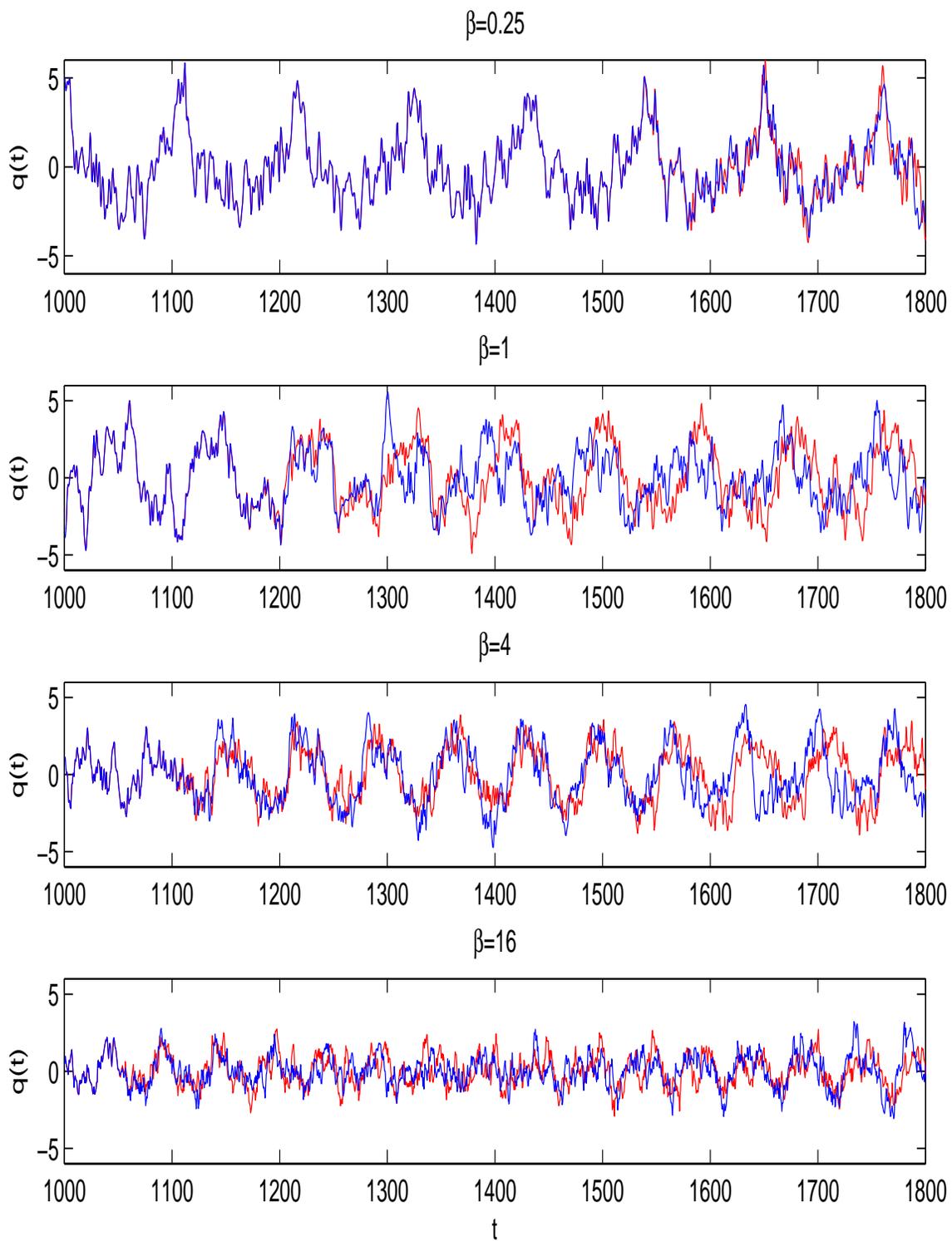}
\caption{Time evolution of $q_2$ for the $\beta$-FPU system with the different value of the nonlinearity parameter $\beta$.}
\label{Fig_qt}
\end{figure}

\section{Shadowing theorem}
\begin{figure}
\includegraphics[scale=1.2]{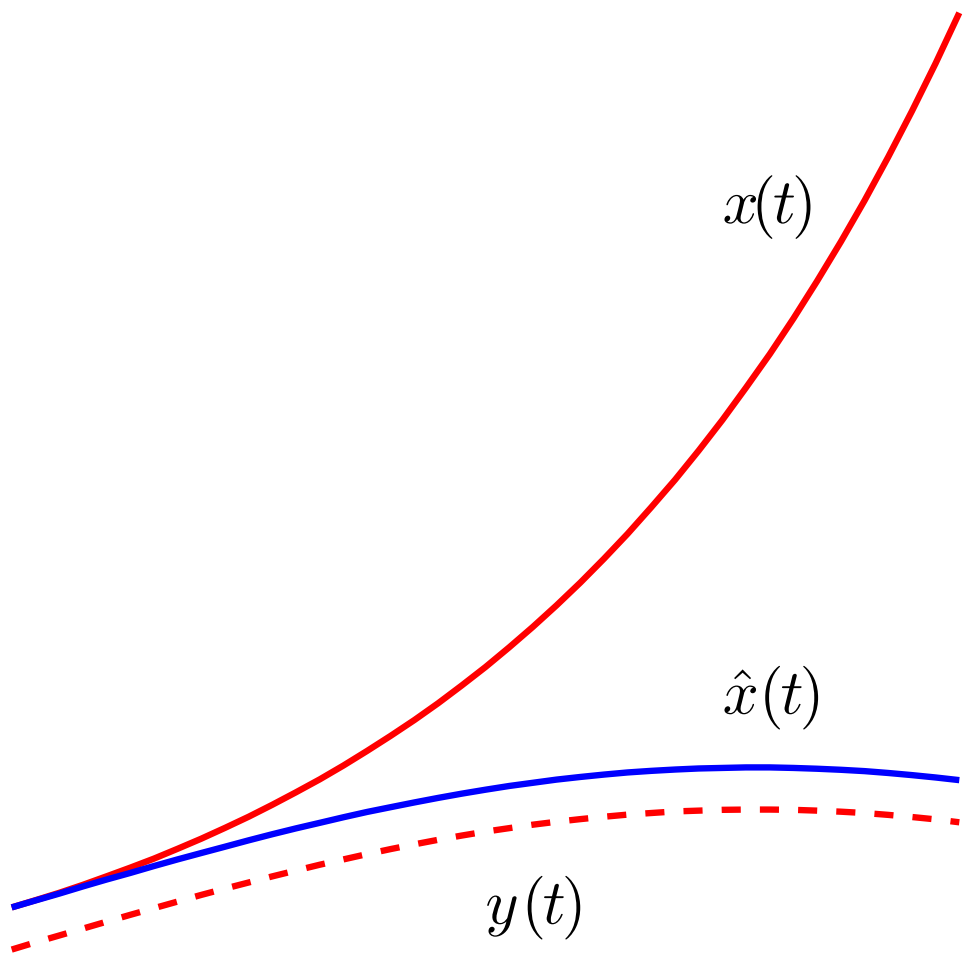}
\caption{Shadowing theorem: $x(t)$ is an exact solution, $\hat{x}(t)$ is numerically obtained solution that diverges from $x(t)$ exponentially. $y(t)$ is an exact
solution of the same system but with slightly different initial conditions. $y(t)$ ``shadows'' $\hat{x}(t)$.}
\label{Fig_shadow}
\end{figure}

Sensitive dependence on the initial condition and numerical noise seems to destroy any hope for solving the chaotic dynamical systems on the computer since any tiny
noise (such as numerical errors) will eventually bring in a huge error into the solution very quickly.
However, the cure to this problem comes from a rigorously proved \textit{shadowing} theorem~\cite{Palmer}.
Suppose we have an exact solution $x(t)$ and a numerical solution $\hat{x}(t)$ and as we have seen their difference grows exponentially with time due to roundoff errors
although $x(0)=\hat{x}(0)$.
The shadowing theorem says that there exists an exact solution $y(t)$ with a slightly perturbed initial condition $y(0)$ (such that $|y(t)-x(t)|$ remain small),
which stays near the numerical solution $\hat{x}(t)$ for sufficiently long time.
We demonstrate the theorem schematically in Fig.~\ref{Fig_shadow}.

To summarize, as we have mentioned in Section~\ref{sect_sensitive}, the sensitive dependence of the solution of the chaotic system on the initial conditions leaves no hope
to compute even approximate solutions with a given initial conditions.
At the times larger than inverse of the Lyapunov exponent the numerical solution will diverge from the exact one.
However, the shadowing theorem provides a different sense of numerical solutions for ODEs.
It assures that numerical solutions of dynamical systems are meaningful.
Usually one is not interested in a particular solution of the dynamical system in the chaotic regime but rather in the structure of attractors that evolve
from a set of initial data.
Since numerically obtained solutions are close to \textit{some} exact solutions with close initial values,
we can use them to understand general characteristics of a given dynamical system such as its phase portrait.

\chapter{Renormalization in FPU chains: analytical description}
\label{sect_renormalization}
\section{Hamiltonian formulation of FPU chains}
Consider a chain of particles coupled via nonlinear springs as shown in Fig.~\ref{Fig_model}.
Suppose the total number of particles is $N$ and the momentum and displacement from the equilibrium position of the $j$-th particle are $p_j$ and $q_j$, respectively.
In this thesis, we consider the systems with only the nearest-neighbor interactions.
Then the chain can be described by the Hamiltonian
\BEA
H=H_2+V.\label{Ham_pq}
\EEA
The quadratic part of the Hamiltonian takes the form
\BEA
H_2=\frac{1}{2}\sum_{j=1}^N{p_j^2}+(q_j-q_{j-1})^2,\label{H2}
\EEA
and the anharmonic potential $V$ is the function of the relative displacement $(q_j-q_{j-1})$
\BEA
V=\sum_{j=1}^Nv(q_j-q_{j-1}),\label{V}
\EEA
where $v(\Delta q)$ is a potential of the spring between two adjacent particles with the distance $\Delta q$ between them.
Here periodic boundary conditions $q_{-1}\equiv q_N$ and $p_{-1}\equiv p_N$ are imposed.
Since the total momentum of the system is conserved, it can be set to zero.
In this thesis, we only consider the potentials of the \textit{restoring} type, i.e., the potentials for which the Gibbs measure exists.
In order to study the
distribution of energy among the wave modes, we transform the Hamiltonian~(\ref{Ham_pq}) to the Fourier variables $Q_k$, $P_k$ via Eq.~(\ref{PQ}).
In Chapter~\ref{sect_mechanics}, it was shown that this transformation is canonical.
The Hamiltonian~(\ref{Ham_pq}) becomes
\BEA
H=\frac{1}{2}\sum_{k=1}^{N-1}|P_k|^2+\omega_k^2|Q_k|^2+V(Q),\label{Ham_PQ}
\EEA
where $\w_k=2\sin(\pi k/N)$ is the linear dispersion relation.
For the case of the $\beta$-FPU chain, when only the fourth order potential is present, we will present the form of $H$ below [Eq.~(\ref{Hamb_PQ})].
Note that, throughout the thesis, for the simplicity of notation, we denote
the periodic wave number space by the set of integers in the range $[0,N-1]$, i.e., we drop the conventional factor, $2\pi/N$.
The zeroth mode of the momentum $P_0$ vanishes due to the fact that the total momentum is zero.
The zeroth mode of the displacement $Q_0$ can also be set to zero, which is the consequence of the fact that the $q_j$ in the Hamiltonian~(\ref{Ham_pq}) are determined
up to a constant. Therefore, we have the following conditions
\BEA
\sum_{j=0}^{N-1}p_j&=&0,\label{p0}\\
\sum_{j=0}^{N-1}q_j&=&0.\label{q0}
\EEA
Now let us demonstrate why the wave description is convenient on the example of noninteracting waves, i.e., when $V\equiv 0$.
Here, we discuss the distribution of energy among the wave modes.
However, in the completely linear system there is no energy exchange among the wave modes and, therefore, the system will never reach energy equipartition state.
Therefore, we imagine that there exist a \textit{weak} nonlinearity that mixes energy and thus we can talk about equilibrium.
But on the other hand, we consider this nonlinearity to be so weak that practically all the energy is in the linear modes.
The other way of thinking about the issue of reaching thermal equilibrium is to imagine a heat bath, which is in thermal contact with the chain.
In this case, the particles of the chain can exchange energy through the thermal bath.
Then, even though the interactions between the particles of the chain are harmonic, it makes perfect sense to discuss thermal equilibrium of the chain.
These two ways of describing thermal equilibrium correspond to microcanonical and canonical (Gibbs) distributions, respectively.
In this Chapter, we will first discuss the microcanonical distribution and obtain the renormalization property of the linear dispersion of the FPU chains.
Then, in Section~\ref{sect_Gibbs}, we will turn our attention to the canonical distribution and obtain similar results about the renormalization.
Then, we will discuss why the canonical distribution is more convenient in practical calculations.
\section{Free noninteracting waves}
\label{sect_free_waves}
If the nonlinear interactions are weak, and can be neglected, Eq.~(\ref{Ham_PQ}) takes form
\BEA
H=\frac{1}{2}\sum_{k=1}^{N-1}|P_k|^2+\omega_k^2|Q_k|^2.\label{Ham_PQ0}
\EEA
In this case, it is convenient to further transform the Hamiltonian~(\ref{Ham_PQ0}) to the complex
normal variables $a_k$ defined by Eq.~(\ref{a}).
As shown in Chapter~\ref{sect_mechanics}, this transformation is also canonical, i.e., the dynamical equation of motion is given by Eq.~(\ref{adot}).
In terms of these normal variables, the Hamiltonian~(\ref{Ham_PQ0}) takes the form
\BEA
H=\sum_{k=1}^{N-1}\w_k|a_k|^2.\label{Ham_lin}
\EEA
For the system of noninteracting waves, it is possible to obtain a standard virial theorem~\cite{LL1} in the form
\BEA
\la K_k\ra\vline_{V=0}=\la U_k\ra\vline_{V=0},\label{free}
\EEA
where
\BEA
K_k\equiv\frac{1}{2}|P_k|^2,\label{Kk}
\EEA
is the kinetic energy of the $k$-th mode and
\BEA
U_k\equiv\frac{1}{2}\w_k^2|Q_k|^2,\label{Uk}
\EEA
is the quadratic potential energy part of the $k$-th mode.
Moreover, in thermal equilibrium  we have equipartition of energy [Eq.~(\ref{equipartition})] in the form
\BEA
\la K_k\ra&=&\la K_l\ra,\label{Kin_equip}\\
\la U_k\ra&=&\la U_l\ra.\label{Pot_equip}
\EEA
To show this, one can notice that the terms $|P_k|^2$ and $\w_k^2|Q_k|^2$ appear in the summation in Eq.~(\ref{Ham_PQ0}) as independent items.
Therefore, when the averages in Eqs.~(\ref{Kin_equip}) and~(\ref{Pot_equip}) are computed using either microcanonical or canonical measure, these
averages hare independent of the wave numbers.
Let us show this for the kinetic energy using the Gibbs measure
\BEA
\la|P_k|^2\ra&=&\int|P_k|^2\exp\left(-\frac{1}{2\T}\sum_{j=1}^{N-1}|P_j|^2+\w_j^2|Q_j|^2\right)dPdQ\nonumber\\
&=&I(Q)\int|P_k|^2\exp\left(-\frac{1}{2\T}|P_k|^2\right)dP_k\int\exp\left(-\frac{1}{2\T}\sum_{j\neq k}|P_j|^2\right)dP\nonumber\\\nonumber\\
&=&I(Q)\int|z|^2\exp(-|z|^2/2\T)dz \left(\int\exp(-|z|^2/2\T)dz\right)^{N-2},\label{P2}
\EEA
where $I(Q)$ is part of the integral that only depends on $Q$.
To obtain the last equality in Eq.~(\ref{P2}), the change of variables $P\rightarrow z$ was made in order to indicate that the resulting expression does not depend
on the wave number $k$.
Therefore, the RHS of Eq.~(\ref{P2}) is independent of the wave number and so is the average kinetic energy of each wave mode.
In a similar way, we can show independence of the average potential energy of the wave mode.
Surprisingly, as we will see below, this property holds in the thermal equilibrium even when nonlinearity is present.
As a consequence of energy equipartition of system~(\ref{Ham_lin}), we have the following properties of free waves,
\BEA
\la a_k^*a_l\ra&=&\la|a_k|^2\ra\d^k_l,\label{asa_lin}\\
\la
a_ka_l\ra&=&0,\label{aa_lin}
\EEA
where $\theta$ is temperature given by
\BEA
\theta=\la|P_k|^2\ra.\label{temp}
\EEA
Here, we have used the equipartition theorem that was discussed in Chapter~\ref{sect_stat_phys} [Eq.~(\ref{p2_equip}].
Note that Eq.~(\ref{asa_lin}) gives the classical Rayleigh-Jeans distribution
for the power spectrum of free waves~\cite{ZLF}
\BEA
n_k=\frac{\T}{\w_k}.\label{PSa}
\EEA
In order to set the grounds for the case of nonlinear FPU chains, we rewrite Eqs.~(\ref{asa_lin}) and~(\ref{aa_lin}) in terms of Fourier modes.
Let us invert Eq.~(\ref{a})
\BEA
P_k&=&\sqrt{\frac{\w_k}{2}}(a_k+a_{N-k}^*),\label{Pa}\\
Q_k&=&\sqrt{\frac{\w_k}{2}}(a_k+a_{N-k}^*).\label{Qa}
\EEA
Then, we can compute the following correlators using Eqs.~(\ref{asa_lin}) and~(\ref{aa_lin})
\BEA
\la P_kP_l\ra&=&\la|P_k^2|\ra\d^{k+l}_N,\nonumber\\
\la Q_kQ_l\ra&=&\la|Q_k|^2\ra\d^{k+l}_N,\nonumber\\
\la P_kQ_l\ra&=&0.\nonumber\\
\label{PQ_corr_lin}
\EEA
Next, we substitute Eq.~(\ref{a}) into Eqs.~(\ref{asa_lin}) and~(\ref{aa_lin}) and then use Eq.~(\ref{PQ_corr_lin}) to obtain
\BEA
\la a_k^*a_l\ra&=&\frac{1}{2\w_k}(\la|P_k|^2\ra+\w_k^2\la|Q_k|^2\ra)\delta^k_l=\frac{\theta}{\w_k}\delta^k_l,\label{asa_lin2}\\
\la
a_ka_l\ra&=&\frac{1}{2\w_k}(\la|P_k|^2\ra-\w_k^2\la|Q_k|^2\ra)\delta^{k+l}_N=0.\label{aa_lin2}
\EEA
The transformations that we just did may seem trivial since in the linear regime Eqs.~(\ref{asa_lin}) and~(\ref{aa_lin}) are sufficient to
give a statistical description of different modes.
However, situation changes when nonlinearity is present.
In particular, the waves $a_k$ and $a_{N-k}$ become correlated, i.e.,
\BEA
\la
a_ka_{N-k}\ra=\frac{1}{2\w_k}(\la|P_k|^2\ra-\w_k^2\la|Q_k|^2\ra)\neq
0,\label{akNk} \EEA since the property~(\ref{free}) is no longer
valid.
However, Eq.~(\ref{PQ_corr_lin}) stays valid even in the nonlinear regime and the proof consequences of this fact will be discussed in the next Sections.
\section{Nonlinear interactions and microcanonical description}
Now, we turn to the case of nonlinearity of any strength.
As we have mentioned above, the waves $a_k$ do not constitute a set of uncorrelated waves.
However, as we will show below, a complete set of new renormalized variables $\at_k$ can be constructed.
Using these new variables, the strongly nonlinear system can be viewed as a system of ``free'' waves.
These waves are free in the sense of vanishing correlations and the power spectrum,
i.e., the new
variables $\at_k$ satisfy the properties of free waves given in Eqs.~(\ref{asa_lin}) and~(\ref{aa_lin}). Next, we show how to construct
these renormalized variables $\at_k$.
As we have discussed in Chapter~\ref{sect_stat_phys}, the microcanonical measure~[Eq.~\ref{ro_micro}] (total energy and the number of particles are fixed),
can be used in order to study the statistics of the system~(\ref{Ham_pq})
\BEA
dw(p,q)=\delta(H(p,q)-E)~\delta\left(\sum_{j=1}^N p_j\right)\delta\left(\sum_{j=1}^N q_j\right)dpdq,\label{dw_micro}
\EEA
where $E$ is the total energy and $dpdq\equiv dp_1\dots dp_N dq_1\dots dq_N$.
Let us make a change of variables
\BEA
(q_1,\dots,q_N)\rightarrow(y_1\equiv q_1,y_2\equiv q_2-q_1,\dots,y_N\equiv q_N-q_{N-1})\label{q2y}
\EEA
This transformation is non-degenerate since it is given by the matrix
\BEA
M=
\left(
\begin{array}{cccccc}
  1 & 0 & . & . & 0 & 0 \\
  -1& 1 & . & . & 0 & 0 \\
  . & . & . & . & . & . \\
  . & . & . & . & . & . \\
  0 & 0 & . & . & 1 & 0 \\
  0 & 0 & . & . & -1 & 1 \\
\end{array}%
\right)\label{matrix}
\EEA
with the unity determinant.
Under the transformation~(\ref{q2y}), the microcanonical measure~(\ref{dw_micro}) becomes
\BEA
dw(p,y)=\delta(H(p,y)-E)dp_2\dots dp_N dy_2\dots dy_N,\label{dw_micro_y}
\EEA
where the Hamiltonian takes the form
\BEA
H&=&\frac{1}{2}\sum_{j=2}^N p_j^2+(p_2+\dots+p_N)^2\nonumber\\
&+&\sum_{j=2}^N \left(\frac{1}{2}y_j^2+v(y_j)\right)+\frac{1}{2}(y_2+\dots+y_N)^2+v(-(y_2+\dots+y_N))\label{Ham_py}
\EEA
Note that the measure~(\ref{dw_micro_y}) does \textit{not} prescribe any probability for $p_1$ and $y_1$, which is a reflection of the
fact that in the coordinates $(p,q)$ the center of mass is at rest at the origin~[Eqs.~(\ref{p0}) and~(\ref{q0})].
Therefore, we have lowered the number of degrees of freedom by $2$.
Although the transformation to the new variables $y_j$ is non-degenerate, it is not canonical.
It means that the pair $(p,y)$ does not a constitute a canonically conjugate pair of variables.
However, we will only use the non-degeneracy condition below.
Next, we will use the properties of Hamiltonian~(\ref{Ham_py}) in order to study the statistical properties of $p_j$ and $y_j$.
\section{Statistics of oscillators}
Since Hamiltonian~(\ref{Ham_py}) is an even function of $p$, we obtain that the correlation between any $p_s$ and any $y_j$ vanishes as an integral of an
odd function over the whole space with respect to measure~(\ref{dw_micro_y}), i.e.,
\BEA
\la p_s y_j\ra_m=\int p_sy_jdw(p,q)=[p\rightarrow-p]=\int -p_sy_jdw(p,q).\label{py01}
\EEA
where $\la\dots\ra_m$ denotes a microcanonical averaging.
If the number is equal to its negative, than it must be zero
\BEA
\la p_s y_j\ra_m=0.\label{py0}
\EEA
In order to study the statistical properties of $y_j$ it is convenient to introduce another variable $\y_1$
\BEA
\y_1\equiv-y_2-\dots-y_N,\label{y1}
\EEA
or, in terms of $q_j$,
\BEA
\y_1=q_1-q_N.\label{y1_q}
\EEA
Since the Hamiltonian~(\ref{Ham_py}) is independent of $y_1$, as we discussed above, let us abuse the notation and drop the ``check'' sign in $\y_1$.
From now on, we have
\BEA
y_1\equiv\y_1.\label{y1_new}
\EEA
However, it is important to emphasize again that the microcanonical measure~(\ref{dw_micro_y}) only prescribes the probabilities for $p_j$ and $y_j$ with $j\in[2,N]$
and $p_1$ and $y_1$ are determined from Eqs.~(\ref{p0}) and~(\ref{q0}).
To symmetrize the measure $dw$, we introduce the microcanonical measure over the whole set of indices $j\in[1,N]$ via
\BEA
dw_1(p,y)=\delta\big(H(p,y)-E\big)\delta\left(\sum_{j=1}^N p_j\right)\delta\left(\sum_{j=1}^N y_j\right)dp_1\dots dp_N dy_1\dots dy_N.\label{dw_micro_y1}
\EEA
We can view the measure $dw$ as a projection of $dw_1$ on to the space of $p_j$ and $y_j$ with $j\in[2,N]$
\BEA
dw=\int_{(p_1,y_1)}dw_1.
\EEA
Using the symmetry properties of $dw_1$, below we will obtain an important property of $p$ and $y$, i.e.,
$\la y_k^2\ra_m$ and $\la y_k y_l\ra_m$ \textit{are independent of $k$ and $l$}.
Therefore, we define
\BEA
\la y^2\ra_m&\equiv&\la y_k^2\ra_m,\\
\la yy\ra_m&\equiv&\la y_k y_l\ra_m,~~~~k\neq l.
\EEA
With these new definitions, we can rewrite the correlation
\BEA
\la y_ky_l\ra_m=\la y^2\ra_m\delta^k_l+\la yy\ra_m(1-\delta^k_l),\label{yy0}
\EEA
for any $k$ and $l$.
Using Eqs.~(\ref{y1}) and~(\ref{y1_new}), we obtain for $k\neq 1$
\BEA
\la yy\ra_m=\la y_1,y_k\ra_m=\la-y_2-\dots-y_N,y_k\ra_m=-\la yy\ra_m(N-2)-\la y^2\ra_m
\EEA
Thus, we find the following relation between $\la y^2\ra_m$ and $\la yy\ra_m$
\BEA
\la yy\ra_m=-\frac{1}{N-1}\la y^2\ra_m.\label{yy_y2}
\EEA
Note that in the thermodynamic limit (canonical ensemble), when $N\rightarrow\infty$, we have
\BEA
\la yy\ra_m=0.\label{yy_bigN}
\EEA
Similarly, we find the following equation for $p$
\BEA
\la pp\ra_m=-\frac{1}{N-1}\la p^2\ra_m.\label{pp_p2}
\EEA
To summarize, we have obtained the following expressions for the correlations of $p$ and $y$
\BEA
\la p_kp_l\ra_m&=&\la p^2\ra_m\delta^k_l-\frac{\la p^2\ra_m}{N-1}(1-\delta^k_l),\label{pp}\\
\la y_ky_l\ra_m&=&\la y^2\ra_m\delta^k_l-\frac{\la y^2\ra_m}{N-1}(1-\delta^k_l),\label{yy}\\
\la p_ky_l\ra_m&=&0.\label{py}
\EEA
We will use these properties in order to study the statistics of the Fourier modes and, after that, of the normal modes.
\section{Statistics of Fourier modes}
The Fourier transform of $y_j$ is defined by
\BEA
Y_k=\displaystyle{\frac{1}{\sqrt{N}}}\sum_{j=1}^Ny_je^{\frac{2\pi \i kj}{N}},\label{Y}
\EEA
with $Y_0=0$ as a consequence Eq.~(\ref{y1}).
Next, let us use Eq.~(\ref{yy}) to obtain the statistical properties of $Y_k$.
\BEA
\la Y_k Y_l\ra_m&=&\left\la\frac{1}{N}\sum_{j=1}^N y_je^{\frac{2\pi \i kj}{N}}\sum_{s=1}^N y_se^{\frac{2\pi \i ls}{N}}\right\ra_m=
\frac{1}{N}\sum_{j,s=1}^N\la y_jy_s\ra_m e^{\frac{2\pi \i (kj+ls)}{N}}\nonumber\\
&=&\frac{1}{N}\sum_{j,s=1}^N\left(\la y^2\ra_m\delta^j_s-\frac{\la y^2\ra_m}{N-1}(1-\delta^j_s)\right)e^{\frac{2\pi \i (kj+ls)}{N}}\nonumber\\
&=&\frac{\la y^2\ra_m}{N}\Big[\sum_{j,s=1}^N \left(1+\frac{1}{N-1}\right)\delta^j_se^{\frac{2\pi\i(k+l)j}{N}}\Big]=\frac{N}{N-1}\la y^2\ra_m\delta^{k+l}_N,\nonumber\\
\label{YY0}
\EEA
for $k,l\neq N$.
Here, we have used the following equality
\BEA
\sum_{n=1}^N\exp\left(-\frac{2\pi\i rn}{N}\right)=\d^r_{jN},\label{exp_formula}
\EEA
for any integer $r$ and $j$. The sum is equal to 1 if $r$ is a multiple of $N$. Otherwise, the sum is zero.
Equation~(\ref{exp_formula}) is just a summation of the geometrical progression with the base $\exp(-2\pi\i r/N)$.
Similarly, using Eq.~(\ref{pp}), we find for $P$
\BEA
\la P_k P_l\ra_m=\frac{N}{N-1}\la p^2\ra_m\delta^{k+l}_N.
\EEA
Finally, from Eq.~(\ref{py}), it immediately follows that
\BEA
\la P_k Y_l\ra_m=0.
\EEA
Next, we find the relation between $Q_k$ and $Y_k$ using Eqs.~(\ref{PQ}) and~(\ref{Y})
\BEA
Y_k&=&\displaystyle{\frac{1}{\sqrt{N}}}\sum_{j=1}^N(q_j-q_{j-1})e^{\frac{2\pi \i kj}{N}}=
\displaystyle{\frac{1}{\sqrt{N}}}\sum_{j=1}^Nq_je^{\frac{2\pi \i kj}{N}}-\displaystyle{\frac{1}{\sqrt{N}}}\sum_{j=1}^Nq_{j-1}e^{\frac{2\pi \i kj}{N}}\nonumber\\
&=&Q_k-Q_ke^{\frac{2\pi \i k}{N}}=-\i \exp\left(\frac{\pi\i k}{N}\right)\w_k Q_k.\nonumber\\
\label{YQ}
\EEA
Now, we have the following statistical properties of the Fourier modes $P_k$ and $Q_k$
\BEA
\la P_kP_l\ra_m&=&\frac{N}{N-1}\la p^2\ra_m\delta^{k+l}_N,\label{corrPP}\\
\la Q_kQ_l\ra_m&=&\frac{1}{\w_k^2}\frac{N}{N-1}\la y^2\ra_m\delta^{k+l}_N,\label{corrQQ}\\
\la P_kQ_l\ra_m&=&0.\label{corrPQ}
\EEA
In the next Section, we will use these properties to study the statistics of the normal modes.
\section{Renormalized waves and their statistics}
As we have seen in Section~\ref{sect_free_waves},
if the anharmonic part of the potential is sufficiently weak, then corresponding waves $a_k$ remain almost free, and Eqs.~(\ref{asa_lin}) and~(\ref{aa_lin})
would be approximately satisfied in the weakly nonlinear regime.
However, when the nonlinearity becomes stronger, waves $a_k$ become strongly correlated, and, in general, the correlations between waves [Eq.~(\ref{aa_lin})]
no longer vanish.
In particular, $\la a_ka_{N-k}\ra\neq 0$.
Naturally, the question arises: can the strongly nonlinear system in thermal equilibrium still be viewed as a system of almost free waves in some statistical sense?
In this thesis, we give an affirmative answer to this question.
It turns out that the system~(\ref{Ham_pq}) can be described by a complete set of \textit{renormalized} canonical variables $\at_k$,
which still possess the wave properties given by Eqs.~(\ref{asa_lin}) and~(\ref{aa_lin}) with a renormalized linear dispersion.
The waves that correspond to these new variables $\at_k$ will be referred to as \textit{renormalized} waves.
We will show that these renormalized waves possess the equilibrium Rayleigh-Jeans distribution~\cite{ZLF} and vanishing correlations between waves.
Therefore, they resemble free, non-interacting waves, and can be viewed as statistical normal modes.
Furthermore, it will be demonstrated that the renormalized linear dispersion for these renormalized waves has the form $\wt_k=\eta(k)\w_k$,
where $\eta(k)$ is the linear frequency renormalization factor.
Moreover and quite surprisingly, this renormalization factor is \textit{independent} of $k$ as a consequence of the microcanonical measure (or Gibbs measure
in the thermodynamic limit).

Consider the generalization of the transformation~(\ref{a}), namely, the transformation from the Fourier variables $Q_k$ and $P_k$ to the
renormalized variables $\at_k$ by
\BEA
\at_k=\frac{P_k-\i\wt_kQ_k}{\sqrt{2\wt_k}},\label{at}
\EEA
where $\wt_k$ is an arbitrary positive function, which satisfies Eq.~(\ref{w_canonical}).
In Chapter~\ref{sect_mechanics} it was shown that transformation~(\ref{at}) is canonical.
From Eqs.~(\ref{corrPP}), (\ref{corrQQ}), and~(\ref{corrPQ}), we obtain the following correlator of $\at_k$
\BEA
\la\at_k\at_l\ra_m&=&\frac{1}{2\sqrt{\wt_k\wt_l}}\left(\la P_kP_l\ra_m-\wt_k\wt_l\la Q_kQ_l\ra_m\right)\nonumber\\
&=&\frac{1}{2\sqrt{\wt_k\wt_l}}\frac{N}{N-1}\Big(\la p^2\ra_m-\frac{\wt_k^2}{\w_k^2}\la y^2\ra_m\Big)\delta^{k+l}_N.\label{aa0}
\EEA
Let us choose $\wt_k$ such that $\la\at_k\at_l\ra_m$ vanishes for \textit{any} $k$ and $l$.
For this, we demand
\BEA
\la p^2\ra_m-\frac{\wt_k^2}{\w_k^2}\la y^2\ra_m=0
\EEA
Thus, we obtain
\BEA
\eta\equiv\frac{\wt_k}{\w_k}=\sqrt{\frac{\la p^2\ra_m}{\la y^2\ra_m}}=\sqrt\frac{\la K\ra_m}{\la U\ra_m},\label{eta}
\EEA
where $\eta$ is the renormalization factor, $K$ is the kinetic energy
\BEA
K=\frac{1}{2}\sum_{j=1}^N p_j^2,\label{K}
\EEA
and $U$ is the quadratic potential energy
\BEA
U=\frac{1}{2}\sum_{j=1}^N y_j^2.\label{U}
\EEA
Note that the RHS of Eq.~(\ref{eta}) is independent of the wave number $k$.
Therefore, we obtain that the renormalization factor $\eta$ is also \textit{independent} of the wave number $k$.
For the power spectrum, we obtain
\BEA
\la\at_k\at_l^*\ra_m=\frac{1}{2\sqrt{\wt_k\wt_l}}\frac{N}{N-1}\Big(\la p^2\ra_m+\frac{\wt_k^2}{\w_k^2}\la y^2\ra_m\Big)\delta^k_l=
\frac{N}{N-1}\frac{\la p^2\ra_m}{\wt_k}\delta^k_l.\label{aas0}
\EEA
To summarize, we have the following properties of the renormalized waves
\BEA
\la\at_k\at_l\ra_m&=&0\label{aa},\\
\la\at_k\at_l^*\ra_m&=&\frac{N}{N-1}\frac{\la p^2\ra_m}{\wt_k}\delta^k_l.\label{aas}
\EEA

In our method, the construction of the renormalized variables $\at_k$ does not depend on a particular form or strength of the anharmonic potential,
as long as it is of the restoring type with only the nearest-neighbor interactions, as in Eq.~(\ref{Ham_pq}).
Therefore, our approach is non-perturbative and can be applied to a large class of systems with strong nonlinearity.
However, in this thesis, we will focus on the $\beta$-FPU chain to illustrate the theoretical framework of the renormalized waves.
We will verify that $\at_k$ effectively constitute normal modes for the $\beta$-FPU chain in thermal equilibrium by showing that
(i) the theoretically obtained renormalized linear dispersion relationship is in excellent agreement with its dynamical manifestation in our numerical simulation,
and
(ii) the equilibrium distribution of $\at_k$ is still a Rayleigh-Jeans distribution and $\at_k$'s are uncorrelated.
Note that similar expressions for the renormalization factor $\eta$ have been previously discussed in the framework of an approximate
virial theorem~\cite{Alabiso} or effective long wave dynamics via the Zwanzig-Mori projection~\cite{Lepri_renorm}.
However, in our theory, the exact formula for the renormalization factor is derived from a \textit{precise} mathematical construction of statistical
normal modes, and is valid for all wave modes $k$ --- no longer restricted to long waves.
\section{Canonical ensemble and Gibbs measure}
\label{sect_Gibbs}
We have used the microcanonical measure in order to describe the statistical properties of the
normal modes of the FPU chain.
However, this approach has at least two disadvantages.
The first one is that the microcanonical measure is hard to deal with, in particular, it is hard to compute
$\la K\ra_m$ and $\la U\ra_m$ and hence to obtain the value of the renormalization coefficient $\eta$~[Eq.~(\ref{eta})].
And the second disadvantage of the microcanonical description is that the isolated system of the coupled oscillators does not necessarily reach an
equilibrium state as it
was observed in the experiment by FPU~\cite{FPU}.
The nonlinear potential $V$ in Eq.~(\ref{Ham_pq}) should be strong enough to ``drive'' the chain to chaos.
Both of these drawbacks can be overcome if we consider the chain of oscillators to be in equilibrium with an imaginary thermal bath, i.e., we describe the chain as a
canonical ensemble.
Then the Gibbs measure provides a statistical description of the chain and the Gibbs measure is much easier to compute with.
Furthermore, we do not have to worry about the system not reaching thermal equilibrium since
the interactions with the thermal bath equilibrate the chain.
As we have noted in Chapter~\ref{sect_stat_phys}, both microcanonical and canonical descriptions coincide for large systems, i.e., when $N\rightarrow\infty$.

Consider the FPU chain in equilibrium with the thermal bath (thermodynamic limit, $N\rightarrow\infty$).
Then, we can write the Hamiltonian~(\ref{Ham_py}) as
\BEA
H=\sum_{j=1}^N\frac{1}{2}(p_j^2+y_j^2)+\sum_{j=1}^N v(y_j).\label{Ham_py_G}
\EEA
The Gibbs measure is given by Eq.~(\ref{Gibbs_def}), with the partition function
\BEA
Z=\int \exp\left(-\frac{1}{\theta}H(p,y)\right)dpdy.\label{Zpy}
\EEA
Here, we integrate over the whole set of variables $p_j$ and $y_j$ for $j\in[1,N]$.
Note that here we use the notation $\T$ for temperature %(see Section~\ref{sect_temperature} for more details)
since we will use the notation $T$ for the interaction coefficient below.
Now, the probability measure $dw(p,y)$ can be written as a product of probability measures for each individual component $p_j$ and $y_j$ since exponential of the
sum is equal to the product of the exponentials of the corresponding items of the sum
\BEA
dw(p,y)=\prod_{j=1}^Ndw_p(p_j)\prod_{j=1}^Ndw_y(y_j),\label{Z_prod}
\EEA
where
\BEA
dw_p(p_j)=\frac{1}{Z_p}\exp\left({-\frac{p_j^2}{2\theta}}\right)dp_j,\label{Zp}
\EEA
and
\BEA
dw_y(y_j)=\frac{1}{Z_y}\exp\left({-\frac{1}{\theta}\left(\frac{y_j^2}{2}+v(y_j)\right)}\right)dy_j.\label{Zy}
\EEA
Random variables $A$ and $B$ are called independent if
\BEA
P(A~\mbox{and}~B)=P(A)P(B).\label{independent}
\EEA
Therefore, from Eq.~(\ref{Z_prod}), we see that all $p_j$ and all $y_j$ form a set of $2N$ independent random variables.

We compute the normalizing constants $Z_p$ and $Z_y$
\BEA
Z_p=\int_{-\infty}^{\infty}\exp\left({-\frac{p^2}{2\T}}\right)dp=\sqrt{2\pi \T}.\label{Zp}
\EEA
Similarly,
\BEA
Z_y=\int_{-\infty}^\infty \exp\left({-\frac{1}{\T}\left(\frac{y^2}{2}+v(y)\right)}\right)~dy.\label{Zy}
\EEA
Note that
\BEA
Z=(Z_pZ_y)^N.
\EEA
From the independence of all momenta and all displacements, we obtain the following relationships
\BEA
\la p_k p_l\ra=\la p^2 \ra\delta^k_l,~~\la y_k y_l\ra=\la y^2 \ra\delta^k_l,~~
\la p_k y_l\ra&=&0,~~\label{py_G}
\EEA
where $\la\dots\ra$ denotes averaging with respect to the Gibbs measure~(\ref{Gibbs_def}).
It is instructive to compare Eq.~(\ref{py_G}) with the similar properties given by Eqs.~(\ref{pp}), (\ref{yy}), and~(\ref{py}) for the microcanonical averaging.
As it is expected, in the limit $N\rightarrow\infty$, both ensembles give the same result.
We then follow the same chain of transformations as we did for the microcanonical ensemble.
First, we transfer to the Fourier modes $P_k$ and $Q_k$ via Eqs.~(\ref{PQ}) and~(\ref{Y}).
Using Eq.~(\ref{py_G}), we obtain the following correlation properties of the Fourier modes
\BEA
\la P_kP_l\ra&=&\la p^2\ra\d^{k+l}_N,\label{PP_G}\\
\la Q_kQ_l\ra&=&\frac{1}{\w_k^2}\la y^2\ra\d^{k+l}_N,\label{QQ_G}\\
\la P_kQ_l\ra&=&0\label{PQ_G}.
\EEA
As a next step, we transfer to the renormalized variables given by Eq.~(\ref{at}).
We obtain the following correlators of the renormalized waves using Eqs.~(\ref{PP_G})-(\ref{PQ_G})
\BEA
\la \at_k\at_l\ra&=&\frac{1}{2\wt_k}\left(\la p^2\ra-\Big(\frac{\wt_k}{\w_k}\Big)^2\la y^2\ra\right)\delta^{k+l}_N,\label{aa_G}\\
\la \at_k\at_l^*\ra&=&\frac{1}{2\wt_k}\left(\la p^2\ra+\Big(\frac{\wt_k}{\w_k}\Big)^2\la y^2\ra\right)\delta^k_l.\label{aas_G}
\EEA
We choose $\wt_k$ to annihilate $\la \at_k\at_l\ra$ for all $k$ and $l$ and obtain the same renormalization condition $\wt_k=\eta\w_k$
as we had for the microcanonical ensemble [Eq.~(\ref{eta})]
\BEA
\eta=\sqrt{\frac{\la p^2\ra}{\la y^2\ra}}=\sqrt{\frac{\la K\ra}{\la U\ra}}.\label{eta_G}
\EEA
However, now we can actually compute $\eta$
\BEA
\eta=\left( \theta Z_y\right)^\frac{1}{2}\left( \int_{-\infty}^\infty y^2e^{-\frac{1}{\T}\left(\frac{y^2}{2}+v(y)\right)}~dy \right)^{-\frac{1}{2}}
\EEA
Furthermore, the power spectrum~(\ref{aas_G}) takes a simple form
\BEA
\la \at_k\at_l^*\ra=\frac{\theta}{\wt_k}\delta^k_l.\label{aas_G1}
\EEA
Again, we compare Eq.~(\ref{aas_G1}) with the similar expression for the microcanonical ensemble given by Eq.~(\ref{aas}).
Note that if we identify $\la p^2\ra_m$ as temperature then, in the thermodynamic limit, both expressions produce the same result.

The immediate consequence of the fact that $\eta$ is independent of
$k$ is that the power spectrum of the renormalized waves possesses
the precise Rayleigh-Jeans distribution, i.e., \BEA
\tilde{n}_k=\frac{\T}{\wt_k},\label{nt} \EEA from Eq.~(\ref{aas_G}),
where $\tilde{n}_k=\la |\at_k|^2\ra$. Combining Eqs.~(\ref{a})
and~(\ref{at}), we find the relation between the ``bare'' waves
$a_k$ and the renormalized waves $\at_k$ to be \BEA
a_k=\frac{1}{2}\left(\sqrt{\eta}+\frac{1}{\sqrt{\eta}}\right)\at_k+\frac{1}{2}\left(\sqrt{\eta}-\frac{1}{\sqrt{\eta}}\right)\at_{N-k}.\label{a2at}
\EEA Using Eq.~(\ref{a2at}), we obtain the following form of the
power spectrum for the bare waves $a_k$ \BEA
n_k=\frac{1}{2}\left(1+\frac{1}{\eta^2}\right)\frac{\T}{\w_k},\label{nkbare}
\EEA which is a modified Rayleigh-Jeans distribution due to the
renormalization factor $(1+1/\eta^2)/2$.
Naturally, if the nonlinearity becomes weak, we have $\eta\rightarrow 1$.
Therefore, all the variables and parameters with tildes reduce to the corresponding ``bare'' quantities.
In particular, $\wt_k\rightarrow\w_k$, $\at_k\rightarrow a_k$,
$\tilde{n}_k\rightarrow n_k$. It is interesting to point out that,
even in a strongly nonlinear regime, the ``free-wave'' form of the
Rayleigh-Jeans distribution is satisfied
\textit{exactly}~[Eq.~(\ref{nt})] by the renormalized waves. Thus,
we have demonstrated that even in the presence of strong
nonlinearity, the system in thermal equilibrium can still be viewed
statistically as a system of ``free'' waves in the sense of
vanishing correlations between waves and the power spectrum.

Note that, in the derivation of the formula for the renormalization
factor [Eq.~(\ref{eta})], we only assumed the nearest-neighbor
interactions, i.e., the potential is the function of $q_j-q_{j+1}$.
One of the well-known examples of such a system is the $\beta$-FPU
chain, where only the forth order nonlinear term in $V$ is present.
In the remainder of the thesis, we will focus on the $\beta$-FPU to
illustrate the framework of the renormalized waves $\at_k$.

\chapter{Numerical study of the $\beta$-FPU chain}
\label{sect_numerical}

The Hamiltonian of the $\beta$-FPU chain is of the form
\BEA
H=\sum_{j=1}^N\frac{1}{2}p_j^2+\frac{1}{2}(q_j-q_{j+1})^2+\frac{\beta}{4}(q_j-q_{j+1})^4,\label{H_FPU}
\EEA
where $\beta$ is a parameter that characterizes the strength of nonlinearity.
The corresponding equations of motion become
\BEA
\BC
\dot{q}_j=p_j,\\
\dot{p}_j=(q_{j-1}-2q_j+q_{j+1})+\beta\big[(q_{j+1}-q_j)^3-(q_j-q_{j-1})^3\big].
\EC\label{dyn_pq_1}
\EEA
To investigate the dynamical manifestation of the renormalized dispersion $\wt_k$ of $\at_k$, we numerically integrate Eq.~(\ref{dyn_pq_1}).
Since we study the thermal equilibrium state~\cite{Ford,Alabiso_fpu,Lepri_fpu,Carati} of the $\beta$-FPU chain,
we have verified that the results discussed in the paper do not depend on details of the initial data:
we have used random initial conditions, i.e., $p_j$ and $q_j$ were selected at random from the
uniform distribution in the intervals $(-p_{\rm{max}},p_{\rm{max}})$ and $(-q_{\rm{max}},q_{\rm{max}})$, respectively, with the two constraints that
(i) the total momentum of the system is zero and (ii) the total energy of the system $E$ is set to be a specified constant.
The results we obtained in the thermal equilibrium state were independent of the initial condition.
\section{Parametrization of the FPU chains}
Note that the behavior of the FPU for fixed number of particles $N$ is fully characterized by only one parameter~\cite{Poggy}.
Let us show this for the Hamiltonian with $N=1$ and then generalize it for any $N$.
Consider the general form of the Hamiltonian
\BEA
H=\frac{1}{2}(p^2+y^2)+\frac{\beta}{r}y^r,\label{Hr}
\EEA
where $r>2$.
Suppose $\g=\beta E^{r/2-1}$, where $E$ is the total energy of the system~(\ref{Hr}).
Consider another system with the same Hamiltonian but different total energy $E_1$ and the nonlinearity parameter $\beta_1$
\BEA
H_1=\frac{1}{2}(p_1^2+y_1^2)+\frac{\beta_1}{r}y_1^r.\label{Hr1}
\EEA
If $\beta_1 E_1^{r/2-1}=\g$ then every trajectory $(p_1,q_1)$  of system~(\ref{Hr1}) can be obtained by scaling a corresponding trajectory of
system~(\ref{Hr}) by a factor $\sqrt{E_1/E}$
\BEA
(p_1,y_1)=\sqrt{\frac{E_1}{E}}(p,y).\label{scale1}
\EEA
In order to prove this, we combine Eq.~(\ref{scale1}) with Eq.~(\ref{Hr1})
\BEA
E_1=\frac{E_1}{2E}(p^2+y^2)+\frac{\beta_1}{r}\left(\frac{E_1}{E}\right)^{r/2}y^r.\label{Hr1_tmp}
\EEA
By comparing Eq.~(\ref{Hr}) with Eq.~(\ref{Hr1_tmp}) we obtain
\BEA
\beta=\beta_1\left(\frac{E_1}{E}\right)^{r/2-1},\label{Hr1_compare}
\EEA
which yields the following equality
\BEA
\beta E^{r/2-1}=\beta_1 E_1^{r/2-1}.\label{betaE}
\EEA
Now we generalize this argument for any number $N$.
In this case, in Hamiltonians~(\ref{Hr}) and~(\ref{Hr1}), we will have summation over each particle number $j$.
Then, we apply exactly the same argument to each item in that summation as we did for $N=1$.
In particular, for $\beta$-FPU chain, we have $r=4$ and, therefore, it can be fully characterized by one parameter $\beta E$.
In the numerical experiments, when we test our results for different strength of nonlinearity, we will hold the total energy $E$ fixed and vary the nonlinearity
parameter $\beta$.
\section{Details of the numerical experiments}
We use the sixth order symplectic Yoshida algorithm~\cite{Yoshida}, which was briefly discussed in Chapter~\ref{sect_symplectic}.
In most of the numerical experiments, the system size was chosen to be $N=128$ or $N=256$ and the total energy $E=100$.
We probed the $\beta$-FPU chain with a wide range of the nonlinearity strength $\beta\in[10^{-3},10^4]$.
The time step is chosen to be $dt=0.01$, which ensures the conservation of the total system energy up to the ninth significant digit for a runtime
$\tau=10^6$ time units.
This runtime was enough for the system to reach thermal equilibrium even for the small nonlinearity $\beta=10^{-3}$.
\section{Thermalization}
\begin{figure}
\includegraphics[scale=0.8]{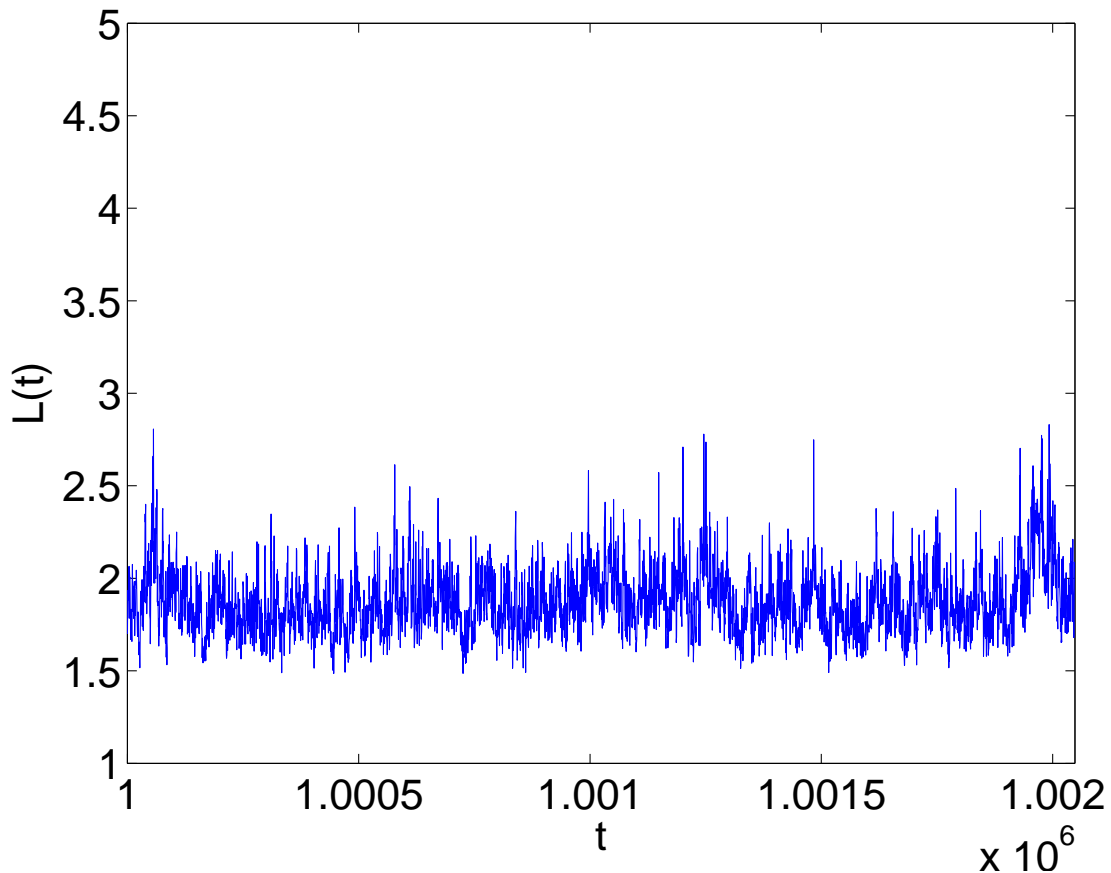}
\vskip 0.2in
\caption{ Energy localization L(t) computed via Eq.~(\ref{Loc}).
The chain was modeled for $N=128$, $\beta=1$, and $E=100$.}
\label{Fig_Lt}
\end{figure}

In order to confirm that the system has reached the thermal equilibrium state~\cite{Parisi}, the value of the energy localization~\cite{Cretegny} was monitored via
\BEA
L(t)\equiv\frac{{N\sum_{j=1}^{N}G_j^2}}{{(\sum_{j=1}^{N}G_j)^2}},\label{Loc}
\EEA
where $G_j$ is the energy of the $j$-th particle defined as
\BEA
G_j&=&\frac{1}{2}p_j^2+\frac{1}{4}\big[(q_j-q_{j+1})^2+(q_{j-1}-q_j)^2\big]\nonumber\\
&+&\frac{\beta}{8}\big[(q_j-q_{j+1})^4+(q_{j-1}-q_j)^4\big].
\EEA
If the energy of the system is concentrated around one site, then $L(t)=O(N)$.
Whereas, if the energy is uniformly distributed along the chain, then $L(t)=O(1)$.
In our simulations, in thermal equilibrium states, $L(t)$ is fluctuating in the range of $1$-$3$.
In Fig.~\ref{Fig_Lt}, we plot the time evolution of $L(t)$ for $N=256$, $E=100$, and $\beta=1$.

The energy localization function $L(t)$ is also used in detecting the appearance of discrete breathers ---
spatially localized periodic solutions of the discrete lattice that we will discuss in Chapter~\ref{sect_breathers}.

Since our simulation is of microcanonical ensemble, we have monitored various statistics of the system to verify that the thermal equilibrium state that is
consistent with the Gibbs distribution~(canonical ensemble) has been reached.
Moreover, we verified that, for $N$ as small as 32 and up to as large as 1024, the equilibrium distribution in the thermalized state in our microcanonical
ensemble simulation is consistent with the Gibbs measure.
We compared the renormalization factor~(\ref{eta}) by computing the values of $\la K\ra$ and $\la U\ra$ numerically and theoretically using the Gibbs measure and
found the discrepancy of $\eta$ to be within $0.1\%$ for $\beta=1$ and the energy density $E/N=0.5$ for $N$ from 32 to 1024.
\section{Spatiotemporal spectrum}
We now address numerically how the renormalized linear dispersion $\wt_k$ manifests itself in the dynamics of the $\beta$-FPU system.
We compute the spatiotemporal spectrum $|\af_k(\w)|^2$, where $\af_k(\w)$ is the Fourier transform of $\at_k(t)$. (Note that, for simplicity of notation, we
drop a tilde in $\af_k$.)
\begin{figure}
\includegraphics[scale=0.8]{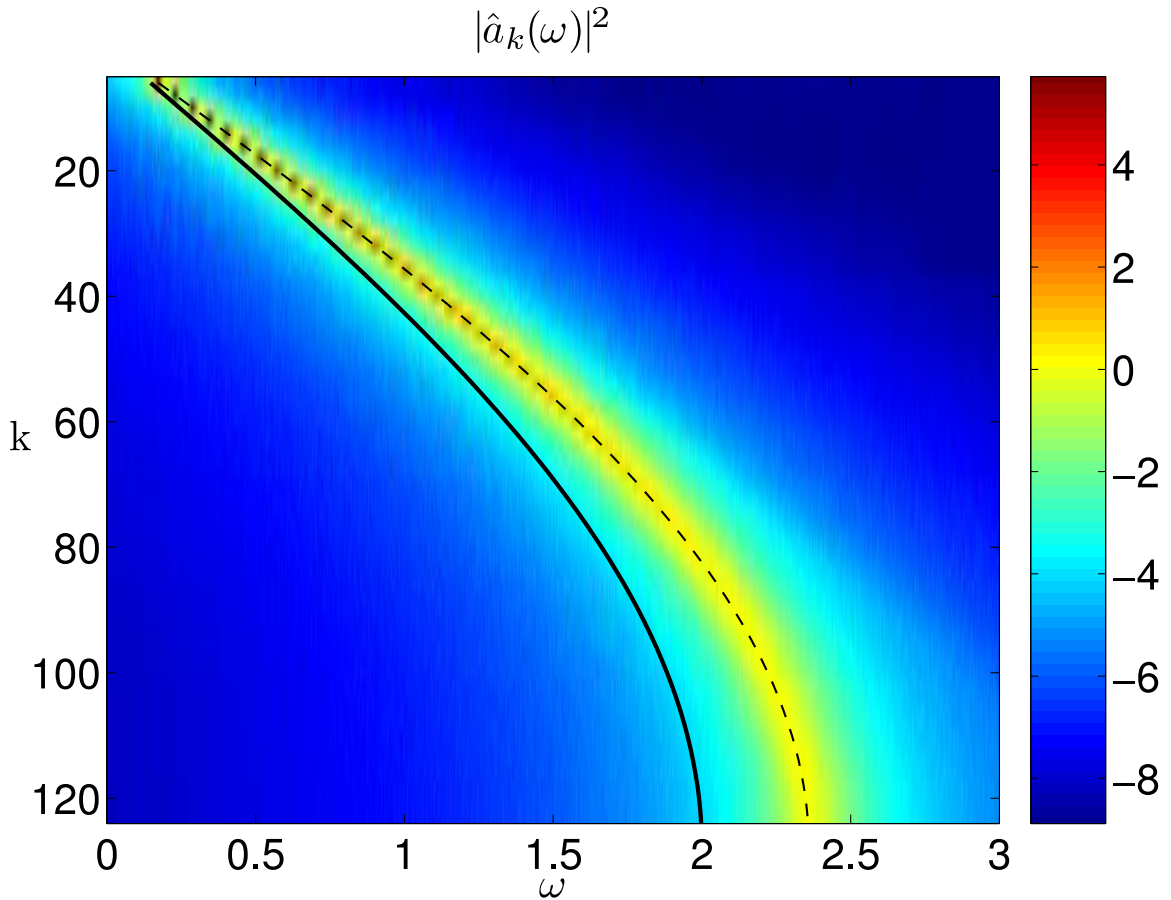}
\caption{The spatiotemporal spectrum $|\af_k(\w)|^2$ in thermal equilibrium.
The chain was modeled for $N=256$, $\beta=0.5$, and $E=100$.
[$\max\{-8,\ln{|\af_k(\w)|^2}\}$, with corresponding gray scale, is plotted for a clear presentation].
The solid curve corresponds to the usual linear dispersion $\w_k=2\sin(\pi k/N)$.
The dashed curve shows the locations of the actual frequency peaks of $|\af_k(\w)|^2$.}
\label{Fig_awk_num}
\end{figure}
Figure~\ref{Fig_awk_num} displays the spatiotemporal spectrum of $\at_k$, obtained from the simulation of the $\beta$-FPU chain for $N=256$, $\beta=0.5$, and $E=100$.
In order to measure the value of $\eta$ from the spatiotemporal spectrum, we use the following procedure.
For the fixed wave number $k$, the corresponding renormalization factor $\eta(k)$ is determined by the location of the center of the frequency spectrum $|\af_k(\w)|^2$,
i.e.,
\BEA
\eta(k)=\frac{\w_{c}(k)}{\w_k},~~\mbox{with}~~
\w_{c}(k)=\frac{{\int} \w|\af_k(\w)|^2~d\w}{\int|\af_k(\w)|^2~d\w}.\nonumber
\EEA
The renormalization factor $\eta(k)$ of each wave mode $k$ is shown in Fig.~\ref{Fig_etak}.
The numerical approximation $\bar{\eta}$ to the value of $\eta$ is obtained by averaging all $\eta(k)$, i.e.,
\BEA
\bar{\eta}=\frac{1}{N-1}\sum_{k=1}^{N-1}\eta(k)\nonumber.
\EEA
The renormalization factor for the case shown in Fig.~\ref{Fig_awk_num} is measured to be $\bar{\eta}\approx1.1824$.
It can be clearly seen in Fig.~\ref{Fig_etak} that $\eta(k)$ is nearly independent of $k$ and its variations around $\bar{\eta}$ are less then $0.3\%$.
We also compare the renormalization factor $\eta$ obtained from Eq.~(\ref{eta}) (solid line in Fig.~\ref{Fig_etak}) with its numerically computed approximation
$\bar{\eta}$ (dashed line in Fig.~\ref{Fig_etak}).
Equation~(\ref{eta}) gives the value $\eta\approx 1.1812$ and the difference between $\eta$ and $\bar{\eta}$
is less then $0.1\%$, which can be attributed to the statistical errors in the numerical measurement.
\begin{figure}
\includegraphics[scale=0.8]{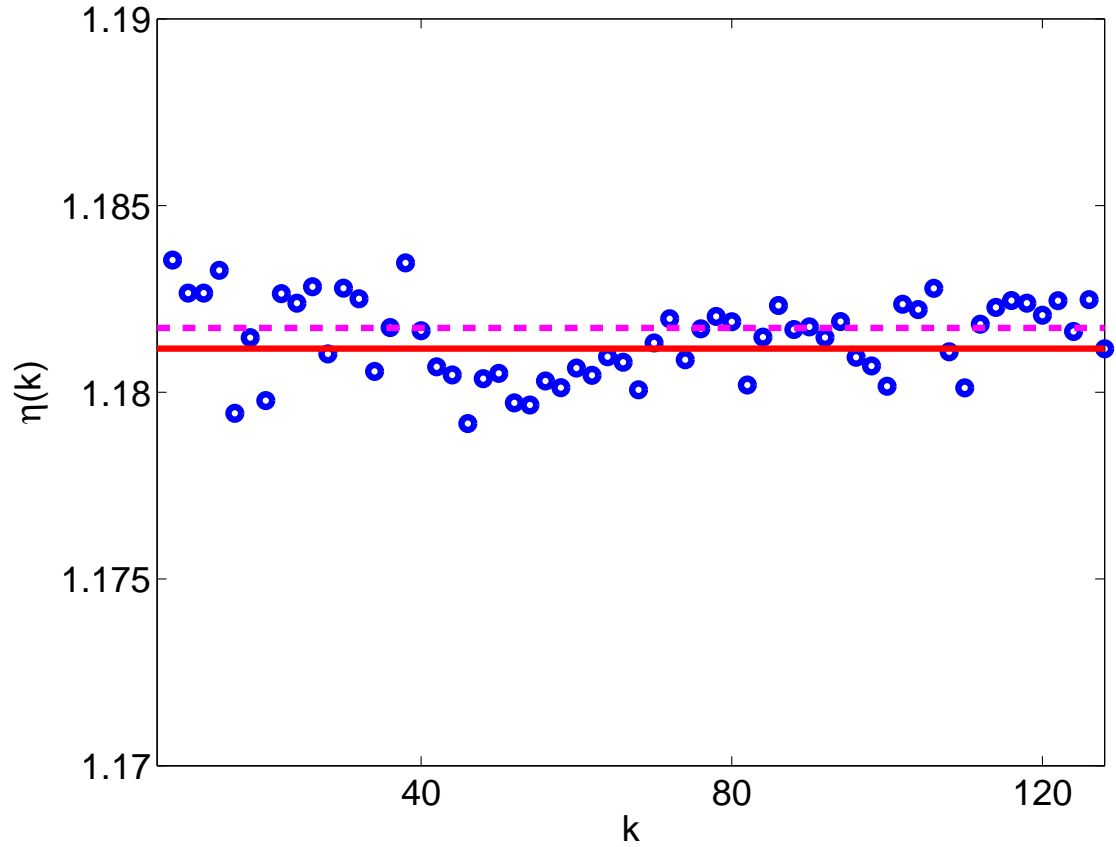}
\caption{Independence of $k$ of the renormalization factor $\eta(k)$.
The circles correspond to $\eta(k)$  obtained from the spatiotemporal spectrum shown in Fig.~\ref{Fig_awk_num}
[only even values of $k$ are shown for clarity of presentation].
The dashed line corresponds to the mean value $\bar{\eta}$.
For $\beta=0.5$, the mean value of the renormalization factor is found to be
$\bar{\eta}\approx1.1824$. The variations of $\eta_k$ around $\bar{\eta}$ are less then $0.3\%$. [Note the scale of the ordinate.]
The solid line corresponds to the renormalization factor $\eta$ obtained from Eq.~(\ref{eta}). For the given parameters $\eta\approx 1.1812$.
}
\label{Fig_etak}
\end{figure}
\begin{figure}
\includegraphics[scale=0.8]{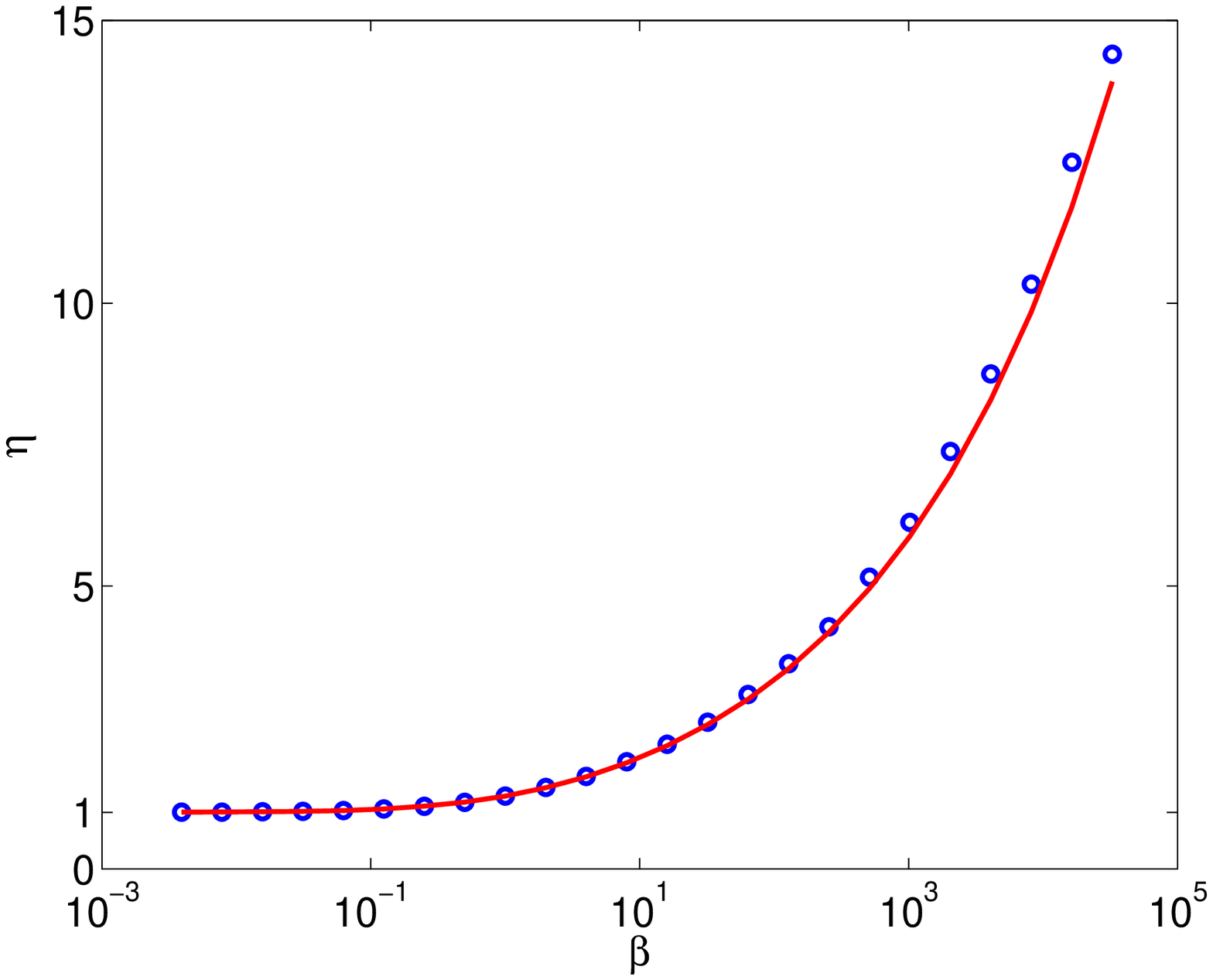}
\caption{The renormalization factor as a function of the nonlinearity strength $\beta$.
The analytical prediction [Eq.~(\ref{eta_G})] is depicted with a solid line and the numerical measurement is shown with circles.
The chain was modeled for $N=256$, and $E=100$.
}
\label{Fig_eta_all}
\end{figure}
In Fig.~\ref{Fig_eta_all}, we plot the value of $\eta$ as a function of $\beta$ for the system with $N=256$ particles and the total energy $E=100$.
The solid curve was obtained using Eq.~(\ref{eta_G}) while the circles correspond to the value of $\eta$ determined via the numerical spectrum $|\af_k(\w)|^2$ as
discussed above.
It can be observed that there is excellent agreement between the theoretic prediction and numerically measured values for a wide range of the nonlinearity
strength $\beta$.

\chapter{Dispersion relation and resonances}
\label{sect_dispersion}
In this chapter, we will discuss how the renormalization of the linear dispersion of the $\beta$-FPU chain in thermal equilibrium can be explained from
the wave resonance point of view.
This formalism is used in the theory of wave turbulence, that we briefly discussed in Chapter~\ref{sect_WT}.
In order to give a wave description of the $\beta$-FPU chain, we rewrite its Hamiltonian in terms of the renormalized variables $\at_k$.
Then, we study the interactions of the renormalized waves.
In order to address how the renormalized dispersion arises from wave interactions, we study the resonance structure of our nonlinear waves.
We will demonstarte that the $\beta$-FPU system is a Hamiltonian system with four-wave interactions.
We will discuss the properties of the resonance manifold associated with the $\beta$-FPU system %described by Eq.~(\ref{H_a})
as a first step towards the understanding
of its long time statistical behavior.
We comment that the resonance structure is one of the main objects of investigation in wave turbulence theory~\cite{ZLF,Newell,Lvov,Cai,MMT,Rink1,Rink2}.
The theory of wave turbulence focuses on the specific type of interactions, namely resonant interactions, which dominate long time statistical properties of the system.
On the other hand, the non-resonant interactions are usually shown to have a total vanishing average contribution to a long time dynamics.

In order to rewrite the Hamiltonian of the $\beta$-FPU chain given by Eq.~(\ref{H_FPU}) in terms of renormalized waves, we first transfer to the Fourier space using
Eq.~(\ref{PQ}). The Hamiltonian becomes
\BEA
H&=&\frac{1}{2}\sum_{k=1}^{N-1}|P_k|^2+\omega_k^2|Q_k|^2\nonumber\\
&+&\frac{\beta}{4N}\sum_{k,l,m,s=1}^{N-1}\omega_k\omega_l\omega_m\omega_s[-(Q_kQ_lQ_mQ_s^*\delta^{klm}_s+c.c.)+Q_kQ_lQ_m^*Q_s^*\delta^{kl}_{ms}].\nonumber\\
\label{Hamb_PQ}
\EEA
Here, we have used Eq.~(\ref{exp_formula}).
Next, we transfer to the renormalized waves $\at_k$ using the inverse of Eq.~(\ref{at})
\BEA
\BC
P_k=\displaystyle{\sqrt{\frac{\wt_k}{2}}}(\at_k+\at_{N-k}^*),\\
Q_k=\displaystyle{\frac{\i}{\sqrt{2\wt_k}}}(\at_k-\at_{N-k}^*).
\EC
\label{PQ_at}
\EEA
In order to give a wave description of the $\beta$-FPU chain, we rewrite Hamiltonian~(\ref{H_FPU}) in terms of
the renormalized variables $\at_k$~[Eq.~(\ref{at})] with $\wt_k=\eta\w_k$,
\BEA
H&=&\sum_{k=1}^{N-1}\frac{\w_k}{2}\left(\eta+\frac{1}{\eta}\right)|\at_k|^2+\frac{\w_k}{4}\left(\eta-\frac{1}{\eta}\right)(\at_k^*\at_{N-k}^*+\at_k\at_{N-k})\nonumber\\
&+&\sum_{k,l,m,s=1}^{N-1}T^{kl}_{ms}\Bigg[\D_{ms}^{kl}\at_k\at_l\at_m^*\at_s^*+\left(\frac{2}{3}\D^{klm}_{s}\at_k\at_l\at_m\at_s^*+c.c.\right)\nonumber\\
&+&\left(\frac{1}{6}\D^{klms}_{0}\at_k\at_l\at_m\at_s+c.c.\right)\Bigg],\label{H_a}
\EEA
where c.c. stands for complex conjugate, and
\BEA
T^{kl}_{ms}=\frac{3\beta}{8N\eta^2}\sqrt{\w_k\w_l\w_m\w_s}\label{T_inter}
\EEA is the interaction tensor coefficient.
Note that, due to the discrete nature of the system of finite size, the wave space is periodic and,
therefore, the ``momentum'' conservation is guaranteed by the following ``periodic'' Kronecker delta functions
\BEA
\D^{kl}_{ms}&\equiv&\delta_{ms}^{kl}-\delta^{klN}_{ms}-\delta^{kl}_{msN},\label{D22}\\
\D^{klm}_{s}&\equiv&\delta^{klm}_{s}-\delta^{klm}_{sN}+\delta^{klm}_{sNN},\label{D31}\\
\D^{klms}_{0}&\equiv&\delta^{klms}_{NN}-\delta^{klms}_{N}-\delta^{klms}_{NNN}.\label{D40}
\EEA
Here, the Kronecker $\delta$-function is equal to 1, if the sum of all superscripts is equal to the sum of all
subscripts, and 0, otherwise.
The periodic delta functions arise when we make the change of indices of the form $N-k\rightarrow k$ in the summation in order to obtain Eq.~(\ref{H_a}).

In analogy with quantum mechanics, where $a^+$ and $a$ are creation and annihilation operators, we can view $\at_k^*$ as the \textit{outgoing} wave with frequency
$\wt_k$ and $\at_k$ as the \textit{incoming} wave with frequency $\wt_k$.
\begin{figure}
\includegraphics[scale=0.8]{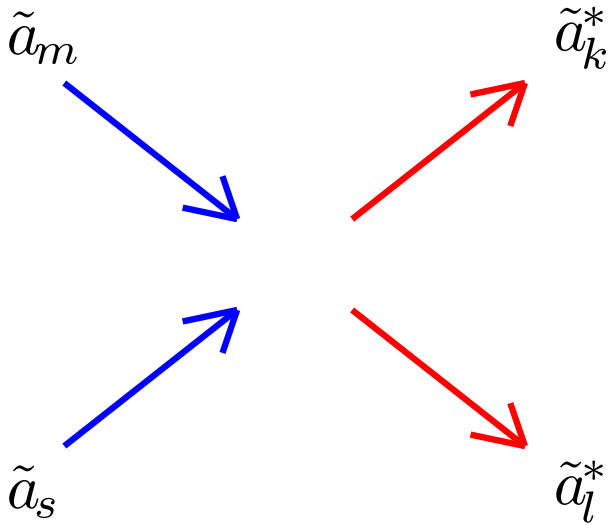}
\caption{Interaction process of type $(2\rightarrow 2)$ given by the term $\at_k^*\at_l^*\at_m\at_s\Delta^{kl}_{ms}$.}
\label{Fig_4wave}
\end{figure}
Then, the nonlinear term
\BEA
\at_k^*\at_l^*\at_m\at_s\Delta^{kl}_{ms}
\EEA
in system~(\ref{H_a}) is schematically shown in Fig.~\ref{Fig_4wave} and can be
 interpreted as the interaction process of the type
$(2\rightarrow 2)$, namely, two outgoing waves with wave numbers $k$ and $l$ are ``created'' as a result of interaction of the two incoming waves with
wave numbers $m$ and $s$.
Similarly, $\at_k^*\at_l\at_m\at_s\Delta_{k}^{lms}$ in system~(\ref{H_a}) describes the interaction process of the type $(3\rightarrow 1)$,
that is, one outgoing wave with wave number $k$ is ``created'' as a result of interaction of the three incoming waves with wave numbers $l$, $m$, and $s$, respectively.
Finally, $\at_k\at_l\at_m\at_s\Delta_{0}^{klms}$ describes the interaction process of the type $(4\rightarrow 0)$, i.e., all four incoming waves
interact and annihilate themselves.
Furthermore, the complex conjugate terms $\at_k\at_l^*\at_m^*\at_s^*\Delta_{lms}^{k}$ and $\at_k^*\at_l^*\at_m^*\at_s^*\Delta^{0}_{klms}$ describe the
interaction processes of the type $(1\rightarrow 3)$ and $(0\rightarrow 4)$, respectively.

Instead of the processes with the ``momentum'' conservation given via the usual $\delta^{kl}_{ms}$, $\delta^{klm}_s$, or $\delta^{klms}_0$ functions for an infinite
discrete system, the resonant
processes of the $\beta$-FPU chain of a \textit{finite} size are constrained to the manifold given by $\D^{kl}_{ms}$, $\D^{klm}_s$, or $\D^{klms}_0$, respectively.
Next, we describe these resonant manifolds in detail.
As will be pointed out in Chapter~\ref{sect_width}, there is a consequence of this \textit{finite} size effect to the properties of the renormalized waves.
\section{Resonance manifolds}
The resonance manifold that corresponds to the $(2\rightarrow 2)$ resonant processes in the discrete periodic system, therefore, is described by
\BEA
\BC
k+l\eqN m+s,\\
\wt_k+\wt_l=\wt_m+\wt_s,
\EC\label{2to2}
\EEA
where we have introduced the notation $g\eqN h$, which means that
$$
\left[
\begin{array}{ll}
         g=h,\\
         g=h+N,\\
         g=h-N,
\end{array}
\right.
$$
for any $g$ and $h$.
The first equation in system~(\ref{2to2}) is the ``momentum'' conservation condition in the periodic wave number space.
This ``momentum'' conservation comes from
\BEA
|\D^{kl}_{ms}|=1.
\EEA
Note that $|\D^{kl}_{ms}|$ can assume only the value of $1$ or $0$.
Similarly, from $|\D^{klm}_{s}|=1$ and $|\D^{klms}_{0}|=1$, the resonance manifolds corresponding to the resonant processes of types
$(3\rightarrow 1)$ and $(4\rightarrow 0)$ are given by
\BEA
\BC
k+l+m\eqN s,\\
\wt_k+\wt_l+\wt_m=\wt_s,
\EC\label{3to1}
\EEA
and
\BEA
\BC
k+l+m+s\eqN 0,\\
\wt_k+\wt_l+\wt_m+\wt_s=0,
\EC\label{4to0}
\EEA
respectively.
For the processes of type $(3\rightarrow 1)$, the notation $g\eqN h$ means that
$$
\left[
\begin{array}{ll}
         g=h,\\
         g=h+N,\\
         g=h+2N.
\end{array}
\right.
$$
For the $(4\rightarrow 0)$ processes, $g\eqN h$ means that
$$
\left[
\begin{array}{ll}
         g=h+N,\\
         g=h+2N,\\
         g=h+3N.
\end{array}
\right.
$$
\section{Trivial resonances of the type $(2\rightarrow 2)$}
\label{sect_umklapp}
To solve system (\ref{2to2}), we rewrite it in a continuous form with
\BEA
\BC
x={k}/{N},\\
y={l}/{N},\\
z={m}/{N},\\
v={s}/{N},
\EC\label{changekx}
\EEA
which are real numbers
in the interval $(0,1)$.
By recalling that $\wt_k=2\eta\sin({\pi k}/{N})$, we have
\BEA
\BC
x+y\eq1 z+v,\\
\sin(\pi x)+\sin(\pi y)=\sin(\pi z)+\sin(\pi v).
\EC\label{2to2cont}
\EEA
Thus, any rational quartet that satisfies Eq.~(\ref{2to2cont}) yields a solution for Eq.~(\ref{2to2}).
There are two distinct types of solutions for Eq.~(\ref{2to2cont}).
The first one is given by
\BEA
\BC
x+y=z+v,\\
\sin(\pi x)+\sin(\pi y)=\sin(\pi z)+\sin(\pi v).
\EC\label{2to2cont_triv}
\EEA
Here we show that the system~(\ref{2to2cont_triv}) has only trivial solutions.
We express $v$ from the first equation in~(\ref{2to2cont}) and insert it into the second equation
\BEA
\sin(\pi x)+\sin(\pi y)=\sin(\pi z)+\sin(\pi (x+y-z)).\label{2to2sin}
\EEA
Using Prosthaphaeresis formulas, we can rewrite Eq.~(\ref{2to2sin}) as
\BEA
\sin\left(\pi\frac{x+y}{2}\right)\sin\left(\pi(x-z)\right)\sin\left(\pi(y-z)\right)=0.\label{2to2sin2}
\EEA
By recalling that $0<x,y,z<1$, we obtain that the only solutions for the system~(\ref{2to2cont_triv}) are given by
\BEA
\BC
x=z,\\
y=v,
\EC \mbox{or}~~
\BC
x=v,\\
y=z,\label{trivial}
\EC
\EEA
i.e., these are \textit{trivial} resonances, as we mentioned above.
\section{Non-trivial resonances of the type $(2\rightarrow 2)$}
\label{sect_umlklapp}
The second type of the resonance manifold of the $(2\rightarrow 2)$-type interaction processes corresponds to
\BEA
\BC
x+y=z+v\pm 1,\\
\sin(\pi x)+\sin(\pi y)=\sin(\pi z)+\sin(\pi v),
\EC\label{2to2cont_nontriv}
\EEA
In order to solve Eq.~(\ref{2to2cont_nontriv}), we pull out $v$ from the first equation and substitute it to the second equation.
Then we use trigonometric addition formulas to obtain the following two branches
\BEA
z_1&=&\frac{x+y}{2}+\frac{1}{\pi}\arcsin(A)+2j,\label{folded1}\\
z_2&=&\frac{x+y}{2}-1-\frac{1}{\pi}\arcsin(A)+2j,\label{folded2}
\EEA
where $A\equiv\tan\left(\pi({x+y})/{2}\right)\cos\left(\pi({x-y})/{2}\right)$ and $j$ is an integer.

The second type of resonances arises from the discreteness of our model of a finite length, leading to \textit{non-trivial} resonances.
For our linear dispersion here, non-trivial resonances are only those resonances that involve wave numbers crossing the first Brillouin zone.
This process is known as the Umklapp scattering in the setting of phonon scattering~\cite{Umklapp}.
Note that, in the previous studies~\cite{Kramer,Kramer2} of the FPU chain from the wave turbulence point of view,
the effects arising from the finite  nature of the chain were not taken into account, i.e., only the limiting case of $N\rightarrow\infty$,
where $N$ is the system size, was considered.
\section{Numerical observations of the resonances}
\begin{figure}
\includegraphics[scale=1]{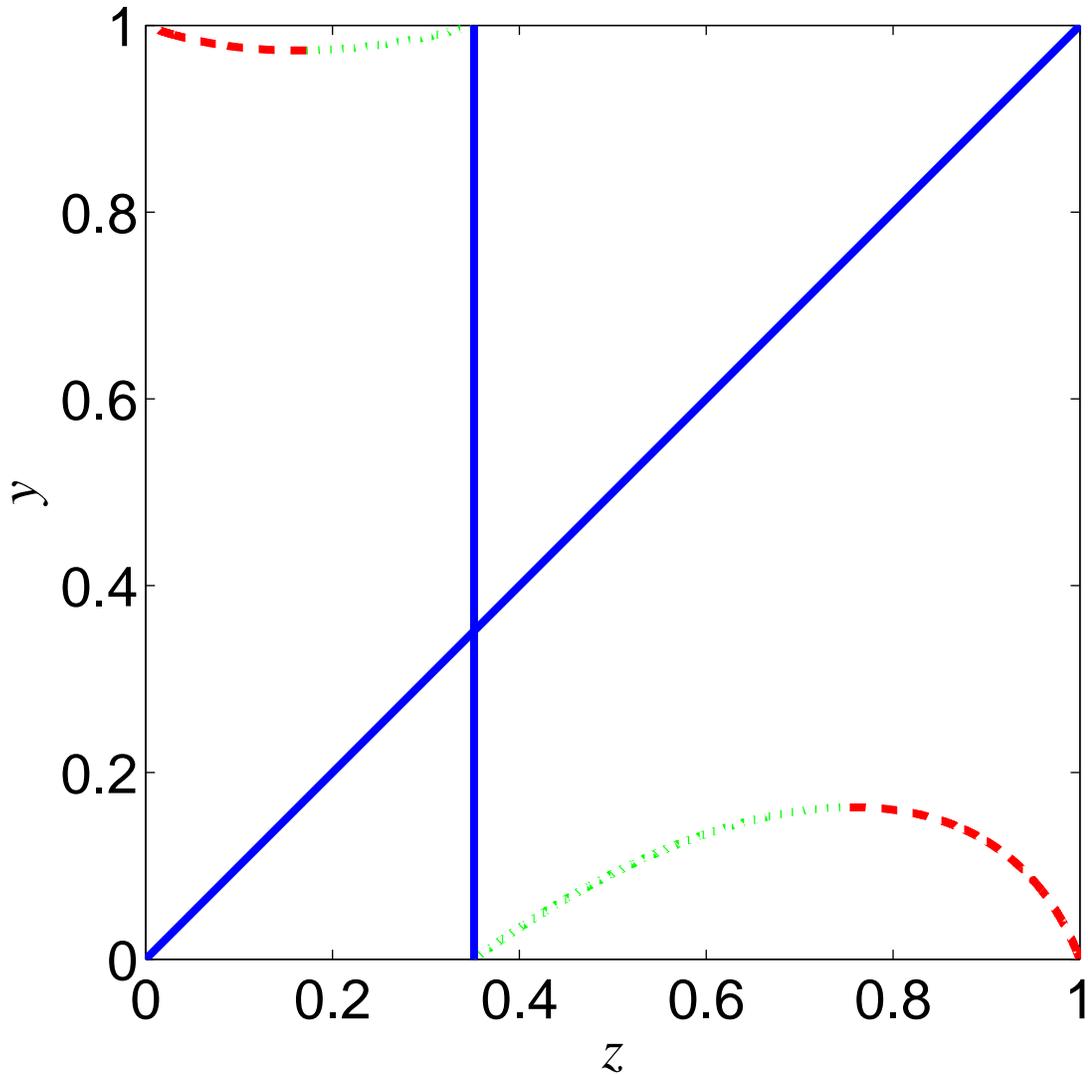}
\caption{The solutions of Eq.~(\ref{2to2cont}). The solid straight lines correspond to the trivial resonances [solutions of Eq.~(\ref{trivial})].
The solutions are shown for fixed $x=k/N$, $k=90$, $N=256$ as the fourth wave number $v$ scans from ${1}/{N}$ to $(N-1)/{N}$ in the resonant quartet
Eq.~(\ref{2to2cont}). The non-trivial resonances are described by the dotted or dashed curves.
The dotted branch of the curves corresponds to the non-trivial resonances described by Eq.~(\ref{folded1}) and the dashed branch corresponds to the
non-trivial resonances
described by Eq.~(\ref{folded2}).}
\label{Fig_resonances}
\end{figure}

\begin{figure}
\includegraphics[scale=0.8]{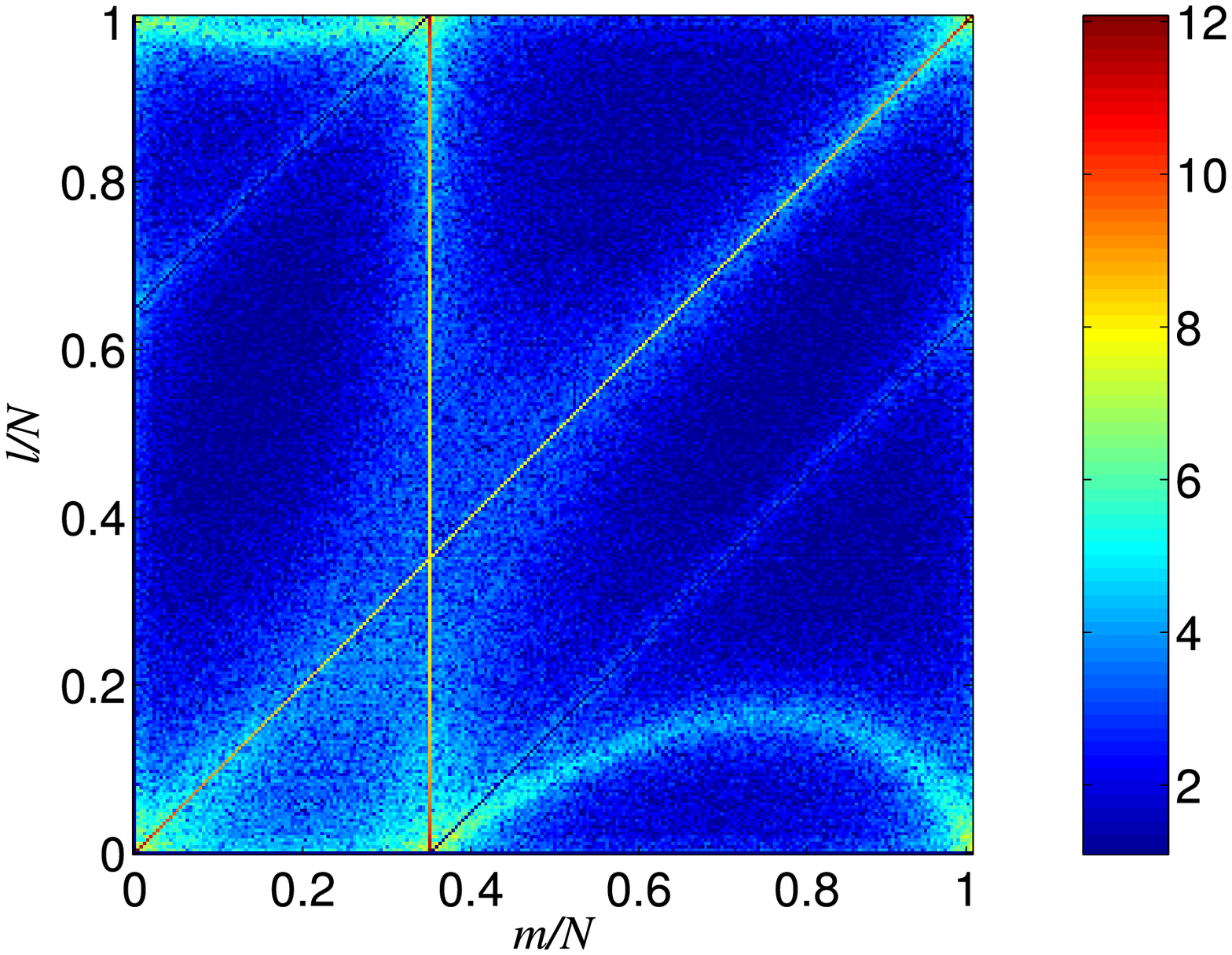}
\caption{The long time average $|\la\at_k^*\at_l^*\at_m\at_s\ra\D^{kl}_{ms}|$ of the $\beta$-FPU system in thermal equilibrium.
The parameters for the FPU chain are $N=256$, $\beta=0.5$, and $E=100$.
$\la\at_k^*\at_l^*\at_m\at_s\ra\D^{kl}_{ms}$ was computed for fixed $k=90$.
The darker grayscale corresponds to the larger value of $\la\at_k^*\at_l^*\at_m\at_s\ra\D^{kl}_{ms}$.
The exact solutions of Eq.~(\ref{2to2cont}), which are shown
in Fig.~\ref{Fig_resonances}, coincide with the locations of the peaks of $|\la\at_k^*\at_l^*\at_m\at_s\ra\D^{kl}_{ms}|$.
Therefore, the lighter areas represent the near-resonance structure of the finite $\beta$-FPU chain.
(The two dark lines show the locations, where $s=0$ and, therefore, $\at_k^*\at_l^*\at_m\at_s\D^{kl}_{ms}=0$.)
[$\max\{2,\ln(|\la\at_k^*\at_l^*\at_m\at_s\ra\D^{kl}_{ms}|)\}$ with the corresponding color-scale is plotted for a clean presentation].}
\label{Fig_resonances_num}
\end{figure}

In Fig.~\ref{Fig_resonances}, we plot the solution of Eq.~(\ref{2to2cont}) for $x={k}/{N}$ with the wave number $k=90$ for the system with $N=256$ particles
(the values of $k$ and $N$ are chosen merely for the purpose of illustration).
We stress that \textit{all} the solutions of the system~(\ref{2to2cont}) are given by the Eqs.~(\ref{trivial}),~(\ref{folded1}), and~(\ref{folded2}), and that
the non-trivial solutions arise only as a consequence of discreteness of the finite chain.
The curves in Fig.~\ref{Fig_resonances} represent the loci of $(z,y)$, parameterized by the fourth wave number $v$, i.e., $x$, $y$, $z$, and $v$ form a resonant
quartet, where $z={m}/{N}$, and $y={l}/{N}$.
Note that the fourth wave number $v$ is specified by the ``momentum'' conservation, i.e., the first equation in Eq.~(\ref{2to2cont}).
The two straight lines in Fig.~\ref{Fig_resonances} correspond to the trivial solutions, as given by Eq.~(\ref{trivial}).
The two curves (dotted and dashed) depict the non-trivial resonances.
Note that the dotted part of non-trivial resonance curves corresponds to branch~(\ref{folded1}), and the dashed part corresponds to branch~(\ref{folded2}),
respectively.
An immediate question arises: how do these resonant structures manifest themselves in the FPU dynamics in the thermal equilibrium?
By examining the Hamiltonian~(\ref{H_a}), we notice that the resonance will control the contribution of terms like $\at_k^*\at_l^*\at_m\at_s\D^{kl}_{ms}$
in the long time limit.
Therefore, we address the effect of resonance by computing long time average, i.e., $\la\at_k^*\at_l^*\at_m\at_s\ra\D^{kl}_{ms}$,
and comparing this average (Fig.~\ref{Fig_resonances_num}) with  Fig.~\ref{Fig_resonances}.
To obtain Fig.~\ref{Fig_resonances_num}, the $\beta$-FPU system was simulated with the following parameters: $N=256$, $\beta=0.5$, $E=100$, and the averaging time
window $\tau=400\tilde{t}_1$, where $\tilde{t}_1$ is the longest linear period, i.e., $\tilde{t}_1={2\pi}/{\wt_1}$.
In Fig.~\ref{Fig_resonances_num}, mode $k$ was fixed with $k=90$ and the mode $s$, a function of $k$, $l$, and $m$, is obtained from the constraint
$k+l\eqN m+s$, i.e., $|\D^{kl}_{ms}|=1$.
Note that we do not impose here the condition $\wt_k+\wt_l=\wt_m+\wt_s$,
therefore, $|\la\at_k^*\at_l^*\at_m\at_s\ra\D^{kl}_{ms}|$ is a function of $l$ and $m$.
The two dark lines show the locations, where $s=0$ and, therefore, $\at_k^*\at_l^*\at_m\at_s\D^{kl}_{ms}=0$.
By comparing Figs.~\ref{Fig_resonances} and \ref{Fig_resonances_num}, it can be observed that the locations of the peaks of the long time average
$|\la\at_k^*\at_l^*\at_m\at_s\ra\D^{kl}_{ms}|$ coincide with the loci of the $(2\rightarrow 2)$-type resonances.
This observation demonstrates that, indeed, there are nontrivial $(2\rightarrow 2)$-type resonances in the finite $\beta$-FPU chain in thermal equilibrium.
Furthermore, it can be observed in Fig.~\ref{Fig_resonances_num} that, in addition to the fact that the resonances manifest themselves as the locations of the peaks
of $|\la\at_k^*\at_l^*\at_m\at_s\ra\D^{kl}_{ms}|$, the structure of \textit{near-resonances} is reflected in the finite width of the peaks around the loci of the exact
resonances.
Note that, due to the discrete nature of the finite $\beta$-FPU system, only those solutions $x$, $y$, $z$, and $v$ of Eq.~(\ref{2to2cont}),
for which $Nx$, $Ny$, $Nz$, and $Nv$ are integers, yield solutions $k$, $l$, $m$, and $s$ for Eq.~(\ref{2to2}).
In general, the rigorous treatment of the exact integer solutions of Eq.~(\ref{2to2}) is not straightforward.
For example, for $N=256$, we have the following two exact quartets $\vec{k}=\{k,l,m,s\}$:
$$
\left[
\begin{array}{ll}
         \vec{k}=\{k,N/2-k,N/2+k,N-k\},\\
         \vec{k}=\{k,N/2-k,N-k,N/2+k\},\\
\end{array}
\right.
$$
for $k<N/2$, and
$$
\left[
\begin{array}{ll}
         \vec{k}=\{k,3N/2-k,k-N/2,N-k\},\\
         \vec{k}=\{k,3N/2-k,N-k,k-N/2\},\\
\end{array}
\right.
$$
for $k>N/2$.
We have verified numerically that for $N=256$ there are no other exact integer solutions of Eq.~(\ref{2to2}).
In the analysis of the resonance width in Chapter~\ref{sect_width}, we will use the fact that the number of exact non-trivial resonances [Eq.~(\ref{2to2})]
is significantly smaller than the total number of modes.
\section{Near-resonances}
The broadening of the resonance peaks in Fig.~\ref{Fig_resonances_num} suggests that, to capture the near-resonances for characterizing long time statistical
behavior of the $\beta$-FPU system in thermal equilibrium, instead of Eq.~(\ref{2to2}), one  needs to consider
the following effective system
\BEA
\BC
k+l\eqN m+s,\\
|\wt_k+\wt_l-\wt_m-\wt_s|<\Delta\w,
\EC\label{2to2near}
\EEA
where $0<\Delta\w\ll\wt_k$ for any $k$, and $\Delta\w$ characterizes the resonance width, which results from the near-resonance structure.
Clearly, $\D\w$ is related to the broadening of the spectral peak of each wave $\at_{\a}(t)$ with $\a=k,l,m$, or $s$ in the quartet,
and this broadening effect will be studied in detail in Chapter~\ref{sect_width}.
Note that the structure of near-resonances is a common characteristic of many periodic discrete nonlinear wave systems~\cite{Lvov2,Colm,Pushkarev}.
\section{Resonances of the type $(3\rightarrow 1)$ and $(4\rightarrow 0)$}
Next, we prove that there are no exact
$(3\rightarrow 1)$-type resonances [Eq.~(\ref{3to1})] in the $\beta$-FPU chain.
We change variables according to Eq.~(\ref{changekx}) to obtain
\BEA
\BC
x+y+z\eq1 v,\\
\sin(\pi x)+\sin(\pi y)+\sin(\pi z)=\sin(\pi v),
\EC\label{app3_3to1_cont}
\EEA
with
$0<\{x,y,z,v\}<1$.
The first equation in~(\ref{app3_3to1_cont}) implies that either $x+y+z=v$, or $x+y+z=v+1$, or $x+y+z=v+2$.
Since the treatment of all three cases is similar, we only consider the second one in details
\BEA
\BC
x+y+z=v+1,\\
\sin(\pi x)+\sin(\pi y)=-\sin(\pi z)+\sin(\pi v),
\EC\label{app3_2}
\EEA
Denote $D\equiv|\sin(\pi(x+y))|$.
Then, using the trigonometric addition formulas, the properties of the modulus, and the fact that $\sin(\pi x)>0$ for $x\in(0,1)$,
we obtain for the right-hand side of Eq.~(\ref{app3_2})
\BEA
D&=&|\sin(\pi(x+y))|\nonumber\\
&\leq&|\sin(\pi x)||\cos(\pi y)|+|\sin(\pi y)||\cos(\pi x)|\nonumber\\
&<&\sin(\pi x)+\sin(\pi y).\label{app3_lhs}
\EEA
Now, consider the left-hand side of Eq.~(\ref{app3_2})
\BEA
& &\sin(\pi v)-\sin(\pi z)=\sin(\pi(x+y+z-1))-\sin(\pi z)\nonumber\\
& &=\sin(\pi(x+y-1))\cos(\pi z)\nonumber\\
& &~~+\sin(\pi z)\cos(\pi(x+y-1))-\sin(\pi z)\nonumber\\
& &\leq|\sin(\pi(x+y-1))||\cos(\pi z)|\nonumber\\
& &~~+|\sin(\pi z)||\cos(\pi(x+y))|-\sin(\pi z)\nonumber\\
& &<|\sin(\pi(x+y))|=D,\label{app3_rhs}
\EEA
where the use is made of $|\sin(\alpha-\pi)|=|\sin(\alpha)|$.
Combining Eqs.~(\ref{app3_lhs}) and (\ref{app3_rhs}), we obtain
\BEA
\sin(\pi v)-\sin(\pi z)<D<\sin(\pi x)+\sin(\pi y).\label{app3_nonequiv}
\EEA
From inequality~(\ref{app3_nonequiv}), it follows that Eq.~(\ref{app3_2}) has no solutions, and, therefore, there are no exact resonances of type $(3\rightarrow 1)$
in the $\beta$-FPU chain.
Now it is apparent that all the nonlinear terms
$\at_k^*\at_l\at_m\at_s\D^{k}_{lms}$ are non-resonant and their long time average $\la\at_k^*\at_l\at_m\at_s\ra\D^{k}_{lms}$ vanishes.
As for the resonances of type $(4\rightarrow 0)$, since the dispersion relation is non-negative, one can immediately conclude that
the solution of the system (\ref{4to0}) consists only of zero modes.
Therefore, the processes of type $(4\rightarrow 0)$ are also non-resonant, giving rise to $\la\at_k\at_l\at_m\at_s\ra\D_{0}^{klms}=0$.
In this thesis, we will neglect the higher order effects of the near-resonances of the types $(3\rightarrow 1)$ and $(4\rightarrow 0)$.

In the following Chapters, we will study the effects of the resonant terms of type $(2\rightarrow 2)$, namely, the linear dispersion renormalization and the
broadening of the frequency peaks of $\at_k(t)$. It turns out that, the former is related to the trivial resonance of type $(2\rightarrow 2)$ and the latter
is related to the near-resonances, as will be seen below.

\chapter{Self-consistency approach to frequency renormalization}
\label{sect_self_consistency}
We now turn to the discussion of how the trivial resonances give rise to the dispersion renormalization.
This question was examined in~\cite{our_prl} before.
There, it was shown that the renormalization of the linear dispersion of the $\beta$-FPU chain arises due to the collective effect of the nonlinearity.
In particular, the trivial resonant interactions of type $(2\rightarrow 2)$, i.e., the solutions of Eq.~(\ref{trivial}),
enhance the linear dispersion (the renormalized dispersion relation takes the form $\wt_k=\eta\w_k$ with $\eta>1$), and effectively weakens the nonlinear interactions.
Here, we further address this issue and present a self-consistency argument to arrive at an approximation $\eta_{sc}$ for the renormalization factor $\eta$.
As will be seen below, the self-consistency argument essentially is of a mean-field type, i.e., the renormalization arises from the scattering of a wave by
a mean-background of waves in thermal equilibrium via trivial resonant interactions.
We note that our self-consistency, mean-field argument is not limited to the weak nonlinearity.
Very good agreement of the renormalization factor $\eta$ and its dynamical approximation $\eta_{sc}$ --- for weakly as well as strongly nonlinear waves ---
confirms that the renormalization is, indeed, \textit{a direct consequence} of the trivial resonances.
\section{Mean-field approximation of the linear dispersion}
As it was mentioned above, the contribution of the non-resonant terms have a vanishing long time effect to the statistical properties of the system,
therefore, in our self-consistent approach, we ignore these non-resonant terms.
By removing the non-resonant terms and using the canonical transformation
\BEA
\at_k=\frac{P_k-\i\eta_{sc}\w_k Q_k}{\sqrt{2\eta_{sc}\w_k}},\label{a_sc}
\EEA
where $\eta_{sc}$ is a factor to be determined,
we arrive at a simplified effective Hamiltonian from Eq.~(\ref{H_a}) for the finite $\beta$-FPU system
\BEA
H_{\rm{eff}}&=&\sum_{k=1}^{N-1}\frac{\w_k}{2}\left(\eta_{sc}+\frac{1}{\eta_{sc}}\right)|\at_k|^2
+\sum_{k,l,m,s=1}^{N-1}T^{kl}_{ms}\Delta^{kl}_{ms}\at_k^*\at_l^*\at_m\at_s.\label{H_a_res}
\EEA
The ``off-diagonal'' quadratic terms $\at_k\at_{N-k}$ from Eq.~(\ref{H_a}) are not present in Eq.~(\ref{H_a_res}),
since $\at_k$ are chosen so that $\la\at_k\at_{N-k}\ra=0$ (see Chapter~\ref{sect_renormalization}, Eqs.~(\ref{asa_lin}) and~(\ref{aa_lin})).
Note that Hamiltonian~(\ref{H_a_res}) is written in a standard four-wave interaction form that is widely used in wave turbulence~\cite{Lvov}.
However, for the discrete system like FPU one should take into account Umklapp processes that we discussed in Section~\ref{sect_umklapp}.

\begin{figure}
\includegraphics[scale=1]{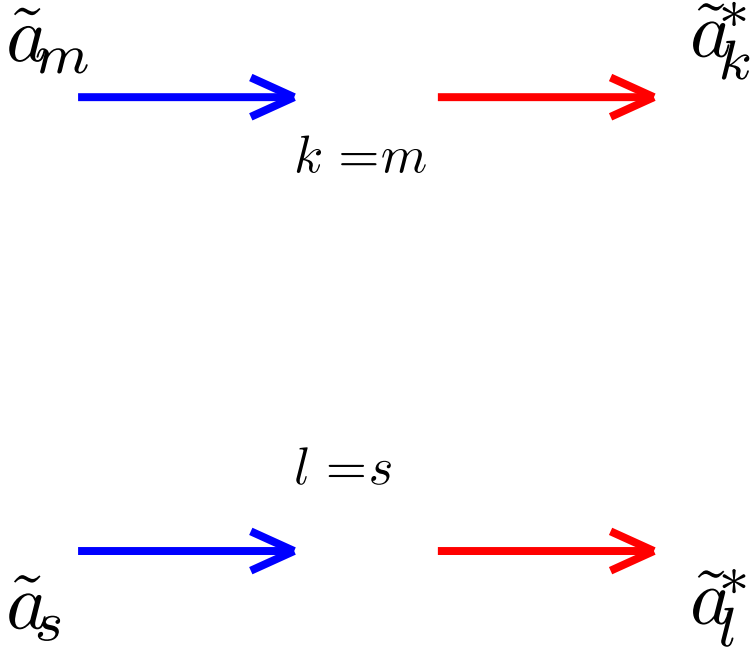}
\caption{Trivial resonant interactions given by Eq.~(\ref{trivial_k}).}
\label{Fig_4wave_trivial}
\end{figure}

Next we will obtain an approximation to the frequency renormalizing factor $\eta$ as a consequence of the trivial \textit{resonance} interactions.
As we have discussed above, the nonlinear terms in Eq.~(\ref{H_a_res})
\BEA
\at_k^*\at_l^*\at_m\at_s\Delta^{kl}_{ms},\nonumber
\EEA
can be interpreted as interaction processes when two waves $\at_k$ and $\at_l$ are created as a result of
the interaction of two other waves $\at_m$ and $\at_s$~[Fig.~\ref{Fig_4wave}].
From Eq.~(\ref{trivial}), we obtain that the trivial resonant interactions are characterized by the conditions
\BEA
\BC
k=m,\\
l=s,
\EC \mbox{or}~~
\BC
k=s,\\
l=m.\label{trivial_k}
\EC
\EEA
In Fig.~\ref{Fig_4wave_trivial}, we show schematically the trivial resonant interaction process.
The contribution of the \textit{trivial} resonances in $H_{\rm{eff}}$ is
\BEA
H_4^{\rm{tr}}=4\sum_{k,l=1}^{N-1}T^{kl}_{kl}|\at_l|^2|\at_k|^2,\label{H4_diag}
\EEA
which can be ``linearized'' in the sense that averaging the coefficient in front of $|\at_k|^2$ in $H_4^{\rm{tr}}$ gives rise to a quadratic form
\BEA
H_2^{\rm{tr}}\equiv\sum_{k=1}^{N-1}\left(4\sum_{l=1}^{N-1}T^{kl}_{kl}\la|\at_l|^2\ra\right)|\at_k|^2.\nonumber
\EEA
Note that the subscript $2$ in $H_2^{\rm{tr}}$ emphasizes the fact that $H_2^{\rm{tr}}$ now can be viewed as a Hamiltonian for the
free waves with the familiar effective linear dispersion~\cite{our_prl,ZLF}
\BEA
\Omega_k=4\sum_{l=1}^{N-1}T^{kl}_{kl}\la|\at_l|^2\ra.\label{Omega}
\EEA
This linearization is essentially a mean-field approximation, since the long-time average of trivial resonances in Eq.~(\ref{H4_diag}) is
approximated by the interaction of waves $\at_k$ with background waves $\la |\at_l|\ra$.
The self consistency condition, which determines $\eta_{sc}$, can be imposed as follows.
The quadratic part of the Hamiltonian (\ref{H_a_res}), combined with the
``linearized'' quadratic part, $H_2^{\rm{tr}}$, of the quartic $H_4^{\rm{tr}}$, should be equal to an effective quadratic Hamiltonian for the renormalized waves, i.e.,
\BEA
\tilde{H}_2=\sum_{k=1}^{N-1}\wt_k|\at_k|^2.\label{Heff}
\EEA
Therefore, the equation for the renormalization parameter $\eta_{sc}$ via the self consistency argument becomes
\BEA
\sum_{k=1}^{N-1}\frac{\w_k}{2}\left(\eta_{sc}+\frac{1}{\eta_{sc}}\right)|\at_k|^2+\sum_{k=1}^{N-1}\left(4\sum_{l=1}^{N-1}T^{kl}_{kl}\la|\at_l|^2\ra\right)|\at_k|^2=
\sum_{k=1}^{N-1}\wt_k|\at_k|^2,\nonumber\\\label{self}
\EEA
where $\wt_k$ is the renormalized linear dispersion, which is used in the definition of our renormalized wave, Eq.~(\ref{a_sc}), and $\wt_k=\eta_{sc}\w_k$.
Equating the coefficients of $\w_k|\at_k|^2$ on both sides of Eq.~(\ref{self}) for every wave number $k$ yields
\BEA
\frac{1}{2}\left(\eta_{sc}+\frac{1}{\eta_{sc}}\right)+4\sum_{l=1}^{N-1}\frac{3\beta}{8N\eta_{sc}^2}\w_l\langle |\at_l|^2\rangle=\eta_{sc},\nonumber
\EEA
where use is made of Eq.~(\ref{T_inter}). After algebraic simplification, we have the following equation for $\eta_{sc}$
\BEA
\eta_{sc}^3-\eta_{sc}=\frac{3\beta}{N}\sum_{l=1}^{N-1}\w_l\langle|\at_l|^2\rangle.\label{eq_eta}
\EEA
Using the property~(\ref{aas_G}) of the renormalized normal variables $\at_k$, we find the following dependence of $\la|\at_k|^2\ra$ on $\eta_{sc}$,
\BEA
\la|\at_l|^2\ra=\frac{1}{2\eta_{sc}\w_l}\left(\la|P_l|^2\ra+\eta_{sc}^2\w_l^2\la|Q_l|^2\ra\right).\label{rhs}
\EEA
Combining Eqs.~(\ref{eq_eta}) and (\ref{rhs}) leads to
\BEA
\eta_{sc}^4-A\eta_{sc}^2-B=0,\label{eq_eta2}
\EEA
where
\BEA
A&=&1+\frac{3\beta}{2N}\sum_{l=1}^{N-1}\w_l^2\langle|Q_l|^2\rangle=1+\frac{3\beta}{N}\la U\ra,\nonumber\\
B&=&\frac{3\beta}{2N}\sum_{l=1}^{N-1}\langle|P_l|^2\rangle=\frac{3\beta}{N}\la K\ra.\nonumber
\EEA
Here, $\la U\ra$ and $\la K\ra$ are the averaged quadratic potential and kinetic energies of the system.
The only physically relevant solution of Eq.~(\ref{eq_eta2}) is
\BEA
\eta_{sc}=\sqrt{\frac{A+\sqrt{A^2+4B}}{2}}.\label{new_eta}
\EEA
The constants $A$ and $B$ can be easily derived using the Gibbs measure
\BEA
A&=&1+\frac{3\b}{2}\la y^2\ra=1+\frac{3\beta\int y^2\exp\left({-\frac{1}{2\T}(y^2+\beta\frac{y^4}{2})}\right)~dy}
                                      {2\int \exp\left({-\frac{1}{2\T}(y^2+\beta\frac{y^4}{2})}\right)~dy},\label{A_G}\\
B&=&\frac{3\b}{2}\la p^2\ra=\frac{3\beta\T}{2}.\label{B_G}
\EEA
See Section~\ref{sect_Gibbs} for more details about computations with the Gibbs measure.
\section{Limiting behavior of $\eta$}
We have studied the frequency renormalization factor $\eta$ and its approximation via the self-consistency argument $\eta_{sc}$.
Here, we compare the behavior of both $\eta$ and $\eta_{sc}$ in the weakly nonlinear (small $\beta$) and strongly nonlinear (large $\beta$) limits.
In order to study the limiting behavior of $\eta$ and $\eta_{sc}$, we use the canonical Gibbs measure as discussed in Section~\ref{sect_Gibbs}.
There, we obtained the following expressions for the pdf's for the momentum $p_j$ and displacement $y_j$.
Any $p_j$ is distributed with the Gaussian pdf
\BEA
dw_p=Z_p\exp\left({-\frac{p^2}{2\T}}\right),
\EEA
and any $y_j$ is distributed with the pdf
\BEA
dw_y=Z_y\exp\left({-\frac{1}{\T}\left(\frac{y^2}{2}+\beta\frac{y^4}{4}\right)}\right),
\EEA
where $Z_p$ and $Z_y$ are the normalizing constants.
The renormalization factor $\eta$ of the $\beta$-FPU system in thermal equilibrium is given by Eq.~(\ref{eta}), and its approximation via
the self-consistency argument $\eta_{sc}$ is given by Eq.~(\ref{new_eta}).
We will use the following expressions for the average density of kinetic, quadratic potential and quartic
potential energies per degree of freedom of the system, defined in Eqs.~(\ref{V}), (\ref{K}) and~(\ref{U})
\BEA
\frac{\la K\ra}{N}&=&\frac{1}{N}\sum_{j=1}^N\frac{\la p_j^2\ra}{2}=\frac{1}{2}\T,\label{Hp2}\\
\frac{\la U\ra}{N}&=&\frac{1}{N}\sum_{j=1}^N\frac{\la y_j^2\ra}{2}=\frac{1}{2}\la y^2\ra,\label{Hy2}\\
\frac{\la V\ra}{N}&=&\frac{\beta}{4N}\sum_{j=1}^N\la y_j^4\ra=\frac{\beta}{4}\la y^4\ra.\label{Hy4}
\EEA
In a canonical ensemble, the temperature of a system is given by the
temperature of the heat bath. By identifying the average energy
density of the system with $\eb=E/N$ in our simulation (a
microcanonical ensemble), we can determine $\T$ as a function of
$\eb$ and $\beta$ by the following equation
\BEA \frac{1}{N}\big(\la
K\ra+\la U\ra+\la V\ra\big)=\eb.\label{app2_eqT}
\EEA
We start with the case of small nonlinearity $\beta\rightarrow 0$.
Assume that in the first order of the small parameter $\beta$ the temperature has the following form
\BEA
\T(\beta)=\T_0+\beta \T_1+O(\b^2)\label{app2_Tsm},
\EEA
where $\T_0=O(1)$ and $\T_1=O(1)$.
We find the values of $\T_0$ and $\T_1$ using the constraint~(\ref{app2_eqT}).
We use the following expansions in the small parameter $\beta$
\BEA
\int_{-\infty}^{\infty}\exp\left({-\frac{1}{2\T(\beta)}(y^2+\beta\frac{y^4}{2})}\right)~dy=
\sqrt\frac{\pi}{8}\sqrt{\T_0}\left(4+\Big(\frac{2\T_1}{\T_0}-3\T_0\Big)\beta\right)+O(\beta^2)\nonumber\\\label{app2_y0}
\EEA
\BEA
\int_{-\infty}^{\infty}y^2\exp\left({-\frac{1}{2\T(\beta)}(y^2+\beta\frac{y^4}{2})}\right)~dy
=\sqrt\frac{\pi}{8}\sqrt{\T_0}\left(4\T_0+(6\T_1-15\T_0^2)\beta\right)+O(\beta^2),\nonumber\\\label{app2_y2}
\EEA \BEA
& &\int_{-\infty}^{\infty}(y^2+\frac{\beta}{2}y^4)\exp\left({-\frac{1}{2\T(\beta)}(y^2+\beta\frac{y^4}{2})}\right)~dy\nonumber\\
&=&\sqrt\frac{\pi}{8}\sqrt{\T_0}\left(4\T_0+(6\T_1-9\T_0^2)\beta\right)+O(\beta^2).\label{app2_y4}
\EEA
Then, in the first order in $\beta$, Eq.~(\ref{app2_eqT})
becomes \BEA & &\T_0+\beta
\T_1+\frac{\sqrt\frac{\pi}{8}\sqrt{\T_0}\left(4\T_0+(6\T_1-9\T_0^2)\beta\right)}{\sqrt\frac{\pi}{8}\sqrt{\T_0}\left(4+
\Big(\frac{2\T_1}{\T_0}-3\T_0\Big)\beta\right)}=2\eb,\nonumber%\label{app2_eqT2}
\EEA and we obtain $\T_0=\eb$ and $\T_1=(3/4)\eb^2$.
Therefore, for the average kinetic energy density, we have
\BEA \frac{\la
K\ra}{N}=\frac{1}{2}\eb+\frac{3}{8}\eb^2\beta+O(\beta^2),\label{app2_Kb}
\EEA and, for the average quadratic potential energy density, we have
\BEA
\frac{\la
U\ra}{N}&=&\frac{1}{2}\frac{\sqrt\frac{\pi}{8}\sqrt{\T_0}\left(4\T_0+(6\T_1-15\T_0^2)\beta\right)}{\sqrt\frac{\pi}{8}\sqrt{\T_0}
\left(4+\Big(\frac{2\T_1}{\T_0}-3\T_0\Big)\beta\right)}
=\frac{1}{2}\eb-\frac{9}{8}\eb^2\beta+O(\beta^2).\label{app2_U2b}
\EEA
Finally, we obtain that for small $\b$
\BEA
\eta=1+\frac{3}{2}\eb\beta+O(\beta^2).\label{small_eta}
\EEA
Similarly, from Eq.~(\ref{new_eta}), we find the small $\beta$ limit
of the approximation $\eta_{sc}$
\BEA
\eta_{sc}=1+\frac{3}{2}\eb\beta+O(\beta^2).\label{small_etasc}
\EEA
Now, we consider the case of strong nonlinearity $\beta\rightarrow\infty$.
From Eq.~(\ref{app2_eqT}), we conclude that temperature in the large
$\beta$ limit, which we denote as $\tinf$, stays bounded, i.e., $
0<\tinf<2\eb, $ and, in the limit of large $\beta$, we obtain for
Eq.~(\ref{app2_eqT}) \BEA
\tinf+\frac{\int_{-\infty}^{\infty}\frac{\beta}{2}y^4\exp\left({-\frac{\beta}{4\tinf}y^4}\right)~dy}{\int_{-\infty}^{\infty}\exp\left({-\frac{\beta}{4\tinf}y^4}\right)~dy}=2\eb.
\label{app2_eqT3} \EEA
After performing the integration, we obtain $
\tinf=(4/3)\eb, $ and the average kinetic energy density becomes $ \la
K\ra/N=(2/3)\eb $.
For the average quadratic potential energy density, we have
\BEA
\frac{\la U\ra}{N}=\frac{1}{2}\frac{\int_{-\infty}^{\infty}y^2\exp\left({-\frac{\beta}{4\tinf}y^4}\right)~dy}{\int_{-\infty}^{\infty}\exp\left({-\frac{\beta}{4\tinf}y^4}\right)~dy}%=\nonumber\\
=\frac{\Gamma(\frac{3}{4})}{\Gamma(\frac{1}{4})}\left(\frac{4\eb}{3\beta}\right)^{\frac{1}{2}}
\EEA
For the renormalization factor, we obtain the following large $\beta$ scaling
\BEA
\eta=\sqrt{\frac{\Gamma(\frac{3}{4})}{\sqrt{3}\Gamma(\frac{1}{4})}}\eb^\frac{1}{4}\beta^\frac{1}{4}.\label{app2_etainf}
\EEA
Similarly, for the approximation of $\eta_{sc}$, we obtain
$A=C\sqrt{\eb\beta}$, $B=4\eb\beta$, and
$C=2\sqrt{3}\Gamma(3/4)/\Gamma(1/4)$.
Therefore, the large $\beta$ scaling
of $\eta_{sc}$ becomes \BEA
\eta_{sc}=\sqrt{\frac{C+\sqrt{C^2+16}}{2}}\eb^\frac{1}{4}\beta^\frac{1}{4},\label{app2_etascinf}
\EEA which yields Eq.~(\ref{large_eta}).
\section{Comparison of $\eta$ and $\eta_{sc}$}
Next, we compare the renormalization factor $\eta$ [Eq.~(\ref{eta})] with its approximation $\eta_{sc}$ [Eq.~(\ref{new_eta})]
from the self-consistency argument.
We have shown that for the case of small nonlinearity, both $\eta$ and $\eta_{sc}$ have the same asymptotic behavior in the first order of the small
parameter $\beta$ given by Eqs.~(\ref{small_eta}) and~(\ref{small_etasc}).
Moreover, in the case of strong nonlinearity $\beta\rightarrow\infty$, both $\eta$ and $\eta_{sc}$ are given by Eqs.~(\ref{app2_etainf}) and~(\ref{app2_etascinf}), i.e.,
\BEA
\eta\sim\eta_{sc}\sim\beta^{\frac{1}{4}}\label{large_eta}
\EEA
Note that, in~\cite{our_prl}, we numerically obtained the scaling $\eta\sim\beta^{0.2}$, which differs from the exact analytical result~(\ref{large_eta})
due to statistical errors in the numerical estimate of the power.
\begin{figure}
\includegraphics[scale=0.8]{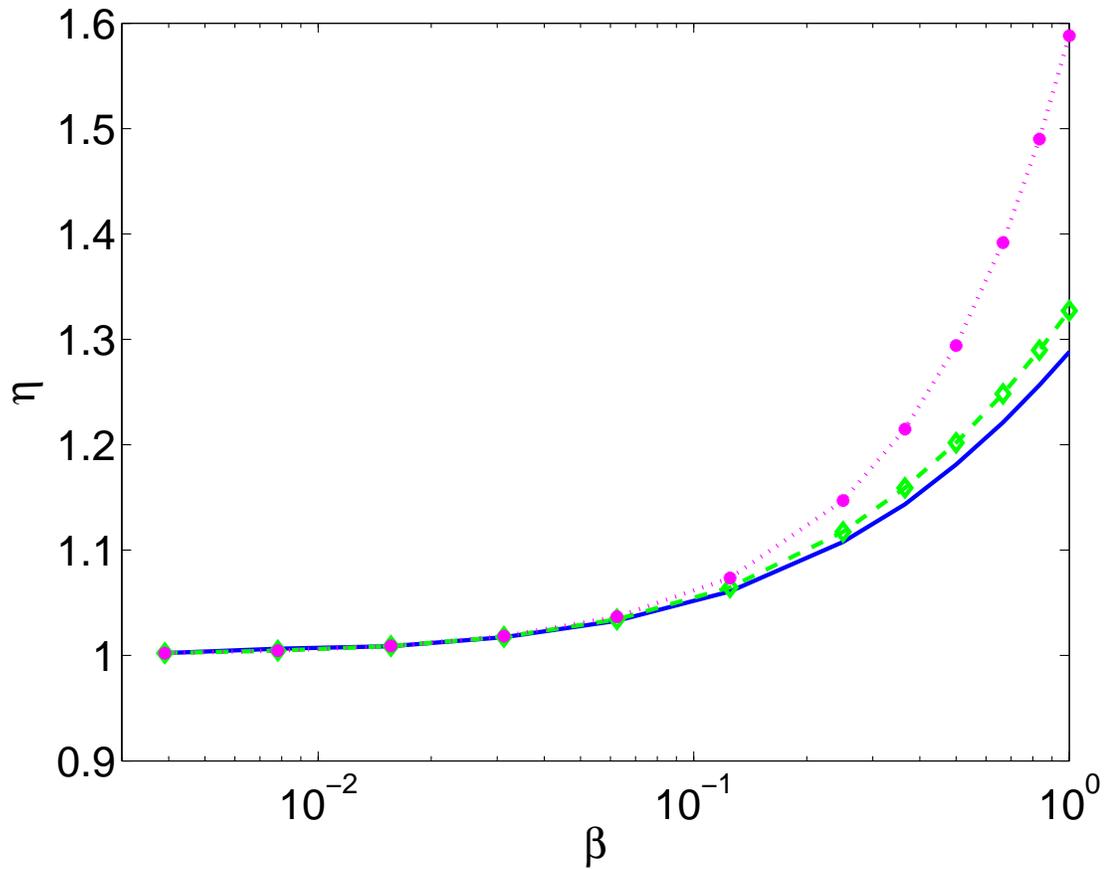}
\caption{The renormalization factor as a function of the nonlinearity strength $\beta$ for small values of $\beta$.
The renormalization factor $\eta$ [Eq.~(\ref{eta})] is shown with the solid line.
The approximation $\eta_{sc}$ [Eq.~(\ref{new_eta})](via the self-consistency argument) is depicted with diamonds connected with the dashed line.
The small-$\beta$ limit [Eq.~(\ref{small_eta})] is shown with the solid circles connected with the dotted line.
Note that, abscissa is of logarithmic scale.
}
\label{Fig_eta_small}
\end{figure}
\begin{figure}
\includegraphics[scale=0.8]{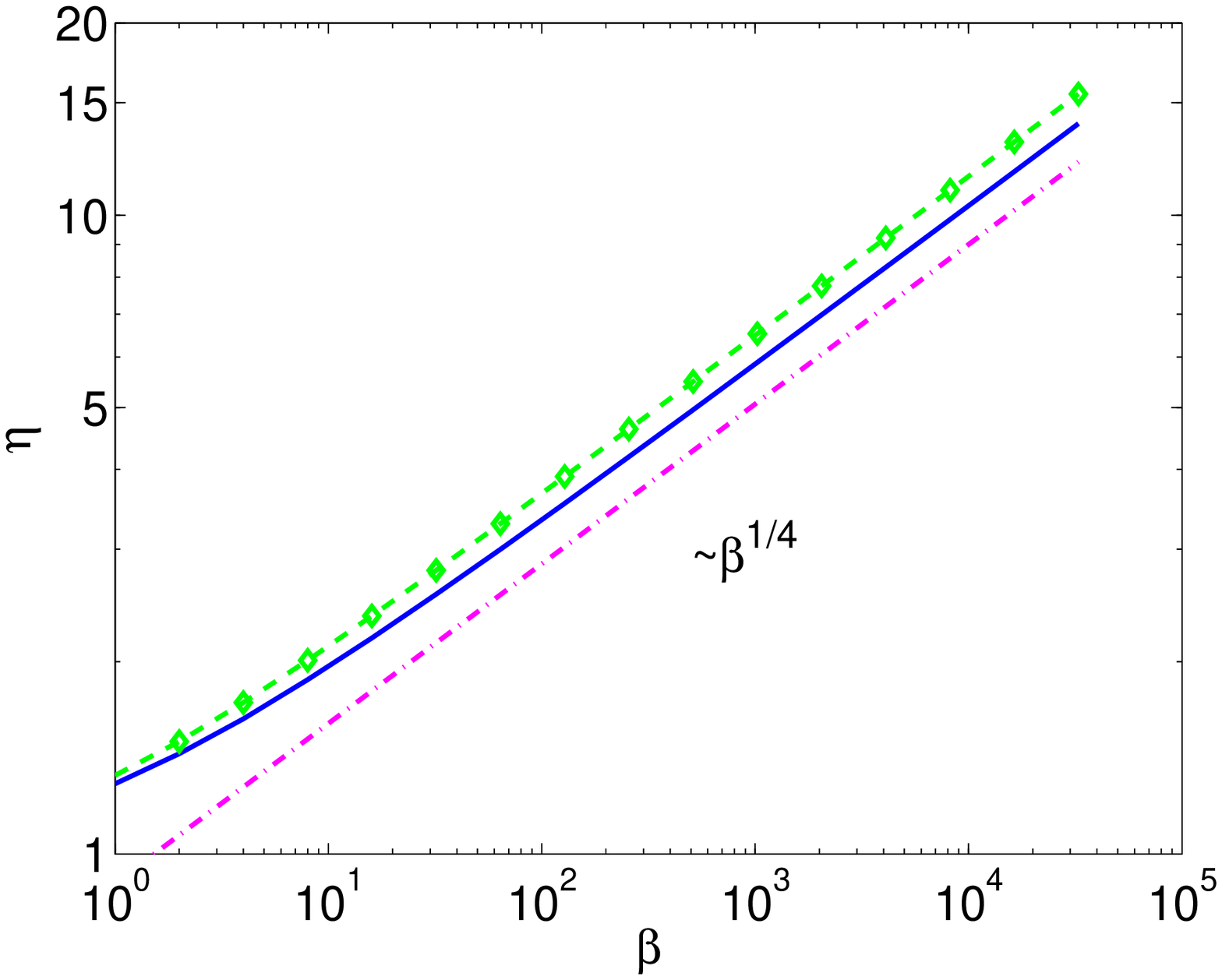}
\caption{The renormalization factor as a function of the nonlinearity strength $\beta$ for large values of $\beta$.
The renormalization factor $\eta$ [Eq.~(\ref{eta})] is shown with the solid line.
$\eta_{sc}$ [Eq.(\ref{new_eta})] is depicted with diamonds connected with the dashed line.
The large-$\beta$ scaling [Eq.~(\ref{large_eta})] is shown with the dashed-dotted line. Note that, the plot is of log-log scale.
}
\label{Fig_eta_large}
\end{figure}
In Fig.~\ref{Fig_eta_small}, we plot the renormalization factor $\eta$ and its approximation $\eta_{sc}$ for the case of small nonlinearity $\beta$ for
the system with $N=256$ particles and total energy $E=100$.
The solid line shows $\eta$ computed via Eq.~(\ref{eta}), the diamonds with the dashed line represent the approximation via Eq.~(\ref{new_eta}),
and the solid circles with the dotted line correspond to the small-$\beta$ limit~(\ref{small_eta}) and~(\ref{small_etasc}).
In Fig.~\ref{Fig_eta_large}, we plot the renormalization factor $\eta$ and its approximation $\eta_{sc}$ for the case of large nonlinearity $\beta$ for
the system with $N=256$ particles and total energy $E=100$.
The solid line shows $\eta$ computed via Eq.~(\ref{eta}), the diamonds with the dashed line represent the approximation via Eq.~(\ref{new_eta}),
and the dashed-dotted line correspond to the large-$\beta$ scaling~(\ref{large_eta}).
Figs.~\ref{Fig_eta_small} and~\ref{Fig_eta_large} show good agreement between the renormalization factor $\eta$ and its approximation $\eta_{sc}$ from the
self consistency argument for a wide range of nonlinearity, from $\beta\sim 10^{-3}$ to $\beta\sim 10^4$.
This agreement demonstrates, that (i) the effect of the linear dispersion renormalization, indeed, arises mainly from the trivial four-wave resonant interactions,
and (ii) our self-consistency, mean-field argument is not restricted to small nonlinearity.
\section{Effective nonlinearity}
Here, we study the decrease of the effective nonlinearity of the system due to the dispersion relation renormalization.
To measure the nonlinearity, we compute the following ratio
\BEA
\ae\equiv\frac{\la H_4\ra}{\la H_2\ra}.\label{self_ea}
\EEA
However, when nonlinearity is strong, $\ae$ does not provide a physically meaningful measure for the strength of nonlinear interactions since the
trivial resonant interactions contribute towards the renormalized linear dynamics, described by Eq.~(\ref{Heff}).
Therefore, we introduce the renormalized measure of effective nonlinearity
\BEA
\tilde{\ae}\equiv\frac{\la\tilde{H}_4\ra}{\la\tilde{H}_2\ra},\label{self_aeren}
\EEA
where $\tilde{H}_4=E-\tilde{H}_2$.
Using the definition of the frequency renormalization factor $\eta$ [Eq.~({\ref{eta})], we obtain
\BEA
\la \tilde{H}_2\ra=\la K\ra+\eta^2\la U\ra=2\la K\ra.
\EEA
Therefore, we can compute $\ae$ and $\tilde{\ae}$ using Eqs.~(\ref{Hp2}) and~(\ref{Hy2}).
In Fig.~\ref{Fig_h4h2}, we present the dependance of both $\ae$ and $\tilde{\ae}$ on the nonlinearity parameter $\beta$.
The total energy was held fixed at the value $E=100$ and $\beta$ was varied.
We observe that as $\beta$ grows, $\ae$ reaches the value of $\sim0.5$ whereas $\tilde{\ae}$ only increases up to $\sim0.25$.
It demonstrates that the effective nonlinearity $\tilde{\ae}$ is still quite small even when the nonlinearity parameter $\beta$ and the bare nonlinearity $\ae$ are
large.
\begin{figure}
\includegraphics[scale=0.8]{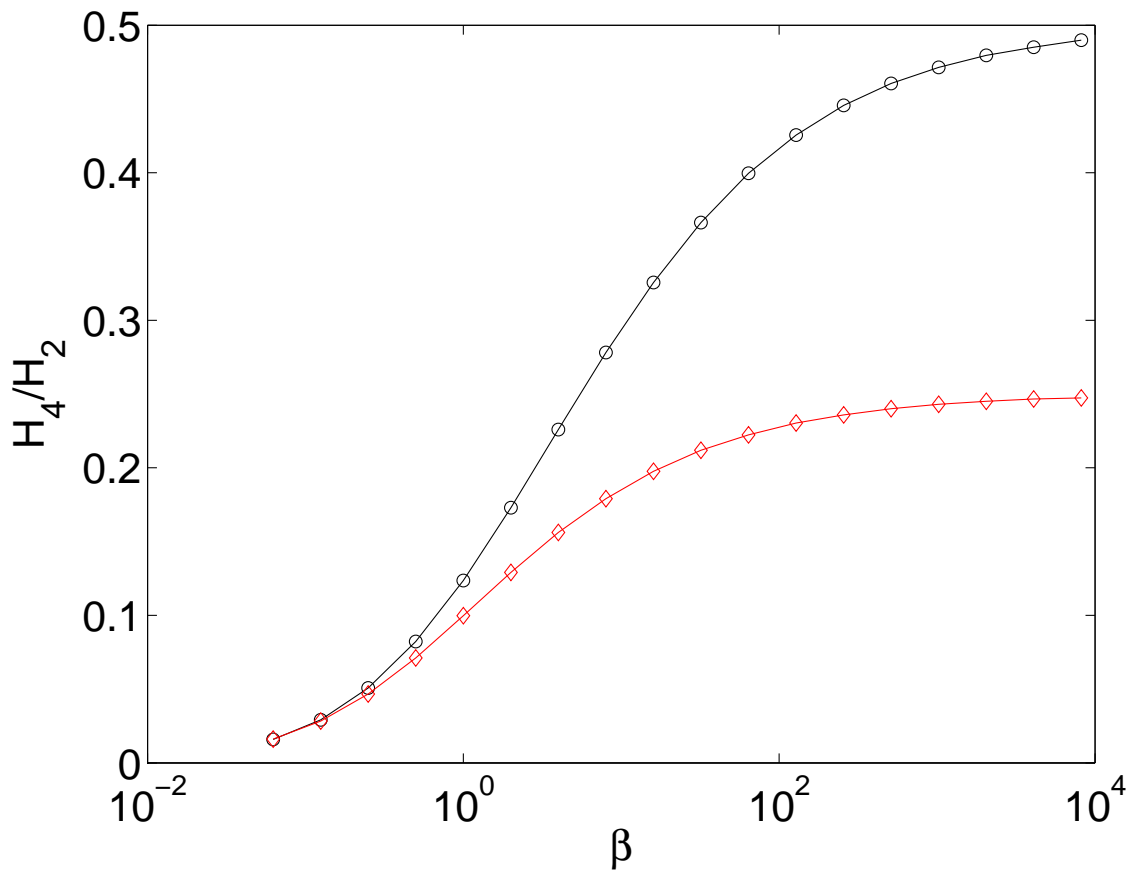}
\caption{
Nonlinearity strength $\ae$ (black circles) and effective nonlinearity strength $\tilde{\ae}$ (red diamonds) as functions of $\beta$.
The number of particles is $N=256$, the total energy is $E=100$.
Note that, the plot is of log-linear scale.
}
\label{Fig_h4h2}
\end{figure}

\chapter{Resonance width}
\label{sect_width}
We further study the properties of these renormalized waves by investigating how long these waves are coherent, i.e., what their frequency widths are.
Therefore, we consider \textit{near-resonant} interactions~\cite{Janssen,Lvov_papa} of the renormalized waves $\at_k$.
These are the interactions that occur in the vicinity of the resonance manifold.
We consider these near-resonances since most of the exact resonant interactions are trivial, i.e., with no momentum exchanges, and they,
cannot effectively redistribute energy among the wave modes.

We will demonstrate that near-resonant interactions of the renormalized waves $\at_k$
provide a mechanism for effective energy exchanges among different wave modes.
Taking into account the near-resonant interactions, we will study analytically the frequency peak broadening of the renormalized waves $\at_k$
by employing a multiple time-scale, statistical averaging method.
Here, we will arrive at a theoretical prediction of the spatiotemporal spectrum $|\af_k(\w)|^2$, where $\af_k(\w)$
is the Fourier transform of the normal variable $\at_k(t)$, and $\w$ is the frequency.
The predicted width of frequency peaks is found to be in good agreement with its numerically measured values.

In addition, for a finite $\beta$-FPU chain, we will mention the consequence of the Umklapp scattering~(see Section~\ref{sect_umklapp}),
to the correlation times of waves.
\section{Effective Hamiltonian}
In the Hamiltonian~(\ref{H_a_res}), the nonlinear terms corresponding to the trivial resonances have been absorbed into the quadratic part via the effective
renormalized dispersion $\wt_k$.
Therefore, the new effective Hamiltonian is
\BEA
\bar{H}=\sum_{k=1}^{N-1}\wt_k|\at_k|^2+\sum_{k,l,m,s=1}^{N-1}\Tt^{kl}_{ms}\Delta^{kl}_{ms}\at_k^*\at_l^*\at_m\at_s,\label{Hef}
\EEA
where
\BEA
\BC
\Tt^{kl}_{ms}=T^{kl}_{ms}=\displaystyle{\frac{3\beta}{8N\eta^2}}\sqrt{\w_k\w_l\w_m\w_s},~k\neq m,~\mbox{and}~k\neq s\\
\Tt^{kl}_{ms}=0,~~\mbox{otherwise.}
\EC\label{Tt}
\EEA
The new interaction coefficient $\Tt^{kl}_{ms}$ ensures that the terms that correspond to the interactions with trivial resonances are
not doubly counted in the Hamiltonian~(\ref{Hef}).
This new interactions in the quartic terms include the exact non-trivial resonant and non-trivial near-resonant as well as non-resonant interactions of the
$(2\rightarrow 2)$-type.

We change the variables to the interaction picture by defining the corresponding variables $b_k$ via~\cite{Lvov,Lvov2}
\BEA
b_k=\at_ke^{\i\wt_kt}.\nonumber
\EEA
Then, the dynamics governed by the Hamiltonian (\ref{Hef}) takes the familiar form
\BEA
\i\dot b_k=2\sum_{l,m,s=1}^{N-1}\Tt^{kl}_{ms}\Delta^{kl}_{ms}b_l^*b_mb_se^{\i\wt^{kl}_{ms} t},\label{b_dyn}
\EEA
where
\BEA
\wt^{kl}_{ms}=\wt_k+\wt_l-\wt_m-\wt_s.\label{wklms}
\EEA
Without loss of generality, we consider only the case of $k<N/2$.
As we have noted before, only for a very small number of quartets does $\wt^{kl}_{ms}$ vanish \textit{exactly}, i.e., $\wt^{kl}_{ms}=0$.
We separate Eq.~(\ref{b_dyn}) into two kinds --- the first kind with $\wt^{kl}_{ms}=0$ that corresponds to exact non-trivial resonances, and the
second kind that corresponds to non-trivial near-resonances and non-resonances.
Since, in the summation, the first kind contains far fewer terms than the second kind, and all the terms are of the same order of magnitude,
we will neglect the first kind in our analysis.
Therefore, Eq.~(\ref{b_dyn}) becomes
\BEA
\i\dot b_k&=&2{\sum_{l,m,s=1}^{N-1}}^{\prime} \Tt^{kl}_{ms}\Delta^{kl}_{ms}b_l^*b_mb_se^{\i\wt^{kl}_{ms} t},\label{b_dyn2}
\EEA
where the prime denotes the summation that neglects the exact non-trivial resonances.

The problem of broadening of spectral peaks now becomes the study of the frequency spectrum of the dynamical variables $b_k(t)$ in thermal equilibrium.
This is equivalent to study the two-point correlation in time of $b_k(t)$
\BEA
C_k(t)=\la b_k(t)b_k^*(0)\ra,
\EEA
where the angular brackets denote the thermal average.
Here we have used the Wiener-Khinchin theorem
\BEA
|b(\w)|^2=\mathfrak{F}^{-1}[C(t)](\w).\label{WKh}
\EEA
\section{Wiener-Khinchin theorem}
Suppose we have a function of time $f(t)$.
Define its spectrum via the inverse Fourier transform
\BEA
\f(\w)\equiv\mathfrak{F}^{-1}[f(t)](\w)=\int f(t)e^{i\w t}dt.\label{f_spectrum}
\EEA
Then, the Fourier transform takes the form
\BEA
f(t)\equiv\mathfrak{F}[\f(\w)](t)=\frac{1}{2\pi}\int \f(\w)e^{-i\w t}d\w.\label{f_inverse}
\EEA
Let us define the auto-correlation function
\BEA
C(t)=\int f(\tau)^*f(t+\tau)d\tau.\label{WKh_corr}
\EEA
Then the Wiener-Khinchin theorem states that
\BEA
C(t)=\mathfrak{F}[|\f(\w)|^2].\label{WKh}
\EEA
\textbf{Proof:}
\
\BEA
C(t)&=&\int f(\tau)^*f(t+\tau)d\tau=\frac{1}{(2\pi)^2}\int\f(\w_1)e^{-\i\w_1(t+\tau)}\f^*(\w_2)e^{\i\w_2\tau}d\w_1d\w_2d\tau\nonumber\\
&=&\frac{1}{(2\pi)^2}\int\f(\w_1)e^{-\i\w_1t}\f^*(\w_2)(2\pi)\delta(\w_1-\w_2)d\w_1d\w_2\nonumber\\
&=&\frac{1}{2\pi}\int|\f(\w)|^2e^{-\i\w t}d\w=\mathfrak{F}[|\f(\w)|^2](t),\label{WKh_proof}
\EEA
\textbf{QED}.
\\
\noindent
Although the proof of the theorem is quite simple, we can not use it for physically relevant signal, i.e., when $f(t)$ is a stationary random process and has no Fourier
transform.
In this situation, we should define the autocorrelation function using expected value (over the Gibbs measure in our situation).
However, for simplicity we will use the version that we have proven.
\section{Effective dynamics of the auto-correlation function $C(t)$}
Now we return to studying the dynamics of the correlation function of the interaction variable $b(t)$.
Under the dynamics (\ref{b_dyn2}), time derivative of the two-point correlation becomes
\BEA
\dot{C}_k(t)&=&\la \dot{b}_k(t)b_k^*(0)\ra\nonumber \\
&=&\la -2\i{\sum_{l,m,s}}^{\prime}\Tt^{kl}_{ms}b_l^*(t)b_m(t)b_s(t)e^{\i\wt^{kl}_{ms}t}\Delta^{kl}_{ms}b_k^*(0)\ra\nonumber\\
&=&-2\i{\sum_{l,m,s}}^{\prime}\Tt^{kl}_{ms}e^{\i\wt^{kl}_{ms}t}J^{kl}_{ms}(t)\Delta^{kl}_{ms}\label{cdot},
\EEA
where
\BEA
J^{kl}_{ms}(t)\equiv\la b_l^*(t)b_k^*(0)b_m(t)b_s(t)\ra.\nonumber
\EEA
In order to obtain a closed equation for $C_k(t)$, we need to study the evolution of the fourth order correlator $J^{kl}_{ms}(t)$.
This correlator is a time-depended generalization of the similar correlator that was used in Chapter~\ref{sect_WT} [Eq.~(\ref{WT_J})].
We utilize the weak effective nonlinearity in Eq.~(\ref{Hef})~\cite{our_prl} as the small parameter in the following perturbation analysis and obtain a closure for
$C_k(t)$, similar to the traditional way of deriving kinetic equation, as in~\cite{ZLF,Benney} and which was discussed in Chapter~\ref{sect_WT}.
We note that the effective interactions of renormalized waves can be weak, as we have shown in~\cite{our_prl}, even if the
$\beta$-FPU chain is in a strongly nonlinear regime.
Our perturbation analysis is a multiple time-scale, statistical averaging method.
Under the near-Gaussian assumption [Chapter~\ref{sect_WT}], which is applicable for the weakly nonlinear wave fields in thermal equilibrium,
for the four-point correlator, we obtain
\BEA
J^{kl}_{ms}(t)\Delta^{kl}_{ms}=C_k(t)C_l(0)(\delta^k_m\delta^l_s+\delta^k_s\delta^l_m).~~\label{4corr}
\EEA

Combining Eqs.~(\ref{Tt}) and~(\ref{4corr}), we find that the right-hand side of Eq.~(\ref{cdot}) vanishes because
\BEA
\Tt^{kl}_{ms}J^{kl}_{ms}(t)\Delta^{kl}_{ms}=0.
\EEA
Therefore, we need to proceed to the higher order contribution of $J^{kl}_{ms}(t)$.
Taking its time derivative yields
\BEA
\dot{J}^{kl}_{ms}(t)\D^{kl}_{ms}&=&\la[\dot{b}_l^*(t)b_m(t)b_s(t)+b_l^*(t)\dot{b}_m(t)b_s(t)\nonumber\\
                &+&b_l^*(t)b_m(t)\dot{b}_s(t)]b_k^*(0)\ra\D^{kl}_{ms}.\label{Jdot}
\EEA
Considering the right-hand side of Eq.~(\ref{Jdot}) term by term, for the first term, we have
\BEA
\la \dot{b}_l^*(t)b_m(t)b_s(t)b_k^*(0)\ra\D^{kl}_{ms}&=&\Bigg\la \Big[2\i{\sum_{\a,\b,\g}}^{\prime}\Tt^{l\a}_{\b\g}b_{\a}(t)b_{\b}^*(t)b_{\g}^*(t)\nonumber\\
& &\times e^{-\i\w^{l\a}_{\b\g}}\D^{l\a}_{\b\g}\Big]b_m(t)b_s(t)b_k^*(0)\Bigg\ra\D^{kl}_{ms}.\label{4b}
\EEA
We can use the near-Gaussian assumption to split the correlator of the sixth order in Eq.~(\ref{4b}) into the product of three correlators of the second order, namely,
\BEA
\la b_k^*(0)b_m(t)b_s(t)b_{\a}(t)b_{\b}^*(t)b_{\g}^*\ra \D^{kl}_{ms}
=C_k(t)n_m n_s\delta^k_{\a}(\delta_{\b}^m\delta^s_{\g}+\delta^m_{\g}\delta^s_{\b}).\nonumber
\EEA
Here, we have used that $n_m=C_m(0)$.
Then, Eq.~(\ref{4b}) becomes
\BEA
\la \dot{b}_l^*(t)b_m(t)b_s(t)b_k^*(0)\ra\D^{kl}_{ms}
=4\i\Tt^{lk}_{ms}C_k(t)n_m n_s e^{-\i\wt^{lk}_{ms}}\D^{kl}_{ms}.\label{4b1}
\EEA
Similarly, for the remaining two terms in Eq.~(\ref{Jdot}), we have
\BEA
\la b_l^*(t)\dot{b}_m(t)b_s(t)b_k^*(0)\ra\D^{kl}_{ms}
=-4\i\Tt^{ms}_{kl}C_k(t)n_l n_s e^{\i\wt^{ms}_{kl}}\D^{kl}_{ms},\label{4b2}
\EEA
and
\BEA
\la b_l^*(t)b_m(t)\dot{b}_s(t)b_k^*(0)\ra\D^{kl}_{ms}=-4\i\Tt^{ms}_{kl}C_k(t)n_l n_m e^{\i\wt^{ms}_{kl}}\D^{kl}_{ms},\label{4b3}
\EEA
respectively.
Combining Eqs.~(\ref{4b1}), (\ref{4b2}), and (\ref{4b3}) with Eq.~(\ref{Jdot}), we obtain
\BEA
\dot{J}^{kl}_{ms}(t)\D^{kl}_{ms}&=&4\i\Tt^{kl}_{ms}C_k(t)e^{-\i\wt^{kl}_{ms}t}\D^{kl}_{ms}(n_m n_s-n_l n_m-n_l n_s).\label{Jdot2}
\EEA
Equation~(\ref{Jdot2}) can be solved for $J^{kl}_{ms}(t)$ under the assumption that the term $e^{-\i\wt^{kl}_{ms}t}$ oscillates much faster than
$C_k(t)$~\cite{Janssen}.
We numerically verify [Fig.~\ref{Fig_corr} below] the validity of this assumption of time-scale separation.
Under this approximation, the solution of Eq.~(\ref{Jdot2})  becomes
\BEA
J^{kl}_{ms}(t)\D^{kl}_{ms}&=&4\Tt^{kl}_{ms}C_k(t)\D^{kl}_{ms}\frac{e^{-\i\wt^{kl}_{ms}t}-1}{-\wt^{kl}_{ms}}(n_m n_s-n_l n_m-n_l n_s ).\label{J}
\EEA
Plugging Eq.~(\ref{J}) into Eq.~(\ref{cdot}), we obtain the following equation for $C_k(t)$
\BEA
\dot{C}_k(t)&=&8\i C_k(t) {\sum_{l,m,s}}^{\prime} \left(\Tt^{kl}_{ms}\right)^2\D^{kl}_{ms}\frac{1-e^{\i\w^{kl}_{ms}t}}{\w^{kl}_{ms}}
(n_m n_s-n_l n_s-n_l n_m).\label{cdot2}
\EEA
Since in the thermal equilibrium $n_k$ is known, i.e., $n_k=\la |b_k(t)|^2\ra=\theta/\wt_k$ [Eq.~(\ref{nt})], Eq.~(\ref{cdot2}) becomes a
closed equation for $C_k(t)$.
The solution of Eq.~(\ref{cdot2}) yields the autocorrelation function $C_k(t)$
\BEA
\ln\frac{C_k(t)}{C_k(0)}&=&8{\sum_{l,m,s}}^{\prime} \left(\Tt^{kl}_{ms}\right)^2\frac{e^{\i\w^{kl}_{ms}t}-1-\i\w^{kl}_{ms}t}{(\w^{kl}_{ms})^2}
(n_l n_s+n_l n_m-n_m n_s)\D^{kl}_{ms}.\nonumber\\
\label{C}
\EEA
Using this observation, together with Eq.~(\ref{Tt}), finally, we obtain for the thermalized $\beta$-FPU chain
\BEA
\ln\frac{C_k(t)}{C_k(0)}&=&\frac{9\beta^2\theta^2}{8N^2\eta^6}\w_k{\sum_{l,m,s}}^{\prime}(\w_m+\w_s-\w_l)\D^{kl}_{ms}
\frac{e^{\i\w^{kl}_{ms}t}-1-\i\w^{kl}_{ms}t}{(\w^{kl}_{ms})^2}.\label{C}
\EEA
Equation~(\ref{C}) gives a direct way of computing the correlation function of the renormalized waves $\at_k$, which, in turn, allows us to predict the
spatiotemporal spectrum $|\af_k(\w)|^2$.
\section{Analytical prediction vs numerical observation}
In Fig.~\ref{Fig_awk_both}(a), we plot the analytical prediction (via Eq.~(\ref{C})) of the
spatiotemporal spectrum
\BEA
|\af_k(\w)|^2\equiv|b_k(\w-\wt_k)|^2=\mathfrak{F}^{-1}[C(t)](\w-\wt_k).\label{WKh_spectrum}
\EEA
By comparing this plot with the one presented in Fig.~\ref{Fig_awk_both}(b), in which the corresponding numerically measured spatiotemporal spectrum is shown,
it can be seen that the analytical prediction of the frequency spectrum via Eq.~(\ref{C}) is in good qualitative agreement with the numerically measured one.
However, to obtain a more detailed comparison of the analytical prediction with the numerical observation, we show, in Fig.~\ref{Fig_aw},
the numerical frequency spectra of selected wave modes with the corresponding analytical predictions.
It can be clearly observed that the agreement is rather good.
\begin{figure}
\includegraphics[scale=0.75]{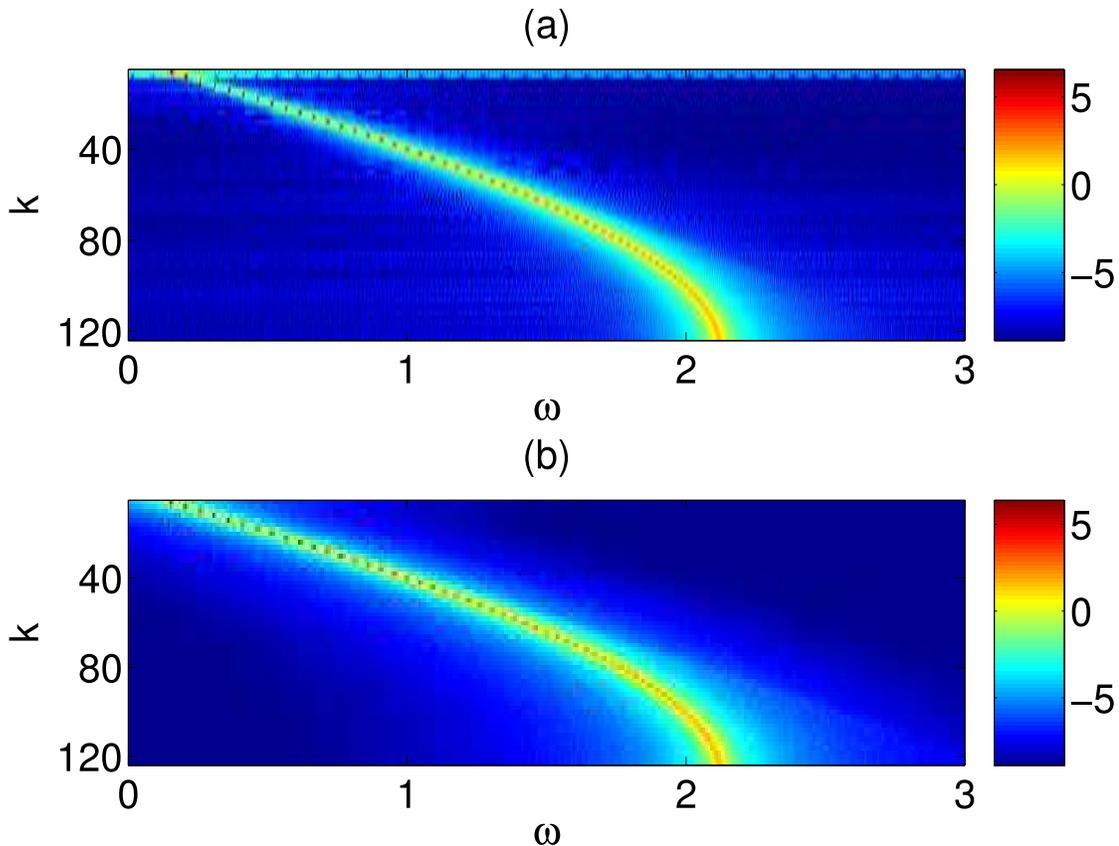}
\caption{ (a) Plot of the analytical prediction for the
spatiotemporal spectrum $|\af_k(\w)|^2$ via Eq.~(\ref{C}). (b) Plot
of the numerically measured spatiotemporal spectrum $|\af_k(\w)|^2$.
The parameters in both plots were $N=256$, $\beta=0.125$, $E=100$
and $\eta=1.06$, $\theta=0.401$. $\eta$ and  $\theta$ were computed
analytically via Gibbs measure. The darker gray scale correspond to
larger values of $|\af_k(\w)|^2$ in $\w$-$k$ space.
[$\max\{-8,\ln{|\af_k(\w)|^2}\}$ is plotted for clear
presentation].} \label{Fig_awk_both}
\end{figure}
\begin{figure}
\includegraphics[width=6in, height=5in]{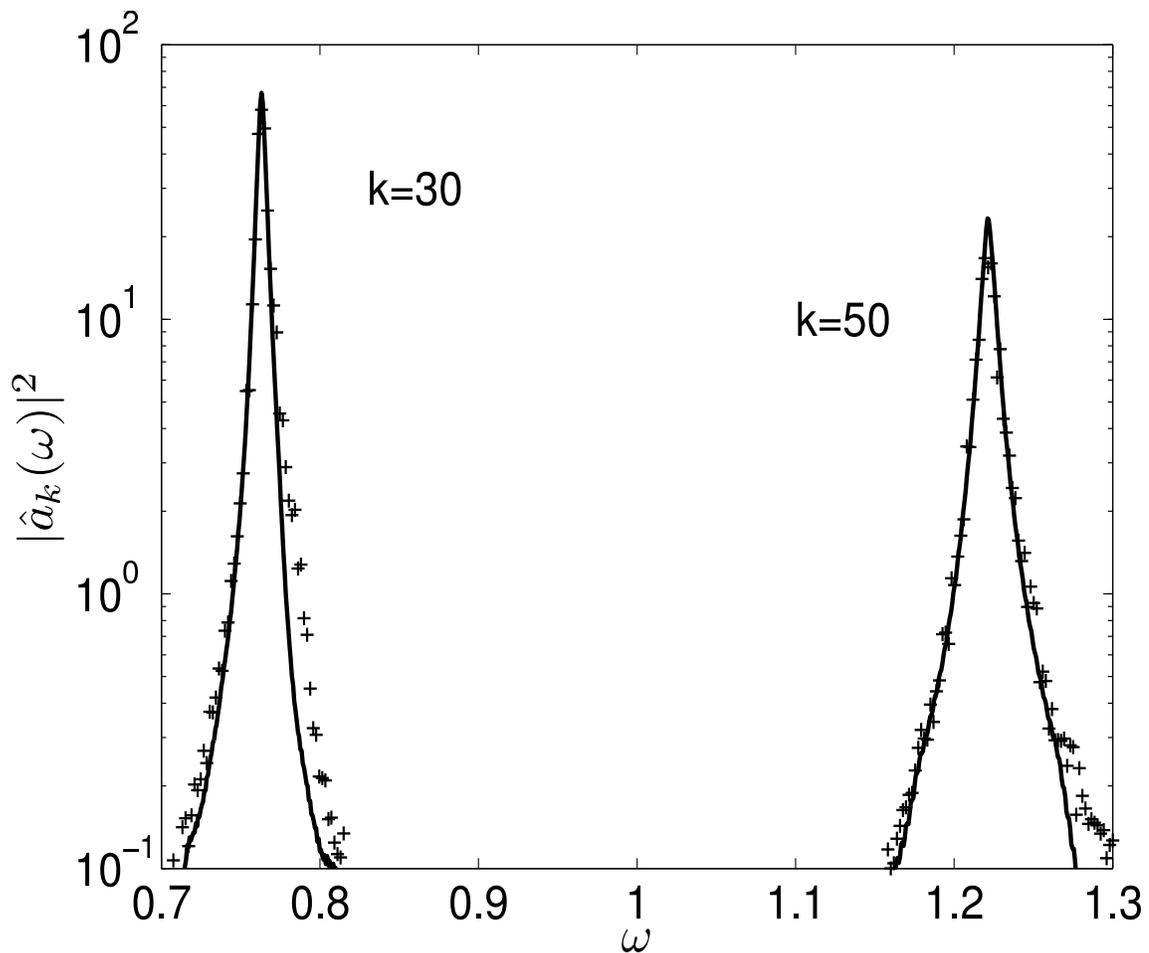}
\caption{Temporal frequency spectrum $|\af_k(\w)|^2$ for $k=30$ (left peak) and $k=50$ (right peak). The numerical spectrum is shown with pluses and the
analytical prediction [via Eq.~(\ref{C})] is shown with solid line. The parameters were $N=256$, $\beta=0.125$, $E=100$.
%Both analytical and numerical peaks for each $k$ are centered at the same frequencies $\w$, which are equal to the numerically observed dispersion $\wt_k$.
}
\label{Fig_aw}
\end{figure}
One of the important characteristics of the frequency spectrum is the width of the spectrum.
We compute the width $W(f)$ of the spectrum $f(\w)$ by
\BEA
W(f)=\frac{\int f(\w)~d\w}{\max_\w f(\w)}.
\EEA
In Fig.~\ref{Fig_widths}, we compare the width, as a function of the wave number $k$, of the frequency peaks from the numerical observation with that obtained from
the analytical predictions.
We observe that, for weak nonlinearity ($\beta=0.125$), the analytical prediction and the numerical observation are in excellent agreement.
In the weakly nonlinear regime, this agreement can be attributed to the validity of (i) the near-Gaussian assumption, and
(ii) the separation between the linear dispersion time scale and the time scale of the correlation $C_k(t)$.
This separation was used in deriving the analytical prediction [Eq.~(\ref{C})].
However, when the nonlinearity becomes larger ($\beta=0.25$ and $\beta=0.5$), the discrepancy between the numerical measurements and the analytical prediction
increases, as can be seen in Fig.~\ref{Fig_widths}.
Nevertheless, it is important to emphasize that, even for very strong nonlinearity, our prediction is still qualitatively valid, as seen in Fig.~(\ref{Fig_widths}).
In order to find out the effect of the Umklapp scattering due to the finite size of the chain, we also computed the correlation [Eq.~(\ref{C})]
with the ``conventional'' $\delta$-function $\delta^{kl}_{ms}$ (i.e., without taking into account the Umklapp processes) instead of our ``periodic'' delta function
$\D^{kl}_{ms}$.
It turns out that the correlation time is approximately $30\%$ larger if it is computed without Umklapp processes taken into account
for the case $N=256$, $\beta=0.5$, $E=100$.
It demonstrates that the influence of the non-trivial Umklapp resonances is important and should be considered when one describes the dynamics of the finite
length chain of particles.
\begin{figure}
\includegraphics[width=6in, height=5in]{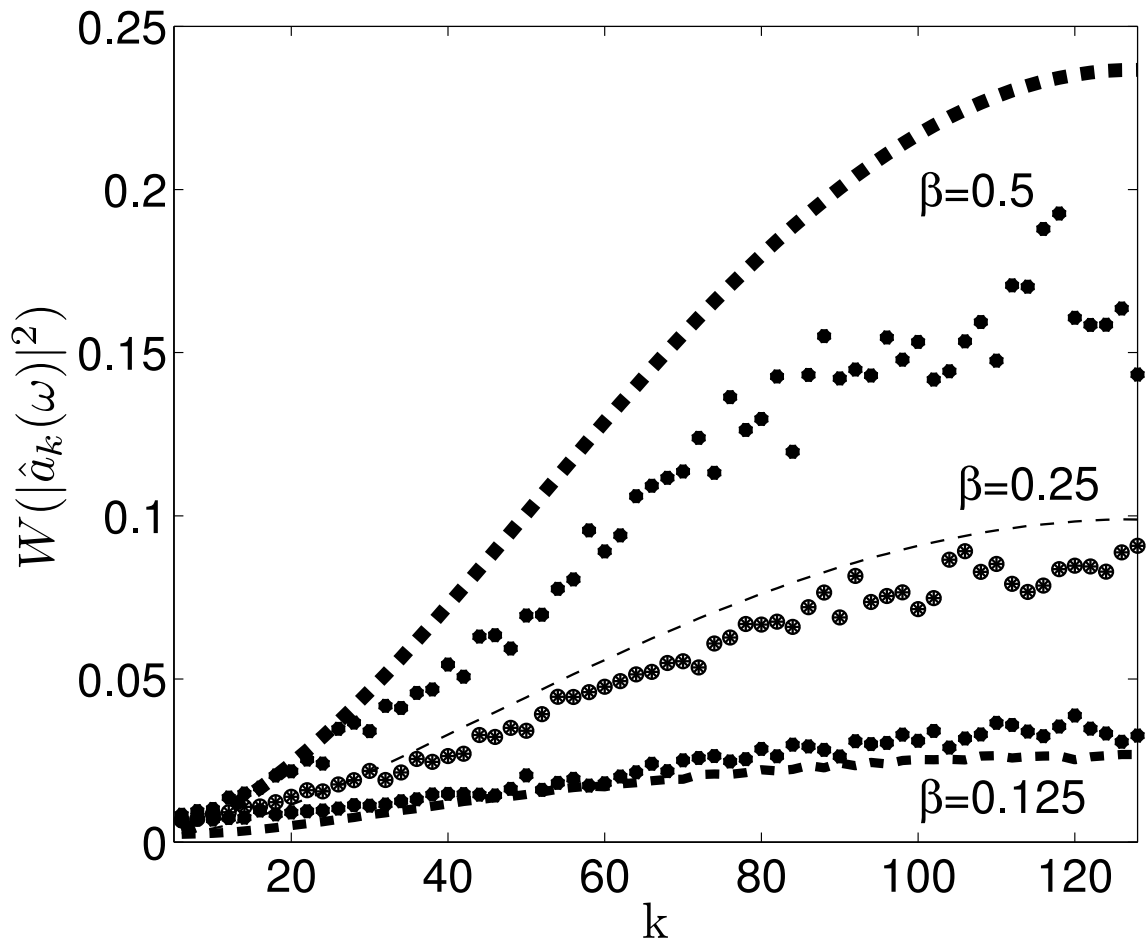}
\caption{Frequency peak width $W(|\af_k(\w)|^2)$ as a function of the wave number $k$.
The analytical prediction via Eq.~(\ref{C}) is shown with a dashed line and the numerical observation is plotted with solid circles.
The parameters were $N=256$, $E=100$.
The upper thick lines correspond to $\beta=0.5$,
the middle fine lines correspond to $\beta=0.25$, and the lower solid circle and dashed line (almost overlap)
correspond to $\beta=0.125$.}
\label{Fig_widths}
\end{figure}
Finally, in Fig.~\ref{Fig_corr}, we verify the time scale separation assumption used in our derivation,
i.e., the correlation time of the wave mode $k$ is sufficiently larger than the corresponding linear dispersion period $\tt_k=2\pi/\wt_k$.
In the case of small nonlinearity ($\beta=0.125$), the two-point correlation changes over much slower time scale than the corresponding linear oscillations
--- the correlation time is nearly two orders of magnitude larger than the corresponding linear oscillations for weak nonlinearity $\beta=0.125$, and
nearly one order of magnitude larger than the corresponding linear oscillations for stronger nonlinearity $\beta=0.25$ and $\beta=0.5$.
This demonstrates that the renormalized waves have long lifetimes, i.e., they are coherent over time-scales that are much longer than their
oscillation time-scales.
\begin{figure}
\includegraphics[width=6in, height=5in]{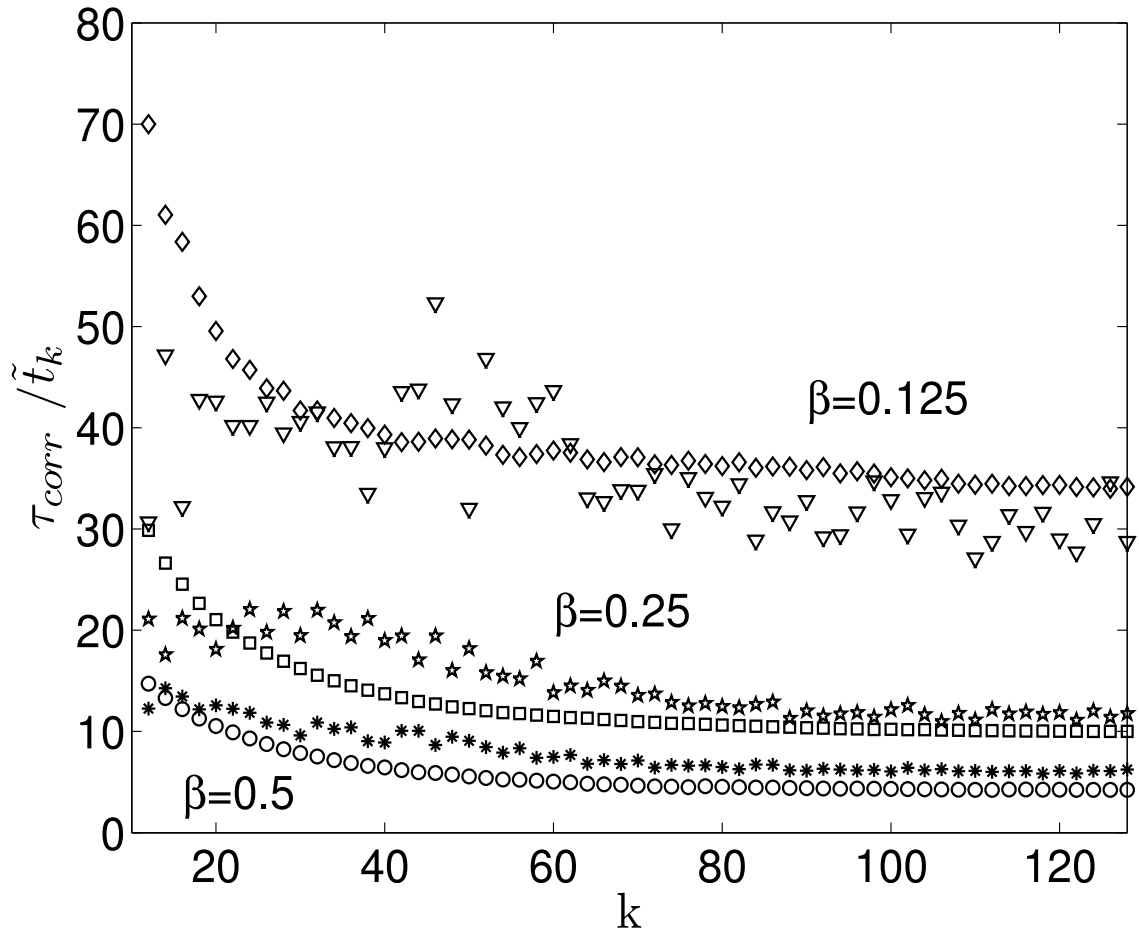}
\caption{ Ratio, as a function of $k$, of the correlation time $\tau_k$ of the mode $k$ to the corresponding linear period $\tt_k=2\pi/\wt_k$.
Circles, squares, and diamonds represent the analytical prediction for $\beta=0.5$, $\beta=0.25$, and $\beta=0.125$ respectively.
Solid circles, pentagrams, and triangles correspond to the numerical observation for $\beta=0.5$, $\beta=0.25$, and $\beta=0.125$ respectively.
The parameters were $N=256$, $E=100$.
The ratio is sufficiently large for all wave numbers $k$ even for relatively large $\beta=0.5$, which validates the time-scale separation assumption used
in deriving Eq.~(\ref{J}).
The comparison also suggests that for smaller $\beta$ the analytical prediction should be closer to the numerical observation, as is confirmed
in Fig.~\ref{Fig_widths}.}
\label{Fig_corr}
\end{figure}

\chapter{Discrete Breathers in $\beta$-FPU chains}
\label{sect_breathers}
We have studied the interactions of renormalized waves in the FPU systems in thermal equilibrium.
As we have seen, under any strength of nonlinearity, the renormalized waves in thermal equilibrium can be regarded as quasiparticles in the \textit{wave-number space}.
The question arises: are there any localized excitations of the FPU lattice in the \textit{physical space}.
In the last decade, discrete breathers (DB) as spatially localized, time periodic lattice excitations were discovered~\cite{Flach}.
Arising from energy localization in nonlinear lattices, they play important roles in many dynamics in fiber optics, condensed matter physics and
molecular biology~\cite{Flach2}.
The existence of DBs has been addressed rigorously~\cite{VarProof}.
Important conceptual issues naturally appear, such as what is the role of DBs on the route to equilibrium~\cite{MaxLyap} and how do they manifest in
thermalization of the FPU system?
Resolution of these issues will certainly provide deep insight into the fundamental understanding of route to thermalization for general nonlinear physical systems.
Most of the results regarding DBs in $\beta$-FPU chains have so far only addressed their behavior in the transient state of \textit{weakly} nonlinear regimes before
thermalization occurs~\cite{Cretegny,SlowRelax}.
Here we present numerical evidence of DBs in the FPU chains in thermal equilibrium.
However, before we do that we will give a short introduction to the large subject of
breather solutions in both continuous and discrete models.
A more complete review of the topics can be found in~\cite{Flach}.
\section{General introduction to breathers}
\label{sect_breathers_general}
Breathers are spatially localized, time-periodic solutions of a PDE or of a lattice system.
Spatial localization usually means exponential decay of the amplitude of a breather from its center.
A well known example of a PDE that possesses breather solutions is a particular case of Klein-Gordon equation
\BEA
\psi_{tt}=C\psi_{xx}-F(\psi),\label{KleinGordon}
\EEA
with the choice $F(\psi)=\sin(\psi)$ and which is referred to as sine-Gordon equation.
The breather solution for the sine-Gordon equation is given by
\BEA
\psi_b=4\tan^{-1}\frac{m\sin(\w t)}{\w \mbox{ch}(mx)},\label{sG_breather}
\EEA
where $\w=\sqrt{1-m^2}$.
However, these breather solutions are structurally unstable in the following sense.
For most of the perturbations of the nonlinear term $F(\psi)$, the breather solutions do not survive.
For instance, it was proved by Segur and Kruskal~\cite{SegurKruskal} that for the choice $F(\psi)=-\psi+\psi^3$ (a so called $\phi^4$ system) breather solutions do not
exist.
The intuitive underlying reason for breather solutions of PDEs to be structurally unstable arises from
the fact that the linear spectrum of continuous PDEs is unbounded and
the frequencies of the breathers would most probably be in resonance with the linear spectrum.
In particular, the linear dispersion relation of the Klein-Gordon model is~\cite{Flach}
\BEA
\w_k=\sqrt{Ck^2+F'(\psi=0)}.\label{KG_spectrum}
\EEA
This dispersion relation is unbounded and, therefore, it is difficult to find breather solutions in general PDEs except for a comparably few non-generic cases.

The situation changes drastically when we turn to the discrete nonlinear models.
Consider a discretization of the Klein-Gordon model
\BEA
\ddot{u}_j+b^2(2u_n-u_{n+1}-u_{n-1})-f(u_n)=0.\label{KleinGordon_discr}
\EEA
Here the linear dispersion relation reads~\cite{Flach}
\BEA
\w_k=\sqrt{1+2b^2(1-\cos(k))}.\label{KG_spectrum_discr}
\EEA
We note, that unlike the spectrum~(\ref{KG_spectrum}) of the continuous model, the spectrum~(\ref{KG_spectrum_discr}) of the discrete model is bounded from both
below and
above, which allows for the existence of breathers if their frequencies lie outside the linear band.
It was demonstrated in~\cite{MacKayAubry} that systems like~(\ref{KleinGordon_discr}) have discrete breather solutions.
The result proved in~\cite{MacKayAubry} is the most general known so far. Indeed, it was actually proved that almost \textit{any} nonlinear lattice
possesses discrete breather solutions if it satisfies some non-resonance conditions.

In Fig.~\ref{Fig_breather}, we can see an idealized form of the discrete breather solution of a nonlinear lattice.
The characteristic properties of the DBs are: (i) spatial localization --- in practice, highly localized DBs usually involve just 4-5 sites of the lattice --- and
(ii) time-periodicity --- the subsequent sites oscillate in an alternating manner with the frequency, higher than the linear band of the spectrum.
The situation when the frequency of the breather lies below the linear band is also possible, however, we will not investigate this kind of breathers in this thesis
since the linear dispersion of the FPU chains satisfies $0\le \w_k\le 2$.
\begin{figure}
\includegraphics[scale=0.8]{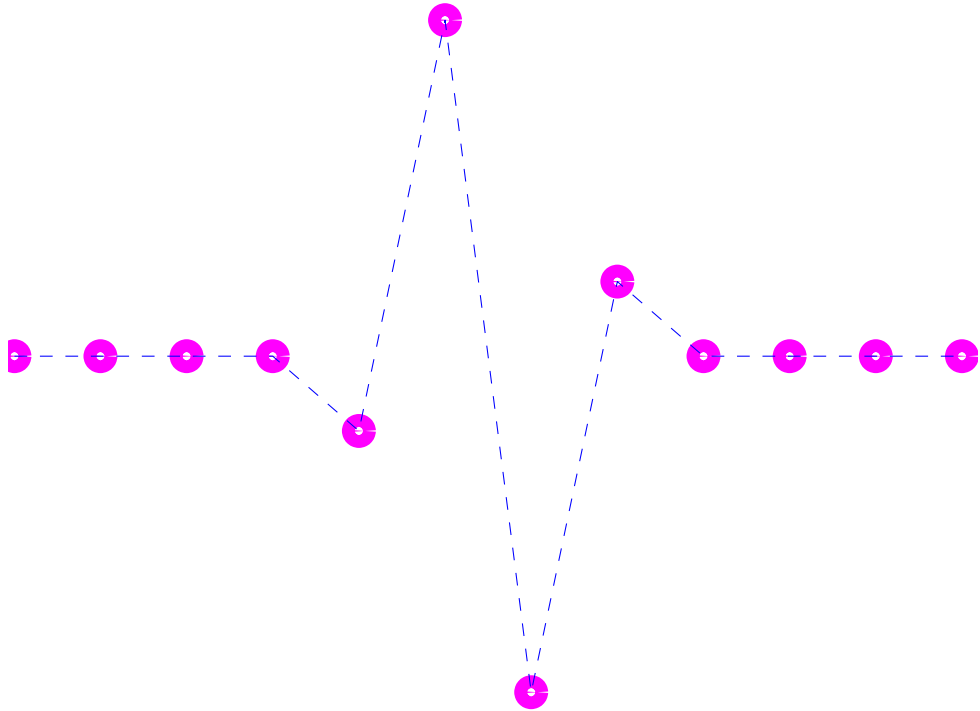}
\caption{Spatial localization and periodicity in time (usually with oscillations of the alternating wings) are characteristic properties of
discrete breathers.}
\label{Fig_breather}
\end{figure}

\section{Discrete Breathers in the $\beta$-FPU chain in transient to the thermal equilibrium}
\label{sect_breathers_transient}
The formation of Discrete Breathers in transient to thermal equilibrium in the weakly nonlinear $\beta$-FPU chain is studied in detail in~\cite{Cretegny}.
Here, we provide some of the results of this work and in the next Section we will discuss the existence of DB's in the $\beta$-FPU system after it has
reached the thermal equilibrium state.

As opposed to the original FPU experiment, the authors of~\cite{Cretegny} considered the initial condition with the highest Fourier mode excited.
It is shown that this mode is modulationally unstable and gives rise to the energy localization in the form of DBs.
In order to observe these DBs numerically, one can integrate the dynamical equations of motion~(\ref{dyn_pq_1}) with the initial condition of a zig-zag form
\BEA
q_j(0)&=&(-1)^ja,\label{DB_plusminus}\\
p_j(0)&=&0,\label{DB_p0}
\EEA
where $a$ is the amplitude of the initial condition.
This initial condition is usually referred to as $\pi$-mode.
$\pi$-mode is an exact solution of the $\beta$-FPU lattice~\cite{Poggy} and, therefore, in order to destabilize it, small noise needs to be introduced to the initial
condition | the random perturbation of the order $10^{-14}$ was used in $p_j(0)$.

It is convenient to detect energy localization using the function $L(t)$ given in Eq.~(\ref{Loc}).
Recall that $L(t)$ is of the order $1$ if energy is nearly uniformly distributed along the chain and of the order $N$ if energy is concentrated
around a few sites of the chain.
In Fig.~\ref{Fig_qLt}, we show the time evolution of the chain $q_j(t)$ (panel (a)), together with the time evolution of $L(t)$ (panel (b)).
First the oscillations of the chain follow the initially given zig-zag pattern, which corresponds to the uniform energy distribution with $L(t)=1$.
Then the random initial perturbation induces modulation instability and energy starts to localize.
At that time we can observe the development of DB's (dark strips in Fig.~\ref{Fig_qLt}(a)) and the increase of the value of $L(t)$ (Fig.~\ref{Fig_qLt}(b)).
Then the localized states disperse their energy and gradually disappear, and the system approaches the energy equipartition state.
At that point $L(t)$ goes to some steady state value.
In Fig.~\ref{Fig_q3t}, we demonstrate the snapshots of the chain, i.e., $q(j)$, at the three different stages of the chain evolution that we have just described.
In Fig.~\ref{Fig_q3t}(a), the initial condition is shown.
In Fig.~\ref{Fig_q3t}(b), we see the energy localization state --- DBs are developed (enclosed in the red rectangle).
Note the amplitude of $q_j$.
And finally, in Fig.~\ref{Fig_q3t}(c), the snapshot of the chain at the thermalized state is shown, when the chain only consists of the renormalized linear waves.

It is known~\cite{Flach} that spatially localized lattice solutions have frequencies that are outside the linear band of the system.
Otherwise, they would resonate with the spatially extended linear waves.
The linear band of the $\beta$-FPU chain is given by the dispersion relation $\w_k=2\sin(\pi k/N)$.
Therefore, frequencies of the DBs must be greater than $2$ as we have discussed in Section~\ref{sect_breathers_general}.
We confirm this fact numerically.
In Fig.~\ref{Fig_awk_tr}, we plot the spatiotemporal spectrum of the $\beta$-FPU chain at the time period when the DBs are developed
(corresponds to $t\in[2000,8000]$ in Fig.~\ref{Fig_qLt}).
It can be seen in Fig.~\ref{Fig_awk_tr} that the spatiotemporal spectrum essentially consists of two parts: (i) the linear band given by $\w_k$ (we also notice
renormalization due to the nonlinearity, see Chapter~\ref{sect_renormalization} for details) and
(ii) the higher frequency region (enclosed in the black rectangle) that corresponds to DBs.
In order to compute the spatiotemporal spectrum, we have used the time series with a length of $16384$ time steps, and each time step was 0.1 time units.

Now that we have observed the existence of DBs in the transient to thermal equilibrium, we turn to
investigating the existence of DBs in the $\beta$-FPU system after it has reached the thermal equilibrium state.

\begin{figure}
\includegraphics[bb=0 0 575 575]{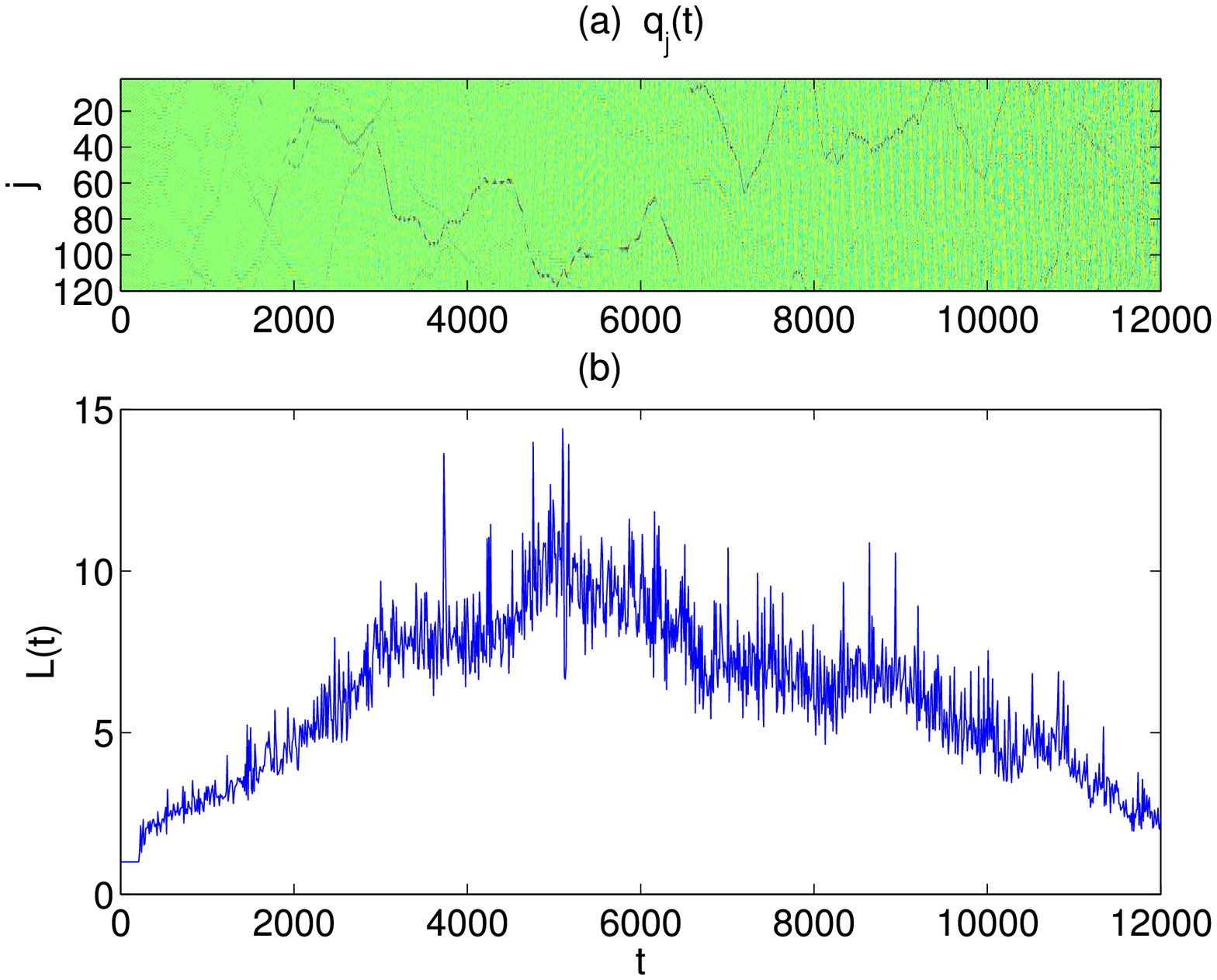}
\caption{$q_j(t)$ as a function of $j$ and $t$ (panel (a)) and
energy localization L(t) (panel (b)) computed via Eq.~(\ref{Loc}).
The chain was modeled for $N=128$, $\beta=0.1$, and $a=0.8$
($E/N\sim1.44$).} \label{Fig_qLt}
\end{figure}
\begin{figure}
\includegraphics[scale=0.8]{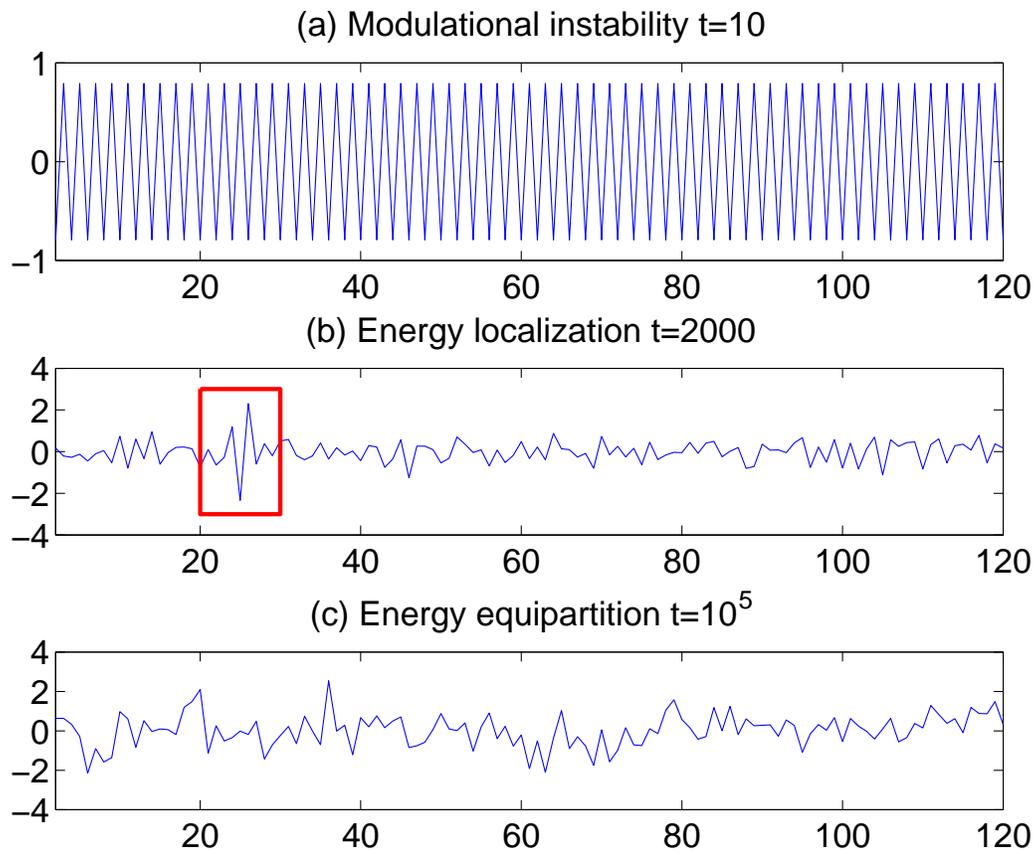}
\caption{ Panel (a): initial state.
\
Panel (b): energy localization state. The breather is enclosed in a red rectangle.
\
Panel (c): energy equipartition state.
\
The chain was modeled for $N=128$, $\beta=0.1$, and $a=0.8$ ($E/N\sim1.44$).}
\label{Fig_q3t}
\end{figure}

\begin{figure}
\includegraphics[scale=0.8]{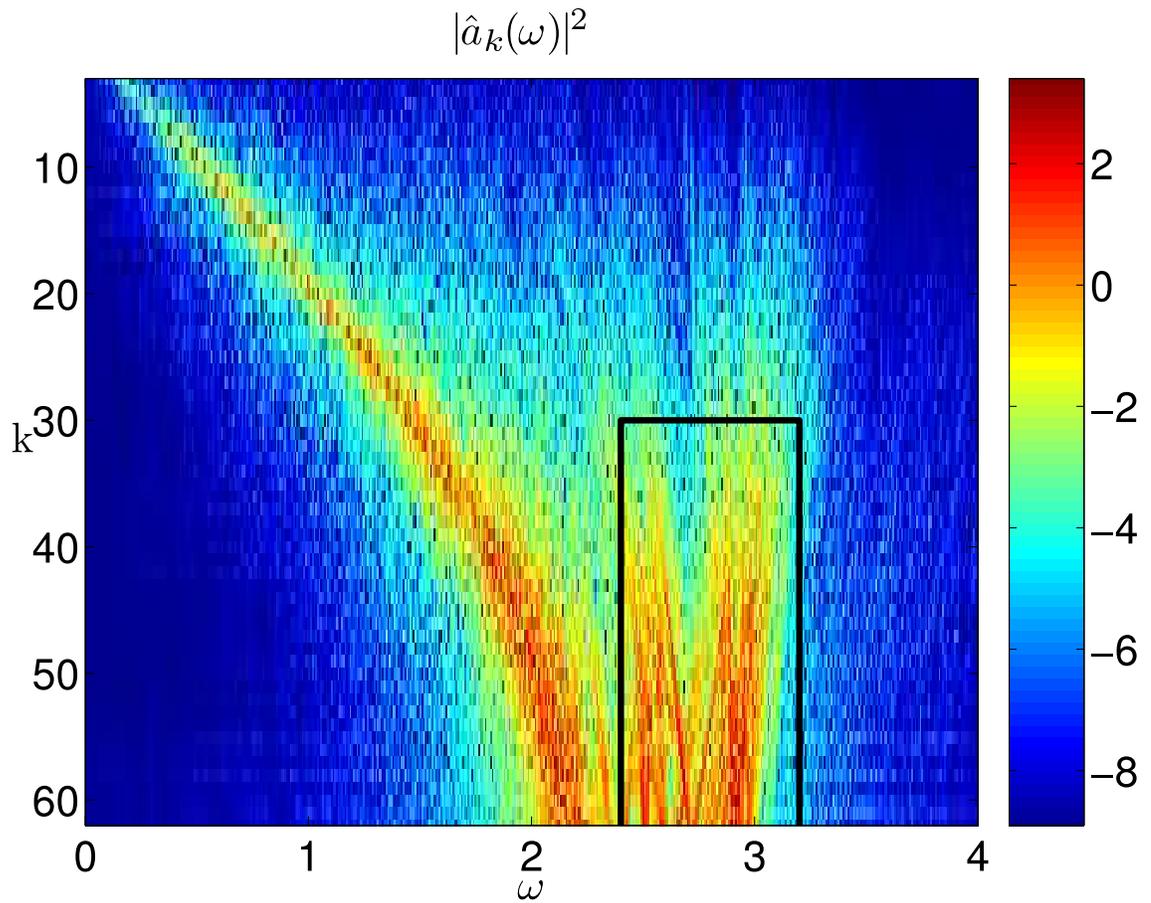}
\caption{Spatiotemporal spectrum demonstrates the oscillations with frequencies  outside the linear band (region enclosed in a black rectangle is separated from the
$\sin$-like linear band.)
The chain was modeled for $N=128$, $\beta=0.1$, and $a=0.8$ ($E/N\sim1.44$) and the spectrum was computed via the time series of the length 16384 time steps with each
time step of the length 0.1 time units.}
\label{Fig_awk_tr}
\end{figure}

\section{Discrete Breathers in the thermal equilibrium}
\label{sect_breathers_thermal}
As we have already seen in Chapter~\ref{sect_renormalization},
the thermalized state of the $\beta$-FPU chain is characterized by the existence of the renormalized waves.
Moreover, in Section~\ref{sect_breathers_transient}, we have discussed the existence of DB excitations that were observed during
transient stages towards thermalization~\cite{Cretegny}.
Here we show via numerical simulation that DBs actually persist and coexist with renormalized waves in the thermalized state.
Thus, in the thermalized $\beta$-FPU, there are two kinds of quasi-particle excitations, one localized in $k$-space as
renormalized nonlinear waves/phonons, and the other, localized in $x$-space as DBs.

As nonlinearity increases, the duration of transient becomes shorter.
After the energy redistributes among all the modes to achieve thermal equilibration, our simulations show that the spatially localized,
high frequency excitations still exist.
These DBs can interact with each other and may be destroyed by collision processes with other DBs or with the renormalized waves.
The spatial structure of these excitations very much resembles the idealized breather oscillations (Fig.~\ref{Fig_breather}) in the absence of spatially
extended waves: they ``live'' above the high frequency edge of the dispersion band and their lifetime is sufficiently long (on the order of 10-100 DB oscillations).
Therefore, they behave like a quasiparticle.
Note that, under certain conditions, supersonic solitons may arise from the $\beta$-FPU system as another kind of localized excitations~\cite{Kink,Zhang}.
However, they were not observed in our thermalized system.

Figure~\ref{fig4_compare} is the energy density plot which shows the time evolution of energy of
each particle for the transient (recording starting time
$T_0=5\times 10^2$) and thermalized ($T_0=5\times 10^5$) states,
respectively. Fig.~\ref{fig4_compare}(c) and (d) display the energy
as a function of site at  $T_0$ corresponding to
Fig.~\ref{fig4_compare}(a) and (b) respectively.  In the transient
case (Fig.~\ref{fig4_compare}(a)) the spatially localized objects
(dark stripes) that carry sufficiently large amount of energy are
clearly observed. Fig.~\ref{fig4_compare}(c) is a snapshot of the
energy density plot (\ref{fig4_compare}(a)) at $T_0$. Here the DBs
are seen as localized peaks~\cite{Cretegny}.  After thermalization the
spatial structure looks different (Fig.~\ref{fig4_compare}(b)).  The
system now consists of the renormalized waves (straight cross-hatch
traces in Fig.~\ref{fig4_compare}(b)). On the top of these waves,
the localized structures similar to DBs manifest themselves
as the wavy dark trajectories (in Fig.~\ref{fig4_compare}(b)).
Although the snapshot (Fig.~\ref{fig4_compare}(d)) of the energy
density plot (Fig.~\ref{fig4_compare}(b)) indicates that in thermal
equilibrium the energy is more evenly distributed among particles,
spatially localized structures are clearly observed.
\begin{figure}
\includegraphics[bb = 50 0 543 603]{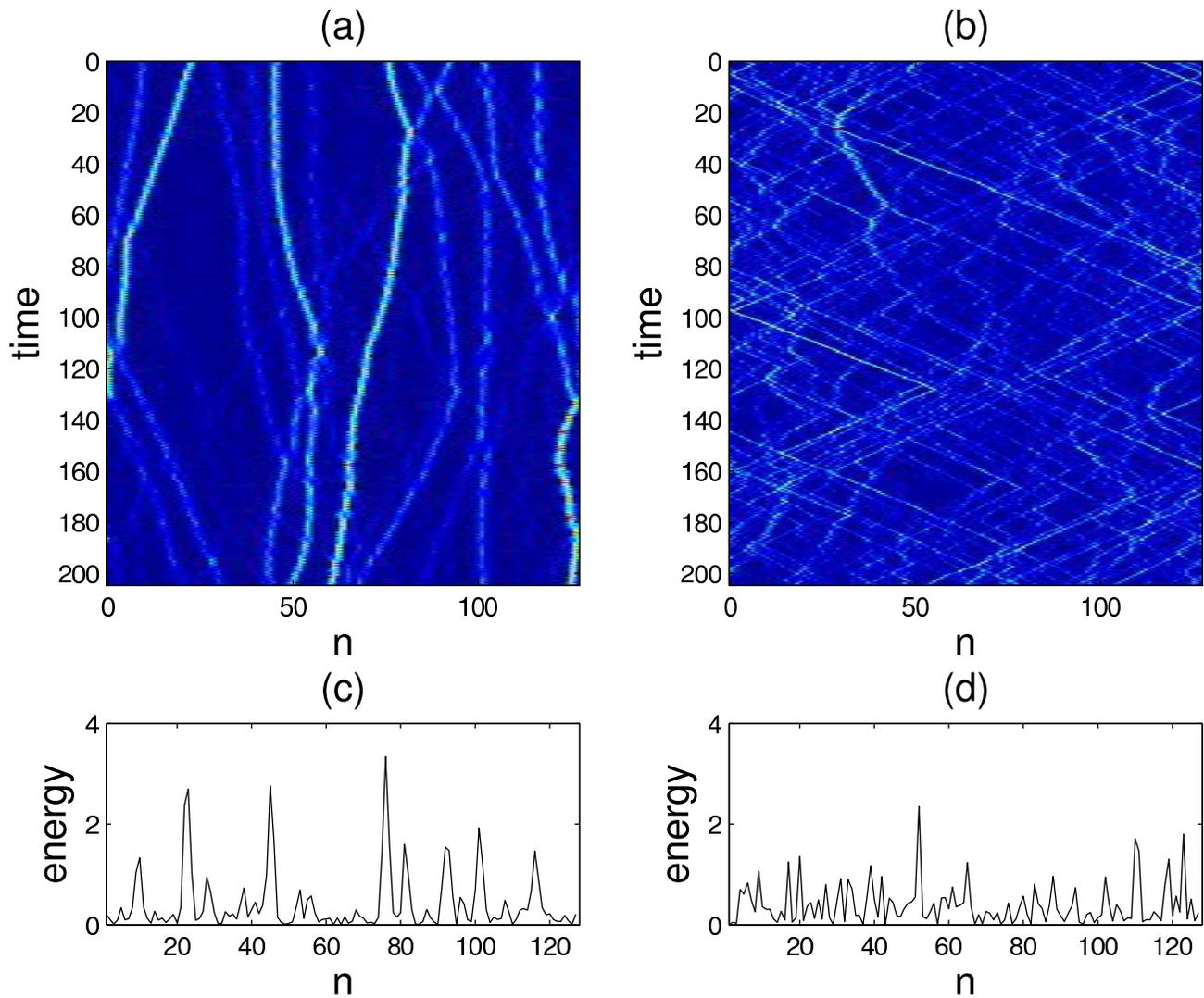}
\caption{Energy density evolution of
 the transient state (a) and (c), of the thermalized state (b) and
 (d), respectively ($\beta=1$ and $H=200$).  In (a) and (b) the darker
 strips correspond to high energy localizations. (c) and (d) are the
 snapshots of energy density.}
 \label{fig4_compare}
\end{figure}

\begin{figure}
\includegraphics[scale=0.8]{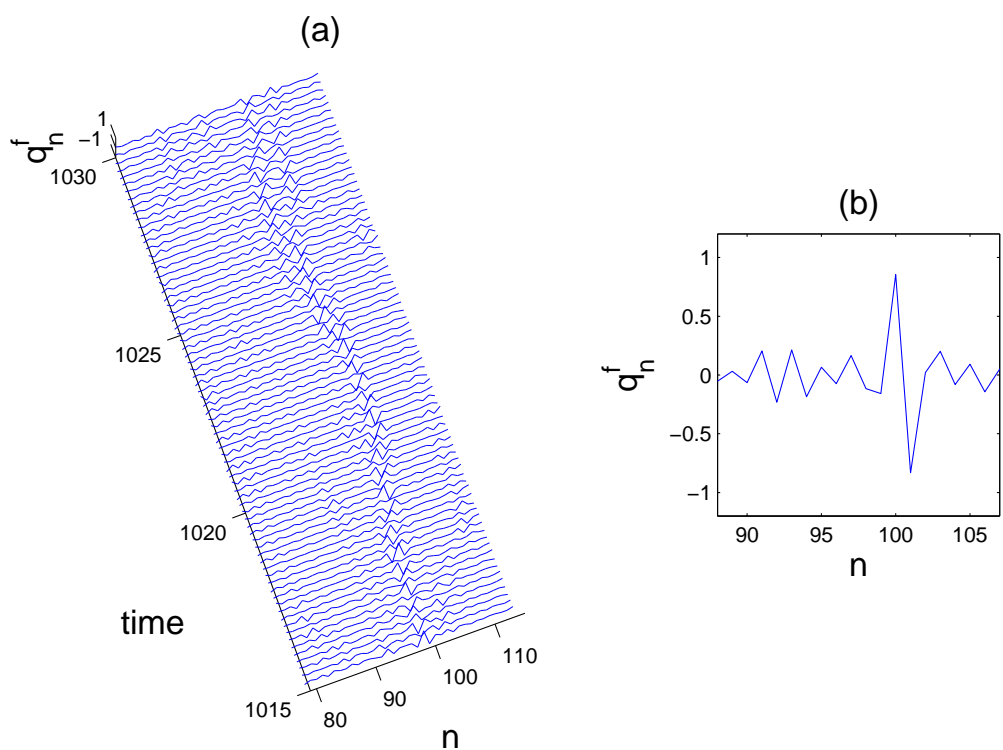}
\caption{(a) Evolution of a discrete breather in thermal
equilibrium. (b) typical snapshot of the breather. $\beta=25$,
$H=200$.} \label{fig5_space_time}
\end{figure}

\begin{figure}
\includegraphics[bb= 20 0 575 575]{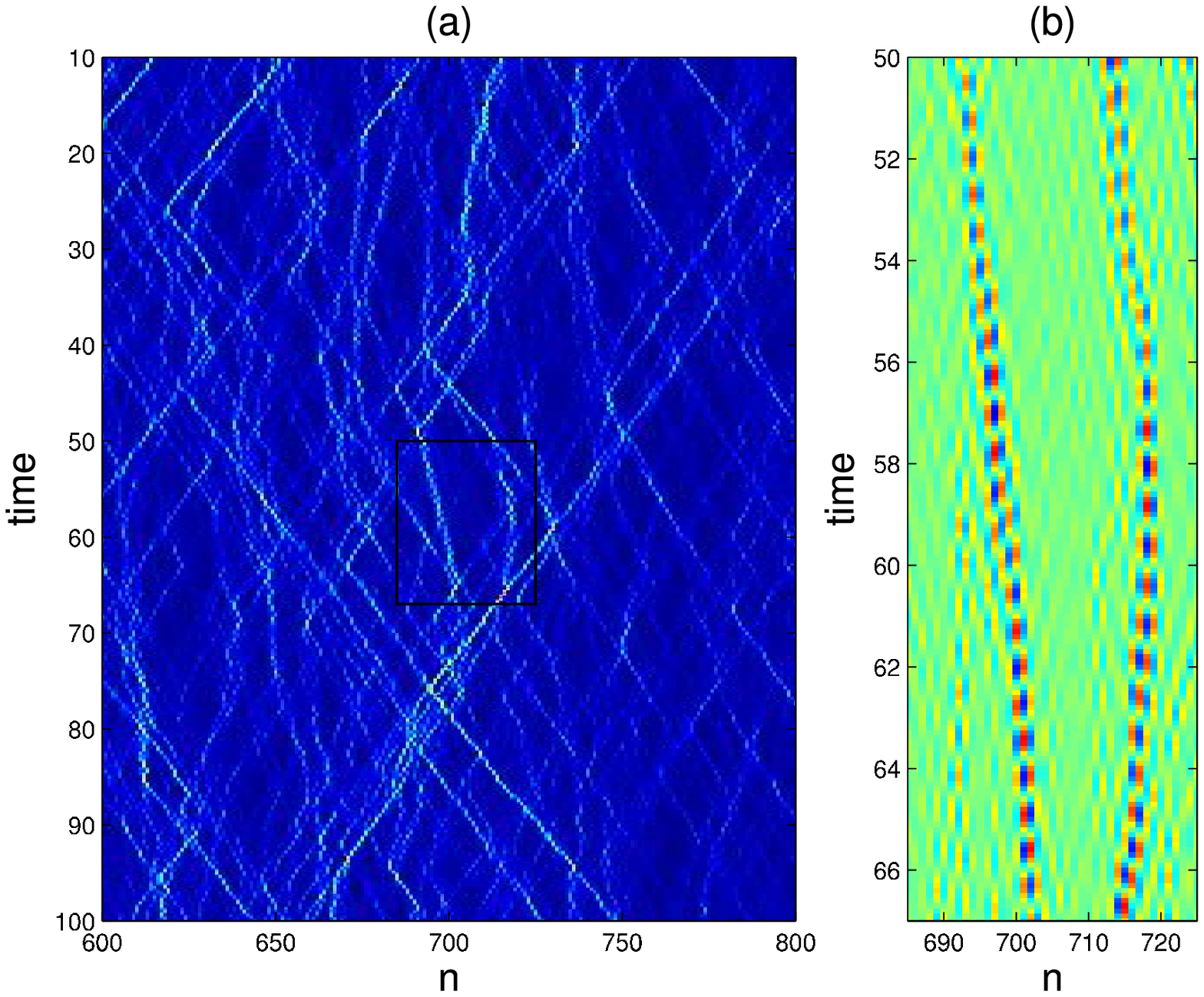}
\caption{Turbulence of discrete breathers (N=1024): (a) evolution of
energy density, (b) zoomed in $q_n^f(t)$ of the area indicated by
the rectangle in (a). } \label{fig7_eqp}
\end{figure}
Since there are renormalized waves in the system, which also carry
energy, we need to find a way to distinguish between these waves and
DBs. We use a frequency filter that cuts out the lower side of the
Fourier spectrum and leaves the high frequency part unmodified,
i.e., $f(g(n,t))=$Re$F^{-1}(H_{\omega}(F(g)))$, where Re denotes the
real part, $F$ is a time Fourier transform, $H_{\omega}$ eliminates
all frequencies below $\omega_{cut}$ and $g(n,t)$ is a dynamical
variable that is being filtered. By applying this filter to the
displacement $q_n$ to obtain $q_n^f\equiv f(q_n)$, we can show the
existence of DBs even for strong nonlinearities, for example,
$\beta=25$. Fig.~\ref{fig5_space_time}(a) shows a clear example of a
DB excitation reconstructed using the filtered $q_n^f$ with
$\omega_{cut}=7$.  Fig.~\ref{fig5_space_time}(b) shows a typical
spatial profile of the DB taken from Fig.~\ref{fig5_space_time}(a),
which strongly resembles the idealized DB~\cite{Cai}. Finally, in
Fig.~\ref{fig7_eqp}, we present the evidence that there is a
\textit{turbulence} of DBs, which chaotically ride on renormalized
waves. The corresponding energy density distribution along with the
distribution of filtered displacement in a zoomed region $q_n^f(t)$
is displayed in Fig.~\ref{fig7_eqp}(a) and (b), respectively.  After
the lower modes from the displacement $q_n$ are filtered, one can
clearly observe that the remaining high frequency oscillations are
spatially highly localized, with the same characteristics as an
idealized breather.  The detailed time dynamics of the DB shows the
main characteristics of breathers: the values of $q_n^f$ change
signs periodically (as indicated by the alternating white and black
spots along the trajectory) as the DB moves in space, with a spatial
span of 2 or 3 sites only, as seen in Fig.~\ref{fig7_eqp}(b).

To summarize, we have demonstrated numerically, that DBs persist in the thermal equilibrium state of the $\beta$-FPU chains.
However, in order to detect them, the linear dynamics of the chain has to be filtered out.
The remaining excitations appear to be localized in space and periodic in time, i.e., they satisfy the definition of DB.

\chapter{Conclusions}
\label{sect_conclusions}

In this thesis, we have investigated one of the most famous problems of the nonlinear science, the celebrated FPU system.
We have studied the problem from both dynamical and statistical angles.

In the beginning of the thesis, we have given some historic notes about the FPU problem as well as
an overview of the topics that are widely used in treating nonlinear problems:
classical mechanics, statistical mechanics, wave turbulence, numerical methods, and chaos.

Then, we turned to the FPU problem and studied the statistical behavior in thermal equilibrium.
We have extended the notion of normal modes to the nonlinear system by showing that regardless of the strength of nonlinearity, the system
in thermal equilibrium can still be
effectively characterized by a complete set of renormalized waves, in the sense that those renormalized waves possess the Rayleigh-Jeans distribution and
vanishing correlations between different wave modes.
In addition, we have studied the property of dispersion relation of the renormalized waves.
The results we obtained in Chapter~\ref{sect_renormalization} are general and can be applied to the large class of nonlinear systems with the nearest-neighbor
interactions in thermal equilibrium.

We have further focused our attention on the $\beta$-FPU chain, which is characterized by the fourth order potential.
We have confirmed that the general renormalization framework that we discussed above is consistent with numerical observations.
In particular, we have shown that the renormalized dispersion of the thermalized $\beta$-FPU chain is in excellent agreement with the numerical
one for a wide range of the nonlinearity strength.
We have further demonstrated that the renormalized dispersion is a direct consequence of the trivial resonant interactions of the renormalized waves.
Using a self-consistency argument, we have found an approximation of the renormalization factor via a mean-field approximation.
In addition, we have used the multiple time-scale, statistical averaging method to obtain the theoretical prediction of the spatiotemporal spectrum
and demonstrated that the renormalized waves have long lifetimes.

Moreover, we studied the FPU system from the particle interaction point of view.
In particular, we have investigated the existence of the discrete breather excitations in the $\beta$-FPU chain.
We have numerically demonstrated that the discrete breather solutions that were observed previously in the transient to the thermal equilibrium still persist
even when the system reaches thermal equilibrium.

\specialhead{LITERATURE CITED}

\appendix
\chapter{KdV equation as a continuous approximation of the FPU chains}
\label{app_kdv}
In this Appendix, we derive the KdV equation as a continuous approximation of the $\a$-FPU chains~\cite{Nizan}.
Consider an $\a$-FPU chain, which is given by a Hamiltonian
\BEA
H=\sum_{j=1}^N\frac{mp_j^2}{2}+P(q_{j+1}-q_j),
\EEA
where the potential is of the form
\BEA
P(s)=\frac{\kappa}{2}s^2+\frac{\a}{3}s^3.
\EEA
Then the equation of motion becomes
\BEA
m\ddot{q}_j=\kappa(q_{j+1}-2q_j+q_{j-1})+\a\left((q_{j+1}-q_j)^2-(q_j-q_{j-1})^2\right).\label{app_eq1}
\EEA
Let us rescale time $t$ and the nonlinearity parameter $\a$ via
\BEA
t&\rightarrow&\sqrt{\frac{k}{m}}t,\\
\a&\rightarrow&\frac{1}{\kappa}\a.
\EEA
Now, Eq.~(\ref{app_eq1}) becomes
\BEA
\ddot{q}_j=(q_{j+1}-2q_j+q_{j-1})+\a\left((q_{j+1}-q_j)^2-(q_j-q_{j-1})^2\right).\label{app_eq2}
\EEA
Next, we denote $y_j=q_j-q_{j-1}$ and rewrite Eq.~(\ref{app_eq2}) as
\BEA
\ddot{y}_j=F(y_{j+1})-2F(y_j)+F(y_{j-1}),\label{app_eq3}
\EEA
where
\BEA
F(s)=s+\a s^2.
\EEA
Suppose the chain has length $L$ and spacing $h$. We consider the continuous limit $N\rightarrow\infty$ and $h\rightarrow 0$ such that $Nh=\mbox{const}=L$.
Denote $x=nh$ and $y(x)\equiv y_n$. Then, $y(x)$ can be regarded as a function of the continuous-valued variable $x$.
Using the Taylor expansion, we have
\BEA
F(y_{j+1})&=&F(y(nh+h))=F(y(x))+h\frac{\p F(y(x))}{\ x}+\frac{1}{2}h^2\frac{\p^2 F(y(x))}{\p x^2}+\dots\nonumber\\
&\equiv&e^{h\p_x}F(y_j).\nonumber\\
\EEA
Using this formal notation, Eq.~(\ref{app_eq3}) can be rewritten as
\BEA
\ddot{y}(x)=(e^{h\p_x}+e^{-h\p_x}-2)F(y(x)),
\EEA
or, equivalently
\BEA
\ddot{y}(x)=4\sinh^2\left(\frac{h}{2}\p_x\right)F(y(x)).\label{app_eq4}
\EEA
Next, we use the Taylor expansion of the $\sinh$ function up to $O(h^4)$ and neglect $O(\a h^4)$ and $O(h^5)$.
Here we use $\a\ll 1$.
Equation~(\ref{app_eq4}) becomes
\BEA
\ddot{y}(x)=4\Big[\left(\frac{h}{2}\p_x\right)^2+\frac{1}{3}\left(\frac{h}{2}\p_x\right)^4\Big](y(x)+\a y(x)^2).\label{app_eq5}
\EEA
After applying the differential operators, Eq.~(\ref{app_eq5}) takes the form
\BEA
\ddot{y}=h^2y_{xx}+\a h^2(y_x^2)_{xx}+\frac{h^4}{12}y_{xxxx}.\label{app_eq6}
\EEA
Let us rescale time, space and displacement variables via
\BEA
t&\rightarrow&\sqrt{12}t,\\
x&\rightarrow&\frac{\sqrt{12}}{h}x,\\
y&\rightarrow&\a y.
\EEA
Then, Eq.~(\ref{app_eq6}) becomes
\BEA
\ddot{y}=(y+y^2+y_{xx})_{xx}.\label{app_eq7}
\EEA
Now, let is return back to the initial variables $u_n$.
Similarly to $y(x)$, we denote $u(x)$ to be a function of a continuous variable $x$
\BEA
u(x)=u(nh)\equiv u_n.
\EEA
Then, using the same rescaling $u\rightarrow\a u$, we obtain the following connection between $y$ and $u$
\BEA
y=\sqrt{12}u_x.\label{app_yu}
\EEA
Combining Eqs.~(\ref{app_eq7}) and~(\ref{app_yu}), we obtain the Boussinesq equation
\BEA
\ddot{u}=(1+2u_x)u_{xx}+u_{xxxx}.\label{app_bous1}
\EEA
Let us first consider the case without the dispersion.
We can rewrite Eq.~(\ref{app_bous1}) as
\BEA
\ddot{u}-H^2(u_x)u_{xx}=0,\label{app_bous2}
\EEA
where $H(s)=\sqrt{1+2s}$.
Next, we introduce new variables
\BEA
\BC
w=u_x,\\
v=u_t.
\EC
\label{app_wv}
\EEA
Then Eq.~(\ref{app_bous2}) becomes a system
\BEA
\BC
w_t-v_x=0,\\
v_t-H^2(w)w_x=0.
\EC
\label{app_sys1}
\EEA
Let us make a linear transformation of Eq.~(\ref{app_sys1}) in the following way.
First, we add the first equation multiplied by $H$ with the second one.
Secondly, we subtract the second equation from the first multiplied by $H$.
Thus, we obtain
\BEA
\BC
Hw_t-Hv_x+v_t-H^2w_x=0,\\
-v_t+Hw_t-Hv_x+H^2w_x=0.
\EC
\label{app_sys2}
\EEA
System~(\ref{app_sys2}) is equivalent to the following system
\BEA
\BC
s_t-Hs_x=0,\\
r_t+Hr_x=0,
\EC
\label{app_sys3}
\EEA
where
\BEA
s=v+\int_0^wH(\eta)d\eta,\nonumber\\
r=-v+\int_0^wH(\eta)d\eta.\nonumber
\EEA
Let us express $w$ in terms of $r$ and $s$
\BEA
r+s=2\int_0^wH(\eta)d\eta\equiv 2G(w).\label{app_G}
\EEA
Therefore, we have
\BEA
w=G^{-1}\left(\frac{r+s}{2}\right).
\EEA
Then, the second equation in~(\ref{app_sys3}) becomes
\BEA
r_t+H\Bigg(G^{-1}\left(\frac{r+s}{2}\right)\Bigg)=0
\EEA
Now, we consider the case with the dispersion.
The same procedure gives us
\BEA
\BC
s_t-Hs_x=w_{xxx},\\
r_t+Hr_x=-w_{xxx},
\EC
\label{app_sr}
\EEA
The LHS of the first equation in~(\ref{app_sr}) is a complete derivative along the characteristics
\BEA
\frac{ds}{dt}=w_{xxx},\label{app_dsdt}
\EEA
along
\BEA
\frac{dx}{dt}=-H.
\EEA
in Eq.~(\ref{app_dsdt}), we assume that the RHS is small for long waves.
Therefore, $s$ is a constant. Without loss of generality, we assume that $s=0$.
Then, we have
\BEA
G(w)=\int_0^w\sqrt{1+2\eta}d\eta=\frac{1}{3}\left((1+2w)^{3/2}-1\right)=\frac{r+s}{2}.
\EEA
Taking into account that $s=0$, we find
\BEA
w=G^{-1}\left(\frac{r}{2}\right)=\frac{1}{2}\left(\Big(\frac{3r}{2}+1\Big)^{2/3}-1\right).
\EEA
Therefore, we obtain
\BEA
F(w)=\left(\frac{3r}{2}+1\right)^{1/3}.
\EEA
For small $u$ we have the following approximations
\BEA
w&=&\frac{r}{2},\nonumber\\
H&=&\frac{r}{2}+1.\nonumber\\
\label{app_wH}
\EEA
Substituting Eqs.~(\ref{app_wH}) into the second equation in~(\ref{app_sr}), we have
\BEA
r_t+\left(\frac{r}{2}+1\right)r_x+\frac{1}{2}r_{xxx}=0.
\EEA
And finally, making the following changes of variables $x\rightarrow x-t$ and then $t\rightarrow t/2$, we obtain the KdV equation
\BEA
r_t+rr_x+r_{xxx}=0.
\EEA
Therefore, we have shown that the KdV equation is indeed a continuous approximation of the $\a$-FPU chains in the small amplitude and long wavelength regime.
Similarly, one can show that the so called modified KdV equation
\BEA
r_t+r^2r_x+r_{xxx}=0
\EEA
is a continuous approximation of the $\b$-FPU chain.
\chapter{Computation of the Lyapunov exponent of the $\b$-FPU chain}
\label{app_lyap}
In order to investigate the chaotic structure of the $\beta$-FPU chain given by Eq.~(\ref{H_FPU}), we measure the Lyapunov exponent.
We consider small perturbations $\d_j$ and $\ve_j$ of the dynamical variables $q_i$ and $p_i$, respectively.
As we have studied in Section~\ref{sect_sensitive}, a system exhibits chaotic behavior if the small perturbations of the dynamical variables grow exponentially
with time.
In order to describe the dynamical behavior of the perturbations $\d_j$ and $\ve_j$, we linearize Eqs.~(\ref{dyn_pq_1}) and then numerically study their evolution
\BEA
\dot{\d}_j&=&\ve_j,\nonumber\\
\dot{\ve}_j&=&(\d_{j+1}-2\d_j+\d_{j-1})-3\beta((q_j-q_{j+1})^2(\d_j-\d_{j+1})+(q_j-q_{j-1})^2(\d_j-\d_{j-1})).\nonumber\\
\label{dyn_ed}
\EEA
Note that we have to solve Eqs.~(\ref{dyn_pq_1}) together with Eqs.~(\ref{dyn_ed}).
The procedure of computing the Lyapunov exponent is the following.
We take some initial condition $(q_i(0),p_i(0))$ and initial perturbation $(\ve_i(0),\d_i(0))$.
We choose the perturbation to be very small, e.g. its $l_2$ norm (which we denote as $d_0$) is of the order $10^{-10}$.
Then, let the dynamical variables $p_j$ and $q_j$ and the perturbations $\d_j$ and $\ve_j$ evolve for $n$ time units according to Eqs.~(\ref{dyn_pq_1}) and
Eqs.~(\ref{dyn_ed}), respectively.
After $n$ time units, we measure the norm of the perturbation again (denote it as $d_1$).
Let us call
\BEA
h_1=\frac{1}{n}\log\frac{d_1}{d_0}.
\EEA
Now, the perturbation has to be rescaled to make it have the norm $d_0$ again.
The direction of the perturbation vector $(\ve,\d)$ has to stay the same, as it was before rescaling, only the absolute value takes its initial value.
After we repeat this procedure $m$ times we obtain the values of $h^{(m)}=\frac{1}{m}\sum_{s=1}^mh_s$.
The Lyapunov exponent can be then estimated as
\BEA
h=\lim_{m\rightarrow\infty}h^{(m)}=\lim_{m\rightarrow\infty}\frac{1}{m}\sum_{j=1}^{m}h_j.
\EEA
Note that the forth order Runge-Kutta method can be used for solving Eqs.~(\ref{dyn_ed}).

Therefore, we have presented a numerical algorithm of computing the Lyapunov exponent of the $\b$-FPU chain. 

\end{document}